\documentclass{aa} 

\pdfoutput=1

\usepackage{subfig}

\usepackage{amsmath}
\usepackage{amsfonts}
\usepackage{amsxtra}
\usepackage[draft]{hyperref}

\usepackage{amssymb}

\usepackage{url}

\usepackage{graphicx,epsfig}
\usepackage{dcolumn}
\usepackage{bm}

\usepackage{color}
\usepackage{longtable}
\usepackage{lscape}
\newcommand{\Fermi}{\textit{Fermi}}

\newcommand{\tot}{\textrm{tot} }

\newcommand{\onepic}{0.4}

\usepackage{array}
\newcolumntype{L}[1]{>{\raggedright\let\newline\\\arraybackslash\hspace{0pt}}m{#1}}
\newcolumntype{C}[1]{>{\centering\let\newline\\\arraybackslash\hspace{0pt}}m{#1}}
\newcolumntype{R}[1]{>{\raggedleft\let\newline\\\arraybackslash\hspace{0pt}}m{#1}}

\begin{document}

\title{Search for Gamma-ray Emission from Galactic Novae with the \Fermi\,-LAT}

   \author{A. Franckowiak
          \inst{1}
          \and
          P. Jean\inst{2,}\inst{3}
          \and          
          M. Wood\inst{4}
          \and
          C. C. Cheung\inst{5}
          \and
          S. Buson\inst{6,}\inst{7}
          }

   \institute{Deutsches Elektronen Synchrotron DESY, D-15738 Zeuthen, Germany\\
              \email{anna.franckowiak@desy.de}
         \and
             CNRS, IRAP, F-31028 Toulouse cedex 4, France
             \and
             GAHEC, Universit\'{e} de Toulouse, UPS-OMP, IRAP, Toulouse, France
             \and
             W.~W.~Hansen Experimental Physics Laboratory, Kavli Institute for Particle Astrophysics and Cosmology, Department of Physics and SLAC National Accelerator Laboratory, Stanford University, Stanford, CA 94305, USA
             \and
             Space Science Division, Naval Research Laboratory, Washington, DC 20375-5352, USA
             \and
             NASA Goddard Space Flight Center, Greenbelt, MD 20771, USA
             \and
             NASA Postdoctoral Program Fellow, USA
                      }

\date{Received MONTH DAY, YEAR; accepted MONTH DAY, YEAR}

  \abstract
   {A number of novae have been found to emit high-energy gamma rays (> 100 MeV). However, the origin of this emission is not yet understood. We report on the search for gamma-ray emission from 75 optically-detected Galactic novae in the first 7.4 years of operation of the  \Fermi\, Large Area Telescope using the Pass 8 data set.}
   {We compile an optical nova catalog including light curves from various resources and estimate the optical peak time and optical peak magnitude in order to search for gamma-ray emission to test if all novae are gamma-ray emitters.}
   {We repeat the analysis of the six novae previously identified as gamma-ray sources and develop a unified analysis strategy which we then apply to all novae in our catalog. We search for emission in a 15-day time window in two-day steps ranging from 20 days before to 20 days after the optical peak time. We perform a population study with Monte Carlo simulations to set constraints on the properties of the gamma-ray emission of novae.}
   {Two new novae candidates have been found at $\sim2\sigma$ global significance. Although these two novae candidates were not detected at a significant level individually, taking them together with the other non-detected novae, we found a sub-threshold nova population with a cumulative $3\sigma$ significance. We report the measured gamma-ray flux for detected sources and flux upper limits for novae without significant detection. Our results can be reproduced by several gamma-ray emissivity models (e.g. a power-law distribution with a slope of 2), while a constant emissivity model (i.e. assuming novae are standard candles) can be rejected.}
   {}

\maketitle

\section{Introduction}
\label{sec:intro}

The Large Area Telescope \citep[LAT;][]{2009ApJ...697.1071A} onboard the \Fermi\ spacecraft has detected gamma-ray emission from six Galactic novae~\citep{2014Sci...345..554A,2016ApJ...826..142C} in the time period from the start of the mission in August 2008 until end of 2015. Novae are runaway thermonuclear explosions on the surface of a white dwarf in a binary system, that accretes matter from its stellar companion. While the first detection was from a symbiotic-like nova \citep{2010Sci...329..817A}, the following five novae were classified as classical novae. Symbiotic novae have an evolved companion (e.g., red giant) with a dense wind as opposed to a main-sequence companion for classical novae. Both leptonic and hadronic models attempt to explain the gamma-ray emission processes. An open question is whether all novae are gamma-ray emitters.

While previous works analyzing gamma-ray emission from novae have been performed with the Pass\,6~\citep{2010Sci...329..817A} and Pass\,7~\citep{2014Sci...345..554A,2016ApJ...826..142C} data sets, this work utilizes the recent Pass\,8 data set, which significantly improves the sensitivity allowing us to detect the gamma-ray emission of novae not seen with previous analyses. For the first time we report gamma-ray fluxes and flux upper limits for a large sample of novae analyzed in a unified way.

In the following we first present the optical novae catalog in Sec.~\ref{sec:NovaCat}, followed by a description of the unified search for gamma rays (Sec.~\ref{sec:gamma}) motivated by the LAT-measured spectral features and durations of the previously detected novae. Results of the unified search for gamma rays are discussed in Sec.~\ref{sec:res}. In Sec.~\ref{sec:pop} we use the results to perform a population study with Monte Carlo simulations to set constraints on the properties of the gamma-ray emission of novae. We conclude in Sec.~\ref{sec:conclusion}.

\section{Optical Nova Catalog}
\label{sec:NovaCat}

We compiled a list of 75 optical novae from Astronomer's Telegrams (ATels)\footnote{\url{http://www.astronomerstelegram.org/}}, Central Bureau for Astronomical Telegrams (CBETs)\footnote{\url{http://www.cbat.eps.harvard.edu/cbet/RecentCBETs.html}}, International Astronomical Union Circulars (IAUC)\footnote{\url{http://www.cbat.eps.harvard.edu/iauc/RecentIAUCs.html}} and a catalog of novae published by the Optical Gravitational Lensing Experiment (OGLE) team~\citep{OGLEREF1,OGLEREF2}
in the time range from August 2008 to end of 2015. Additional light curve information have been collected from the American Association of Variable Star Observers (AAVSO)\footnote{\url{http://www.aavso.org/lcg}} and the Small and Medium Aperture Telescope System (SMARTS)\footnote{\url{http://www.astro.sunysb.edu/fwalter/SMARTS/NovaAtlas/atlas.html}}~\citep{2012PASP..124.1057W}. 

\begin{figure*}[htp]
  \centering
 \includegraphics[scale=0.5]{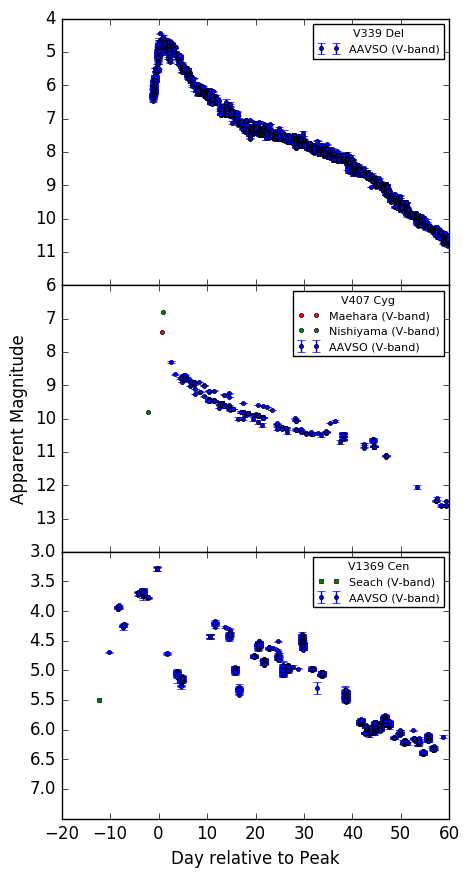}
 \includegraphics[scale=0.5]{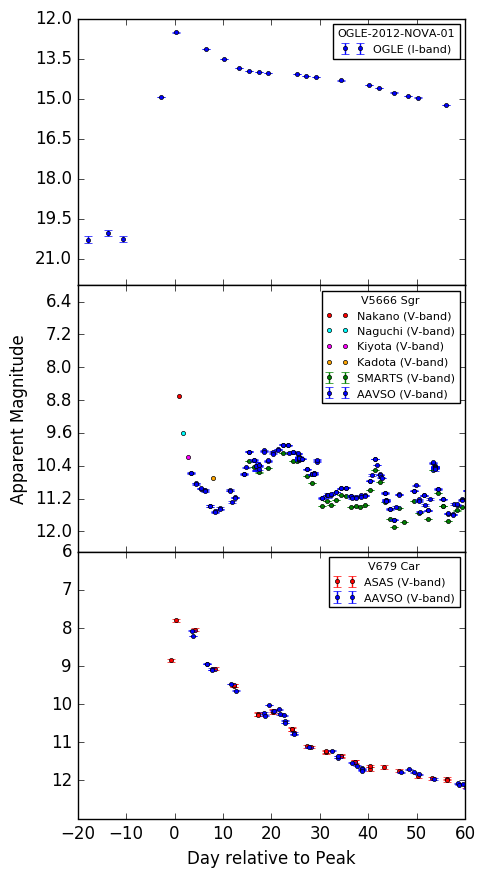}
  \caption{\small Example optical light curves for 6 out of the 75 inspected novae. Observations reported by different telescopes and observers are shown in different colors.}
 \label{fig:novaLC}
\end{figure*}

We use the light curves to estimate the optical peak time and optical peak apparent magnitude for each novae (see Tab.~\ref{tab:novaeList}). There are many cases with a single optical peak around the discovery, and smooth subsequent decline. Many cases also have multiple peaks, with some cases where later peaks are modestly brighter, thus the latter dates are quoted (e.g., V1369 Cen 2013; V5668 Sgr 2015). Example light curves are shown in Fig.~\ref{fig:novaLC}. A full list of novae including the optical peak information is shown in Tab.~\ref{tab:novaeList}. Other nova lists by Bill Gray\footnote{\url{http://projectpluto.com/galnovae/galnovae.txt} (accessed on March 1, 2017)} and Koji Mukai\footnote{\url{https://asd.gsfc.nasa.gov/Koji.Mukai/novae/novae.html} (accessed on March 1, 2017)} have been checked for consistency with our list for the time window of August 2008 to end of 2015. All novae in those lists are included in our list. Our list has a few additional sources, e.g., from the OGLE survey.

\begin{figure}[htp]
  \centering
 \includegraphics[scale=0.4]{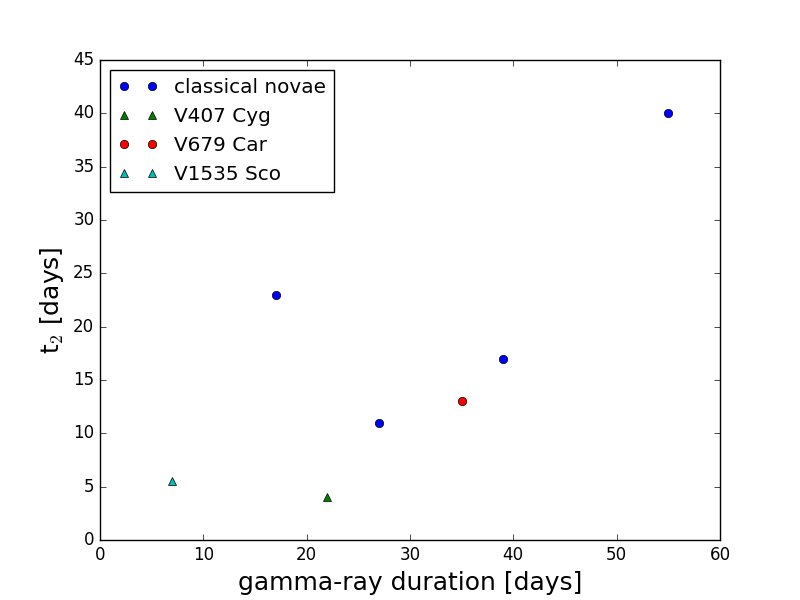}
  \caption{\small Duration of $>100$\,MeV gamma-ray emission as a function of $t_2$ for the gamma-ray detected novae (candidates).}
 \label{fig:t2VsDur}
\end{figure}

Not all light curves are sufficiently well-sampled to allow an accurate determination of the peak time and optical peak apparent magnitude. Those are excluded in some of our later studies. In addition to the peak time we also estimate $t_2$, the time during which the visual light curve fades by 2 mag from the maximum~\citep{1964gano.book.....P}. Again, not all light curves allowed the estimate of $t_2$. For the known gamma-ray novae we do not find a correlation between $t_2$ and the duration of the gamma-ray emission (see Fig.~\ref{fig:t2VsDur}).

\begin{figure*}
\begin{center}
\includegraphics[scale=0.55]{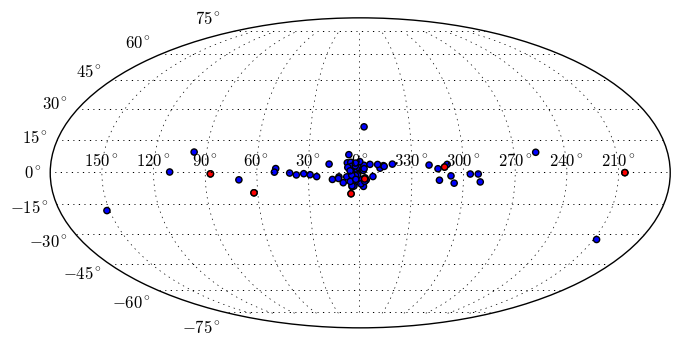}
\includegraphics[scale=0.35]{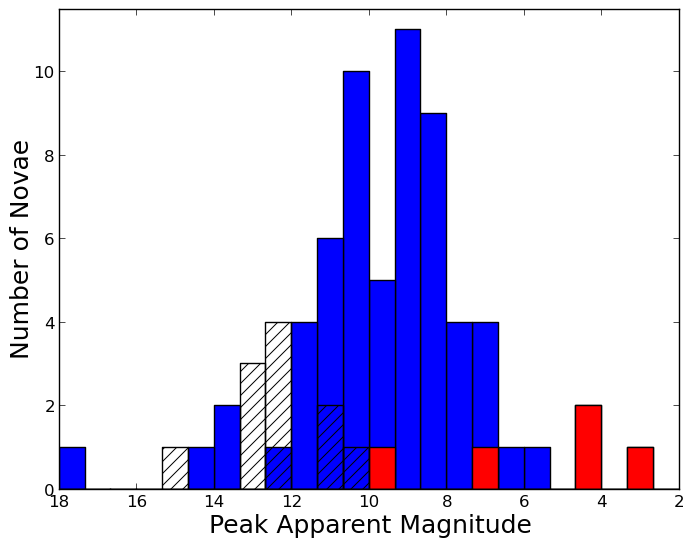}
\noindent
\caption{\small Novae selected in a time range from August 2008 to the end of 2015. Left: Spatial distribution in Galactic coordinates of the novae listed in Tab.~\ref{tab:novaeList}. Previously detected gamma-ray novae are shown in red. Right: Optical peak apparent magnitude distribution of the novae, magnitudes of the five out of the six previously detected gamma-ray novae are overlayed in red. The peak magnitude of V959\,Mon is not included because it has not been measured since the nova was too close to the Sun to be observed during its outburst. Most magnitudes are in V-band. Peak magnitudes of novae which were only observed in I-band are displayed by striped bars.}
\label{fig:novaDist}
\end{center}
\end{figure*}

As expected the spatial distribution of novae peaks in the Galactic bulge region and along the Galactic plane (see Fig.~\ref{fig:novaDist}, left). 
A slight asymmetry shows up in the spatial distribution of novae in the bulge. More novae are discovered at negative than at positive latitudes, which is likely due to the non-uniform exposure of the OGLE experiment~\citep[see Fig. 9 of][]{OGLEREF2}. There are three outliers with Galactic latitude of $|b|>15^\circ$ -- KT\,Eri 2009, U\,Sco 2010 and V965\,Per 2011 at latitudes of $-32.0^\circ$, $21.9^\circ$ and $17.9^\circ$ respectively.

The optical peak apparent magnitude distribution is shown in the right panel of Fig.~\ref{fig:novaDist}.

Note that more gamma-ray novae have been detected in 2016, but are not included in this work~\citep[see e.g.][]{ATel9736,ATel9594}.

\section{Gamma-ray Data Analysis}
\label{sec:gamma}

The \Fermi-LAT is a pair-conversion telescope sensitive to gamma rays with energies from $20\,$MeV to greater than $300\,$GeV~\citep{2009ApJ...697.1071A}. It has a large field of view and scans the entire sky every three hours. The LAT has been operated continuously since August 2008. Thus it is very well suited for searches for transient gamma-ray signals on the timescale of days to weeks. 

In this analysis we use 7.4 years of \Fermi-LAT data recorded from the start of operations in 2008 August 4 to 2015 December 31 (\textit{Fermi} Mission Elapsed Time 239557418--473212804 s),
restricted to the Pass\,8 Source class\footnote{\url{
http://fermi.gsfc.nasa.gov/ssc/data/analysis/documentation/Pass8_usage.html}}. We select the standard good time intervals (e.g., excluding time intervals when the field of view of the LAT intersected the Earth). The Pass\,8 data benefit from improved reconstruction and event selection algorithms with respect to the previous data release Pass\,7 leading to a significantly improved angular resolution and sensitivity~\citep{2013arXiv1303.3514A}. To further increase the sensitivity Pass\,8 allows the division of the data set in four point spread function (PSF) classes that can be used in a joint likelihood analysis (see~\citet{2015arXiv150302641F} for more details). The PSF classes are called PSF0, PSF1, PSF2 and PSF3, where PSF3 has the best and PSF0 the worst angular resolution.

\subsection{Method}
We perform a binned analysis (i.e., binned in space and energy) using the standard \Fermi-LAT ScienceTools package version v10r01p01 available from the \textit{Fermi} Science Support Center\footnote{\url{http://fermi.gsfc.nasa.gov/ssc/data/analysis/}} (FSSC) and the P8R2\_SOURCE\_V6 instrument response functions. We analyze data in the energy range of 100\,MeV to 100\,GeV binned into 24 logarithmic energy intervals equally spaced in log. We restrict the class PSF0 to the energy range of 1\,GeV to 100\,GeV to avoid using a very large region of interest (ROI) that contains events with very poor angular resolution. Note that those events do not contribute significantly to the detection significance. To minimize the contamination from the gamma rays produced in the Earth's upper atmosphere, we apply a zenith angle cut of $\theta<100^\circ$ for PSF0 (note that the restricted energy range for PSF0 allows us to use a relaxed zenith cut), $\theta<85^\circ$ for PSF1, $\theta<85^\circ$ for PSF2 and $\theta<90^\circ$ for PSF3 following in the recommendations of the \Fermi-LAT collaboration. The effect of energy dispersion is included in the fits performed with the \Fermi-LAT ScienceTools.

For each source and each time window we select a $10^\circ \times 10^\circ$ region of interest (ROI) centered on the source position binned in $0\fdg1$ size pixels. The binning is applied in celestial coordinates and an Aitoff projection was used. We construct a model whose free parameters are fitted to the data in the ROI. This model includes a point source at the nova position; its gamma-ray spectrum is represented as a power-law function with exponential cutoff, which was found to describe the spectra of the gamma-ray novae well~\citep{2014Sci...345..554A},

\begin{equation}
\frac{dN}{dE} = N \left(\frac{E}{E_0}\right)^{-\Gamma} \exp \left(-\frac{E}{E_c}\right)
\label{eq:PLExp}
\end{equation}
with normalization $N$, energy cutoff $E_c$ and photon index $\Gamma$. The energy scale is fixed to $E_0 = 1$\,GeV. In addition we model the background point sources in the ROI and the isotropic and Galactic diffuse gamma-ray emission. We are using preliminary versions of the Pass 8 Galactic interstellar emission model and the isotropic spectral template, which slightly differ from the versions published by the \textit{Fermi}-LAT collaboration. 
The spatial model of the used Galactic interstellar emission model is similar to the published one, while the spectral shape of both the Galactic interstellar emission model and the isotropic spectral template shows small relative deviation of $\sim5\%$. We have verified that the results are unchanged if the official versions for the Pass 8 Galactic interstellar emission model \textsf{gll\_iem\_v06.fits}\footnote{\url{https://fermi.gsfc.nasa.gov/ssc/data/access/lat/BackgroundModels.html}}~\citep{Acero:2016qlg} and the isotropic spectral template \textsf{iso\_P8R2\_SOURCE\_V6\_PSF0\_v06.txt}, for PSF0 and the equivalent models for the other PSF classes, are used. We include in the model 3FGL sources~\citep{2015ApJS..218...23A} within a larger region of $15^\circ \times 15^\circ$, to account for the contribution of sources outside the ROI that leak into our selection due to the broad PSF. First we refit the spectral parameters of all background sources. Then we fix all sources except the normalization of sources within $3^\circ$ from the nova position.

\subsection{Analysis of Known Gamma-Ray Novae -- Unified Spectrum}
\label{subsec:knownNovae}

We repeat the analysis of the six known gamma-ray novae with the Pass\,8 data set to develop a unified analysis strategy that can then be applied to the complete catalog of optical novae described in Sec.~\ref{sec:NovaCat}.
We use a similar gamma-ray search time window as was used in the Pass\,7 analysis~\citep{2014Sci...345..554A,2016ApJ...826..142C}. 

\begin{table*}
\caption{Analysis of known Gamma-ray Novae with Pass 8}
\label{tab:knownNovae}
\centering          
\begin{tabular}{L{0.14\linewidth} p{0.08\linewidth} p{0.08\linewidth} p{0.08\linewidth} p{0.04\linewidth} p{0.1\linewidth} p{0.08\linewidth} p{0.13\linewidth}}
\hline\hline 
Nova & $t_{\textrm{start}}$ &  $t_{\textrm{stop}}$ & Duration & TS & Index & Cutoff  & Flux \\
         &        [MET]               & [MET]                      & [days]     &      &  & [GeV]  & [$10^{-7}$\,cm$^{-2}\,$s$^{-1}$] \\
\hline\\
V407\,Cyg\,2010                        & 289872002 & 291772802 & 22 & 526.6 &	1.27 $\pm$ 0.18 &	2.0 $\pm$ 0.5    & $3.47\pm0.44$ \\
V1324\,Sco\,2012                      & 361411202 & 362880003 & 17 & 185.8 &	1.92 $\pm$ 0.16 &	7.7 $\pm$ 4.7    & $4.40\pm0.85$\\
V959\,Mon\,2012                       & 361756802 & 363657603 & 22 & 193.7 &	1.50 $\pm$ 0.28 &	1.3 $\pm$ 0.5    & $2.64\pm0.45$\\
V339\,Del\,2013                         & 398304003 & 400636803 & 27 & 364.0 &	1.68 $\pm$ 0.22 &	3.0 $\pm$ 1.8    & $1.45\pm0.19$\\
V1369\,Cen\,2013                      & 407894403 & 411264003 & 39 & 129.3 &      2.00 $\pm$ 0.26 &	2.0 $\pm$ 1.0    & $2.51\pm0.52$\\
V5668\,Sgr\,2015                       & 448588803 & 453340803 & 55 & 87.5   &      2.11 $\pm$ 0.12 &    --   & $0.61\pm0.13$\\  
\hline
combined (all)                            &                    &                    &      &      &       1.71 $\pm$ 0.08 &    3.2 $\pm$ 0.6    & \\
combined (all, exc. V407\,Cyg)  &                    &                    &       &     &       1.90 $\pm$ 0.08   &    4.3 $\pm$ 1.2    & \\
\hline                  
\end{tabular}
\tablefoot{We adopted the time windows from previous analyses performed by~\citet{2014Sci...345..554A} and ~\citet{2016ApJ...826..142C}, which differ from the 15-days time window used for the results presented in Appendix~\ref{appendixA}. The energy cutoff for V5668\,Sgr is not constrained by the fit. We therefore present the results of a simple power-law fit without a cutoff for which we find that the TS value decreases only insignificantly from 89.0 to 87.5.} 
\end{table*}

We model the nova energy spectrum with a power-law function with an exponential cutoff (see Eq.~\ref{eq:PLExp}), where we allow all spectral parameters (except $E_0$) to be free in the fit. The resulting best fit spectral parameters and test statistic value for the six novae are summarized in Tab.~\ref{tab:knownNovae}. The test statistic, $TS$, is defined as follows~\citep{1996ApJ...461..396M}:

\begin{equation}
TS=-2\Delta \log{\mathcal{L}} = -2 (\log{\mathcal{L}_0} - \log{\mathcal{L}} ),
\label{eq:TS}
\end{equation}
where $\mathcal{L}_0$ is the likelihood evaluated at the best-fit parameters under a background-only, null hypothesis, i.e.~a model that does not include a point source at the nova position, and $\mathcal{L}$ the likelihood evaluated at the best-fit model parameters when including a candidate point source at the nova position. If including a candidate source adds one degree of freedom to the model one can interpret $\sqrt{TS}$ as the number of $\sigma$ of a normal distribution.

The gamma-ray spectral parameters are the same for all novae within the statistical errors, except for the symbiotic nova V407Cyg, which has a harder spectrum (i.e.\ a smaller photon index). In symbiotic novae the dense wind of the red giant companion might influence the process of particle acceleration explaining a different spectral behavior from the classical novae. All the Pass\,8 derived spectral parameters are consistent with the previous Pass\,7 results.

\begin{figure*}[htb!]
\begin{center}
\includegraphics[scale=0.43]{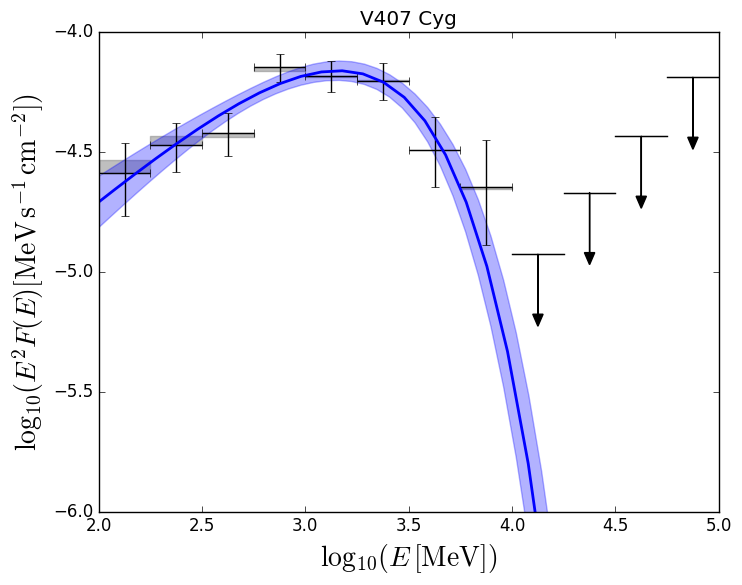}
\includegraphics[scale=0.43]{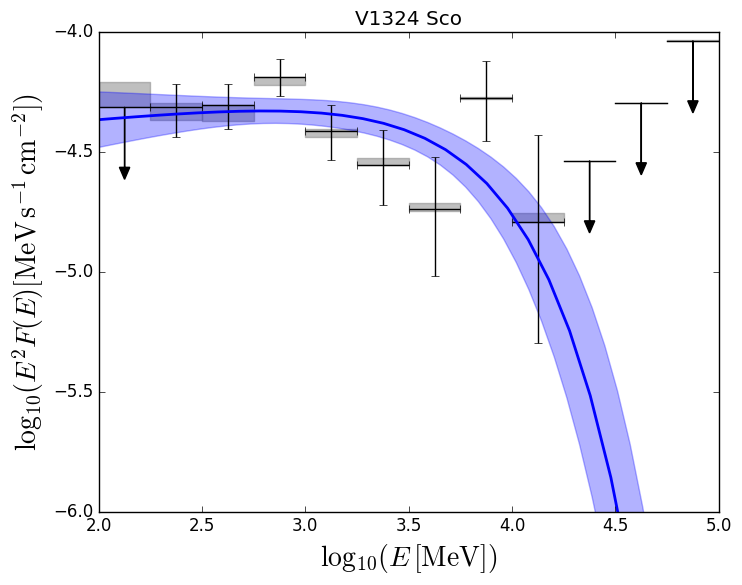}
\includegraphics[scale=0.43]{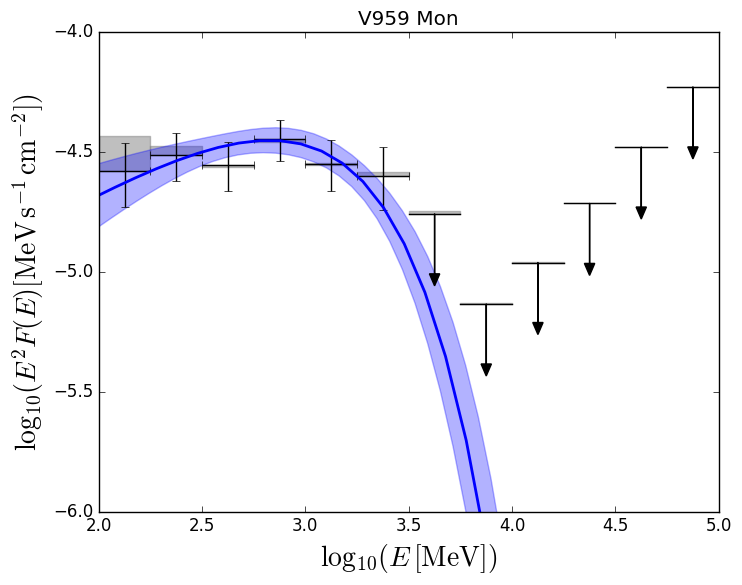}
\includegraphics[scale=0.43]{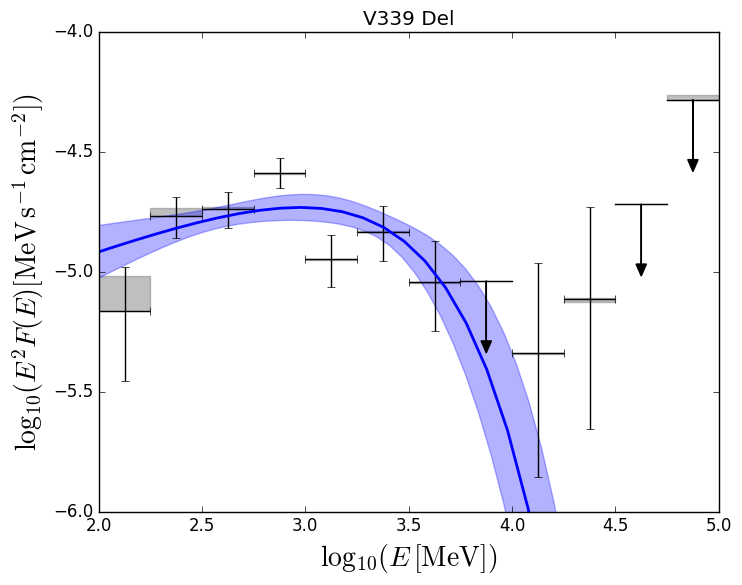}
\includegraphics[scale=0.43]{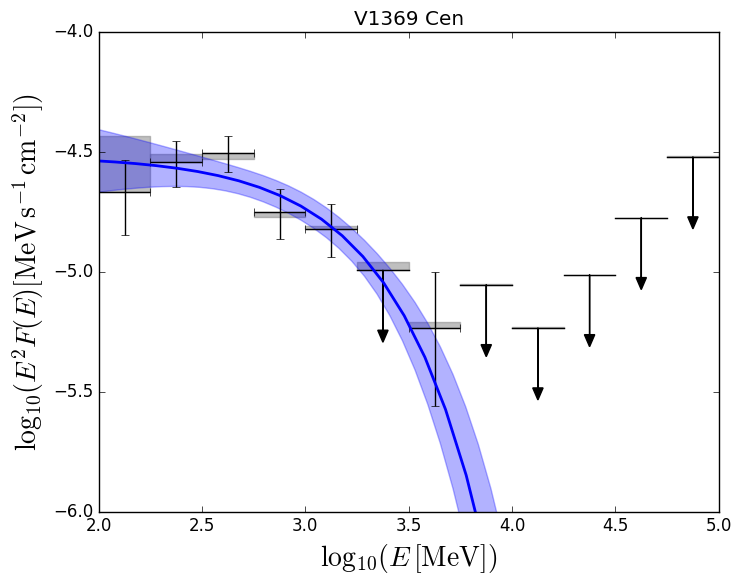}
\includegraphics[scale=0.43]{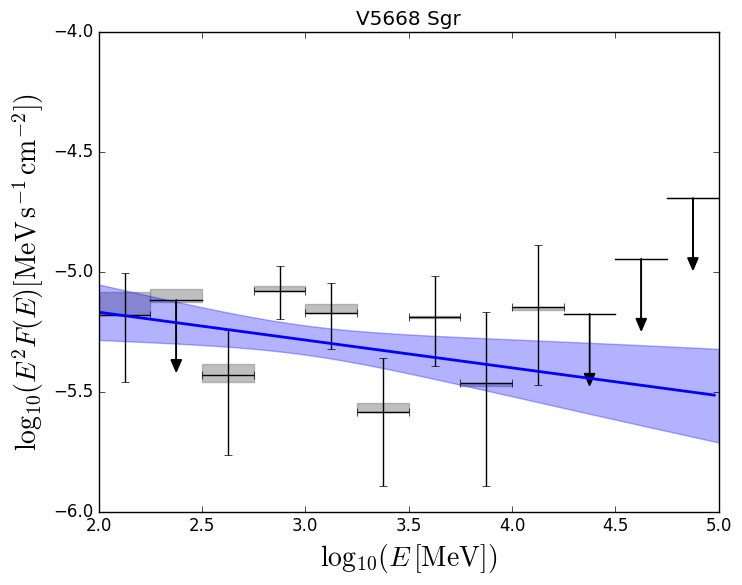}
\noindent
\caption{\small Spectra from the six previously-detected gamma-ray novae. The spectral energy distribution (SED) points including statistical uncertainties derived with the standard Galactic diffuse model are shown as black crosses and a power-law with exponential cutoff fit to that data is shown in blue. We show data points for bins with TS$>4$ otherwise we show 95\% upper limits. The systematic uncertainties introduced by modeling of the Galactic diffuse emission are estimated by repeating the analysis with alternative diffuse models. The envelope of the results using the alternative models are shown as grey bands for each energy bin. For V5668 Sgr the fit did not constrain the cutoff energy and we thus present the results of a simple power-law fit.}
\label{fig:spectraKnownNovae}
\end{center}
\end{figure*}

At low energies systematic uncertainties introduced by modeling the Galactic diffuse emission become important. To estimate those uncertainties we repeat our analysis using the alternative diffuse models introduced in~\citet{2015arXiv150703633B}. We found that the choice of diffuse emission model has a very mild impact on the best-fit spectral parameters shown in Tab.~\ref{tab:knownNovae}, which is negligible within the relatively large statistical errors. The resulting spectra are shown in Fig.~\ref{fig:spectraKnownNovae}.

Since the spectral parameters are similar, we can assume that the spectral shape is the same for all novae. In the following we combine all novae (including and excluding V407\,Cyg) in a joint likelihood analysis tying the photon index and the cutoff energy across all novae, i.e., while the normalization for each nova is a free parameter only one common index and one common cutoff are fitted. The results of the combined fit with a tied index and cutoff are also shown in Tab.~\ref{tab:knownNovae}. 
For the classical novae, the combined fit resulted in $\Gamma = 1.90\pm0.08$ and $E_{\rm c} = 4.3\pm1.2$\, GeV. To reduce the degrees of freedom in the unified analysis which will be applied to all novae from the catalog, we will fix the index and cutoff energy to 1.9 and 4\,GeV respectively in the following.

\begin{figure*}[htb!]
\begin{center}
\includegraphics[scale=0.4]{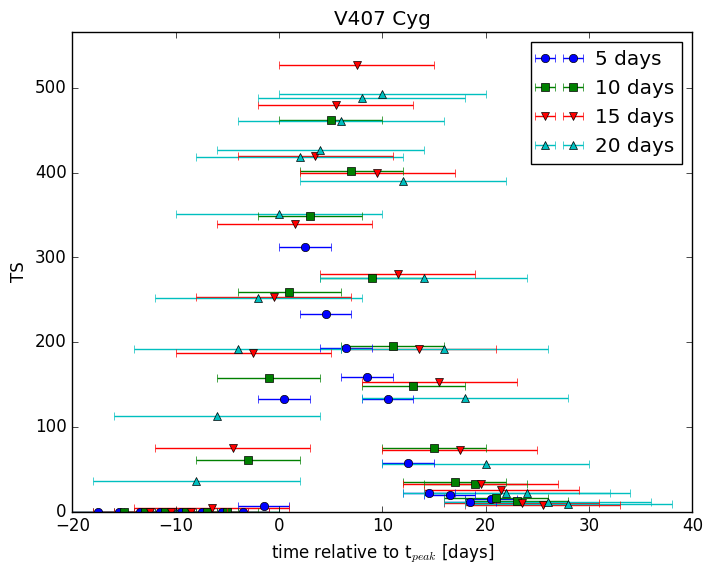}
\includegraphics[scale=0.4]{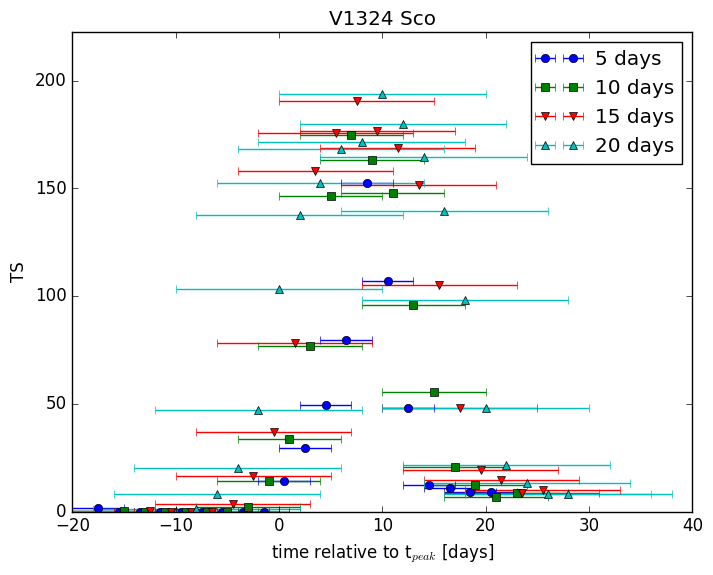}
\includegraphics[scale=0.4]{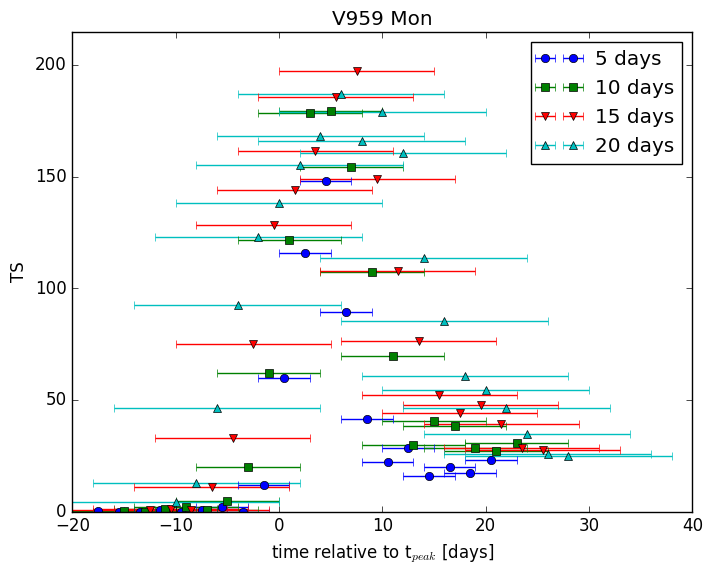}
\includegraphics[scale=0.4]{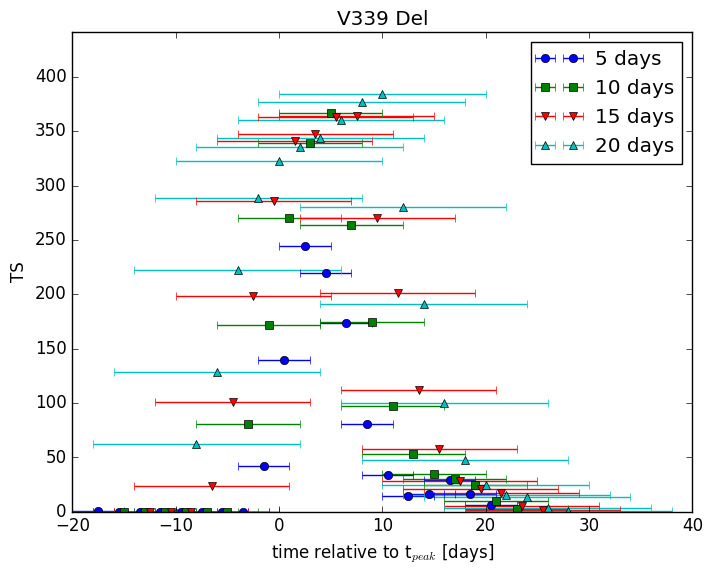}
\includegraphics[scale=0.4]{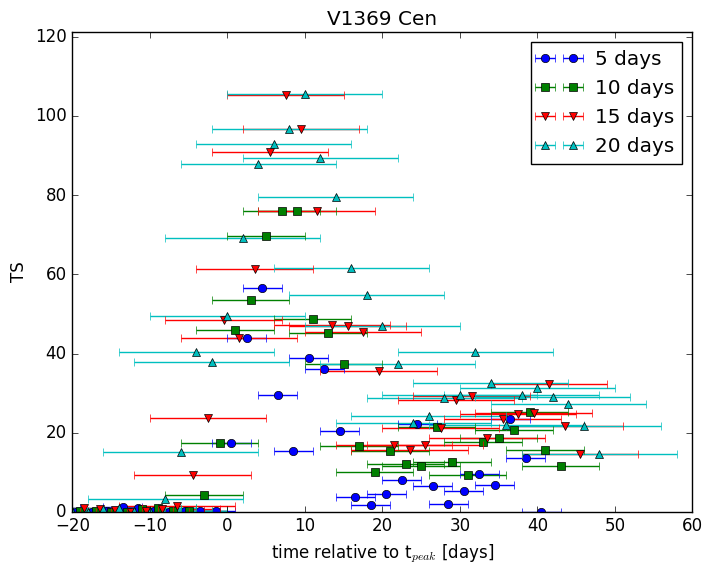}
\includegraphics[scale=0.4]{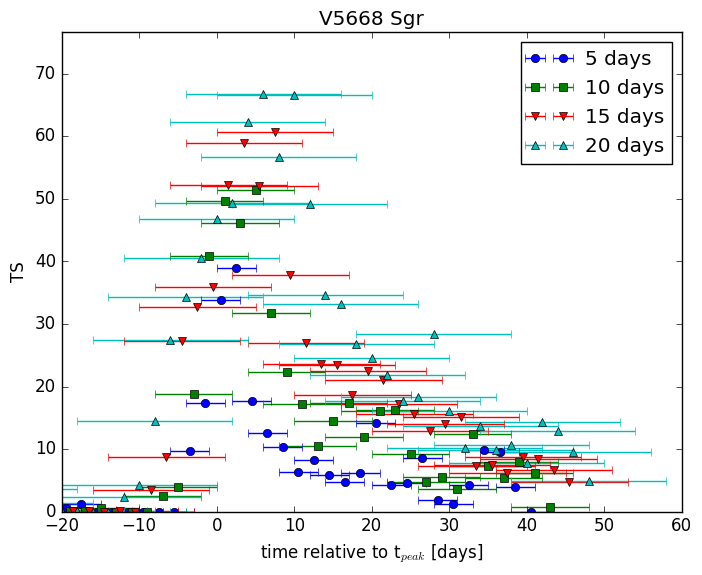}
\noindent
\caption{\small TS for sliding time windows of different length for the six known gamma-ray novae. The x-axis shows the time relative to the optical peak time $t_{peak}$. In the case of V959\,Mon the gamma-ray peak was used.}
\label{fig:slidingTWKnownNovae}
\end{center}
\end{figure*}

We apply a sliding time window analysis to the known novae using four different time windows of length 5, 10, 15 and 20 days to evaluate the optimal time window length, which will then be applied in the unified approach to all novae of the catalog. The analysis is performed in a time window of given length. The time window is then moved by 2 days and the analysis is repeated. One ends up with a light curve with overlapping time bins. For each of the 6 novae we repeat a sliding time window analysis for the 4 different time windows. The results are shown in Fig.~\ref{fig:slidingTWKnownNovae}. The 15-day or 20-day time windows generally yield the largest TS. We decide to perform the unified analysis with a 15-day time window.

\subsection{Search for Gamma-Ray Emission from All Novae in the Catalog -- Sliding Time-Window Analysis}

Given the optimized time window of 15 days and the spectral parameter given by the joint likelihood analysis of the known gamma-ray novae, we then apply a sliding time window analysis to all novae of our catalog to search for gamma-ray emission in a unified approach.

We model the novae gamma-ray spectrum with a power-law function with an exponential cutoff with the normalization free to vary and fixed index and energy cutoff to the parameters obtained above ($\Gamma = 1.9$ and $E_c = 4$\,GeV).  

\begin{figure*}[htb!]
\begin{center}
\includegraphics[scale=\onepic]{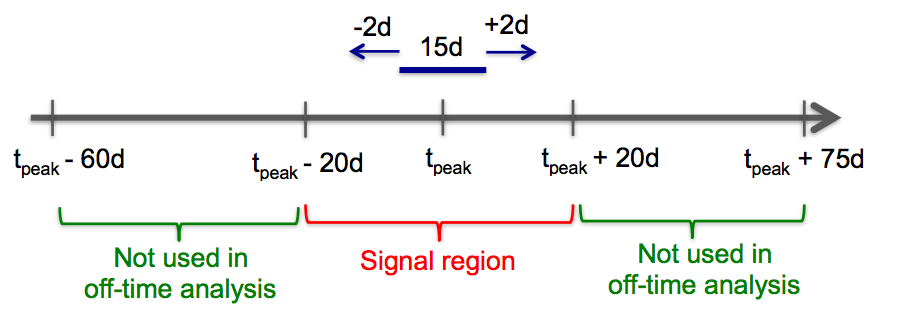}
\noindent
\caption{\small Sketch of the sliding time window analysis. The start time of the time window is shifted in 2 days steps from 20 days before the peak time to 20 days after the peak time, i.e. the time window center moves from $-12.5$ to $27.5$\,days relative to the peak time.}
\label{fig:slidingTW}
\end{center}
\end{figure*}

For each nova we search for gamma-ray emission in a sliding time window of 15-days length, motivated by the study of different time window lengths of the known gamma-ray novae presented above. The time window is moved in two-day steps from starting 60 days before the peak time to ending 75 days after the peak time. We consider the time windows starting 20 days before and 20 days after the peak time the \textit{signal region}. Time windows more than 60 days before and more than 75 days after the peak time are considered our \textit{off-time region}, where no signal is expected, and are used to study background fluctuations (see Fig.~\ref{fig:slidingTW}). To avoid a contamination of the off-time region with signal (e.g. in cases where the peak time could not be determined accurately from the optical light curve) we do not include the time windows 60 to 20 days before the peak and 20 to 75 days after the peak time.

\begin{figure*}[htb!]
\begin{center}
\includegraphics[scale=0.45]{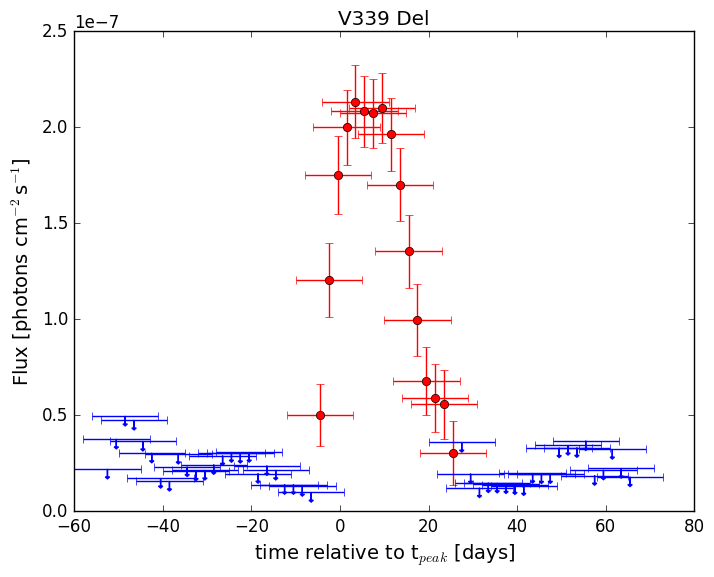}
\includegraphics[scale=0.45]{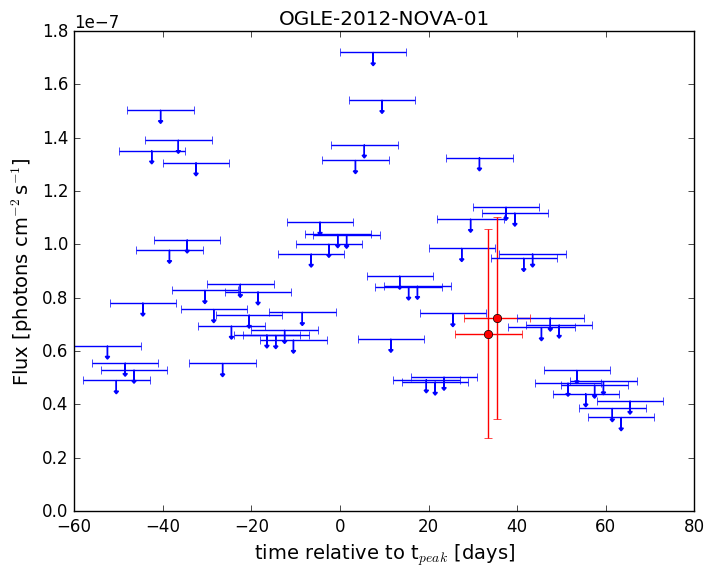}
\noindent
\caption{\small Measured flux ($TS>4$, shown in red) above 100 MeV and flux upper limits (shown in blue) for the various sliding time windows for the two novae, V339\,Del (left) and OGLE-2012-NOVA-01 (right). V339 Del is one of the previously known gamma-ray novae, while no gamma-ray emission was found for OGLE-2012-NOVA-01.}
\label{fig:novaeTW}
\end{center}
\end{figure*}

In Fig.~\ref{fig:novaeTW} we show the sliding time window results for two cases: V339\,Del, which is one of the already known gamma-ray novae, and OGLE-2012-NOVA-01 for which no gamma-ray emission was found. We display flux measurements with a TS larger than 4 as data points in Fig.~\ref{fig:novaeTW} while less significant detections are shown as flux upper limits at $95\%$ confidence level (CL) indicated by blue arrows as adopted by~\citet{2014Sci...345..554A} and~\citet{2016ApJ...826..142C}. 

\section{Results}
\label{sec:res}

\subsection{Nova Ensemble}

All six previously known gamma-ray novae are found in the analysis of the individual sources and due to increased sensitivity with the Pass 8 data set are predominantly discovered with higher significance. However, no significant gamma-ray emission was found for any additional nova except for two new candidates at $2\sigma$  level (see Sec.~\ref{subsec:novaCand}). 

\begin{figure*}[htb!]
\begin{center}
\includegraphics[scale=0.4]{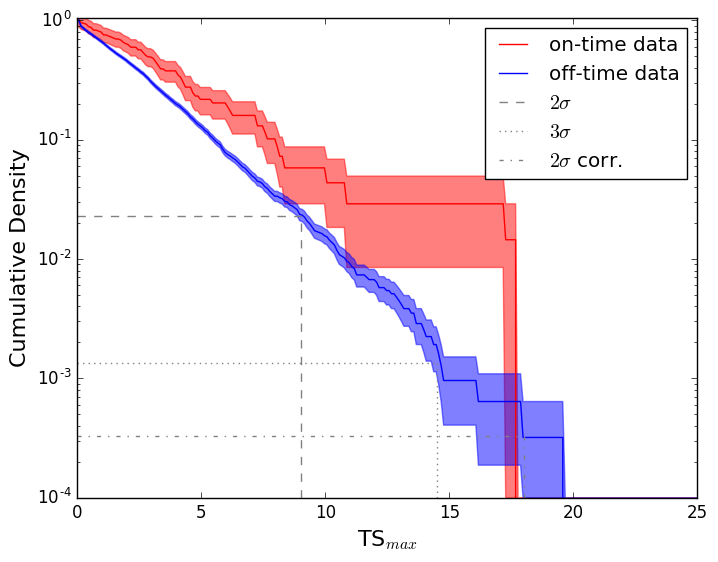}
\includegraphics[scale=0.4]{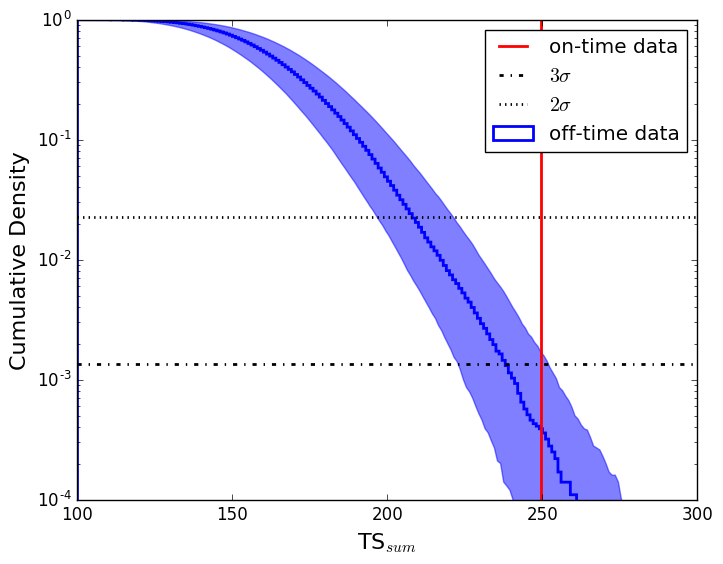}
\noindent
\caption{\small Left: Cumulative density of the $TS_{\textrm{max}}$ distribution for the on-time (red) and off-time regions (blue) including only sources with $TS_{\textrm{max}} <25$. Dotted (dashed) lines show the Gaussian equivalent one-sided 3$\sigma$ (2$\sigma$) probability of finding a larger TS then the TS indicated by the intersection of the dashed line with the blue distribution. The dashed-dotted lined shows the 2$\sigma$ probability after correction for trials. Right: Sum of TS values for 69 novae. The value the on-time data is shown as red line, while the off-time data is shown in blue. The shaded blue band reflects the uncertainty introduced by limited statistics of the off-time sample.}
\label{fig:TSDist}
\end{center}
\end{figure*}

To evaluate the significance of an excess we repeat the sliding time-window analysis in the off-time region. Similar to the on-time analysis we define a random peak time in the off-time region and perform a sliding time window analysis in two-day steps starting 20 days before the assigned peak time and ending 20 days after. In each case we report the time window with the largest TS (defined according to Eq.~\ref{eq:TS}) out of all tested time windows, $TS_{\textrm{max}}$. We compare the $TS_{\textrm{max}}$ distribution from the on-time and off-time region in Fig.~\ref{fig:TSDist} (left).

We use the off-time distribution to determine the $2\sigma$ and $3\sigma$ levels: the probability to get a $TS_{\textrm{max}} > 14.5 (9.0)$ from background fluctuations is $0.13\% (2.3\%)$ (corresponding to $3 (2)\sigma$ in a normal distribution). Testing several novae introduces a trials factor, which we account for by dividing the needed probability by the number of novae in our catalog (we have used 69 excluding the six known gamma-ray novae). The corrected probability corresponding to the 2$\sigma$ level is 0.03\%. A $TS_{\textrm{max}} > 18.0$ is needed to reach the trials corrected 2$\sigma$ level.

In an attempt to find a sub-threshold population of novae we perform a stacking analysis of all sub-threshold novae. Since the novae are fitted independently the stacking becomes a simple sum of TS values, $TS_{sum}$. We remove the 6 known gamma-ray novae and sum the TS of the remaining 69 novae. We repeat this analysis 100000 times on the off-time data by randomly picking 69 TS values from the blue distribution shown in Fig.~\ref{fig:TSDist} (left) and adding them. The distribution of the background $TS_{sum}$ values are shown in Fig.~\ref{fig:TSDist} (right) compared to on-time value of $TS_{sum} = 249.5$. To estimate the uncertainties introduced by limited statistics of the off-time sample we split the off-time sample in half. For each half we repeat the procedure of randomly picking 69 novae 100000 times. The results of the two halves represent the envelope of the shaded blue band shown in Fig.~\ref{fig:TSDist} (right). Including these uncertainties we find a $3\sigma$ effect indicating the existence of a sub-threshold gamma-ray nova population. If we exclude the three symbiotic novae (V745\, Sco, V1534\,Sco and V1535\,Sco, see also Sec.~\ref{subsec:recsymnovae}), the significance decreases to 2.5$\sigma$.

\begin{figure}[htb!]
\begin{center}
\includegraphics[scale=0.45]{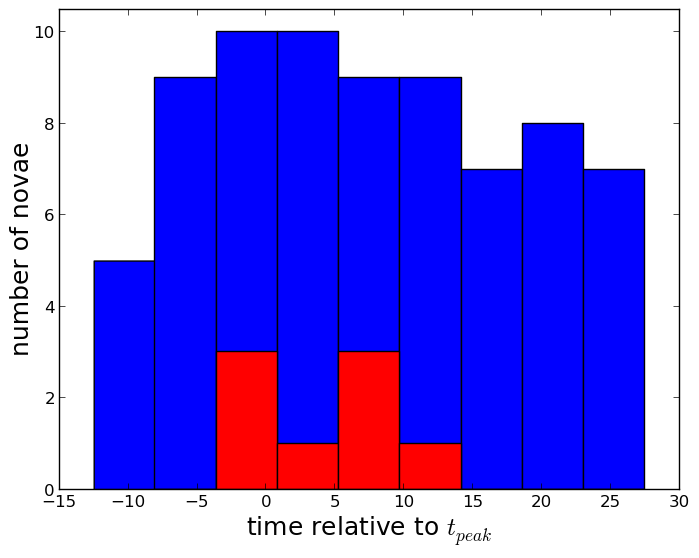}
\noindent
\caption{\small Histogram of the time difference between the central time of the time window with maximal TS and the optical peak time (including all novae in our sample). The time difference can vary from $-12.5$ to $27.5$ days relative to the peak. Novae with a maximal TS smaller 14.5 are shown in blue, sources with larger TS in red.}
\label{fig:TimeDiff}
\end{center}
\end{figure}

The known gamma-ray detected novae indicate an onset of the gamma-ray emission coincident or delayed by a few days with respect to the optical peak. A delay could be caused by absorption of the gamma rays via photon-atom interactions in an initially dense ejecta. Gamma rays appear once the density drops and the ejecta becomes transparent. Figure~\ref{fig:TimeDiff} shows a histogram of the time difference between the central time of the time window with maximal TS and the optical peak time. The time difference can vary from $-12.5$ to $27.5$ days relative to the peak. For background the distribution is expected to be flat. Note that novae with TS larger than 14.5 tend to reach the maximal TS closer to the optical peak time.

\subsection{Gamma-ray Nova Candidates}
\label{subsec:novaCand}
\begin{figure*}[htb!]
\begin{center}
\includegraphics[scale=0.35]{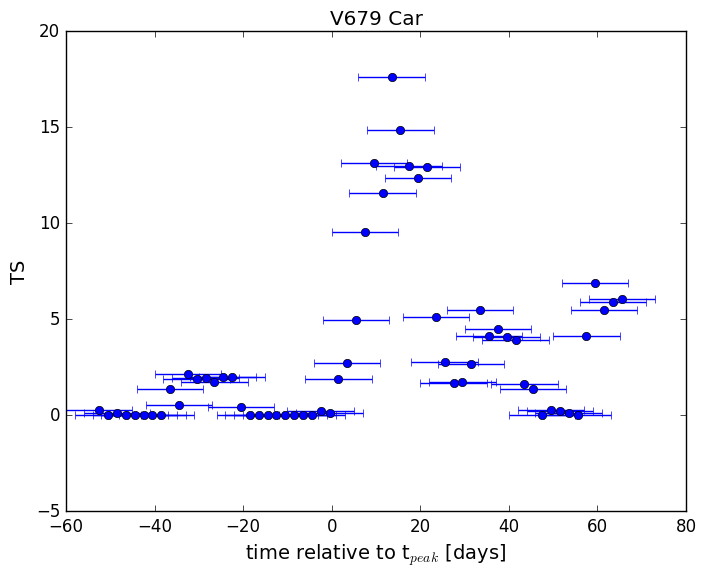}
\includegraphics[scale=0.35]{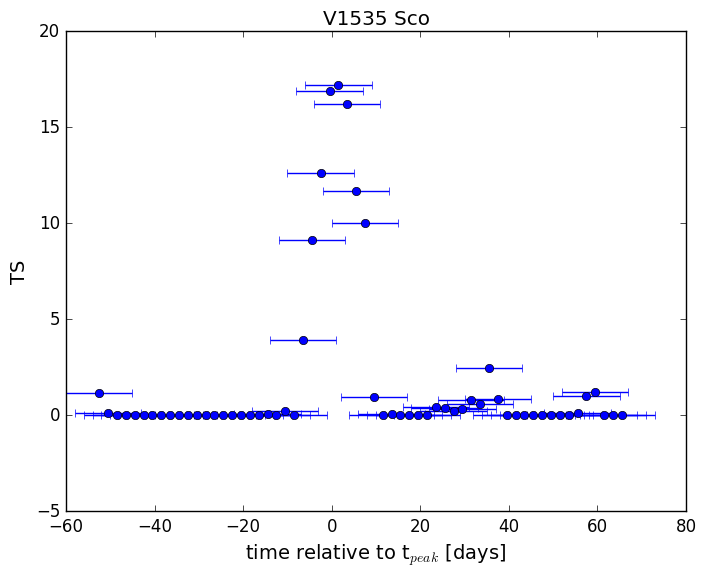}
\includegraphics[scale=0.35]{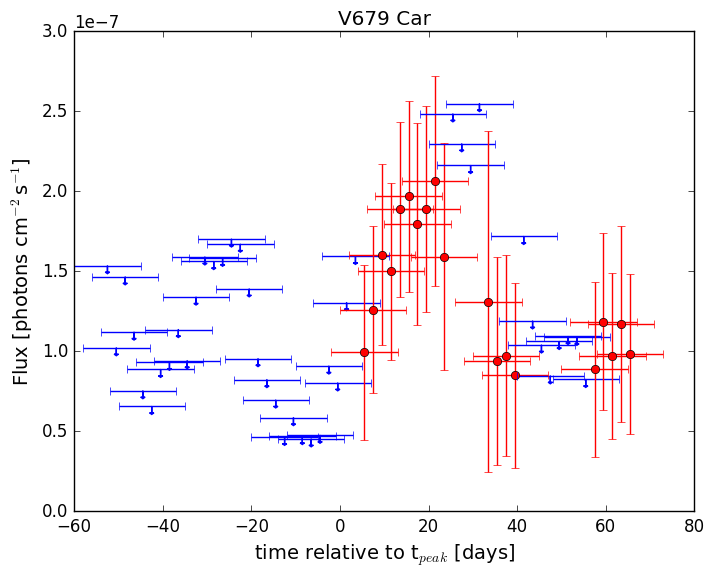}
\includegraphics[scale=0.35]{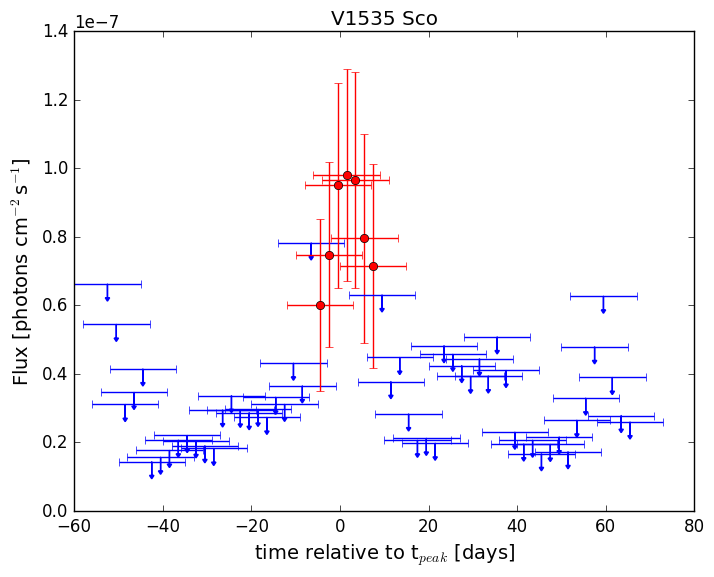}
\noindent
\caption{\small Sliding time window results for V679\,Car (left) and V1535\,Sco (right). The upper row shows the TS value for each tested time window, while the lower row shows the measured flux above 100\,MeV (TS$>4$, red) or 95\% flux upper limits (blue).}
\label{fig:TSLC}
\end{center}
\end{figure*}

\begin{figure*}[htb!]
\begin{center}
\includegraphics[scale=0.35]{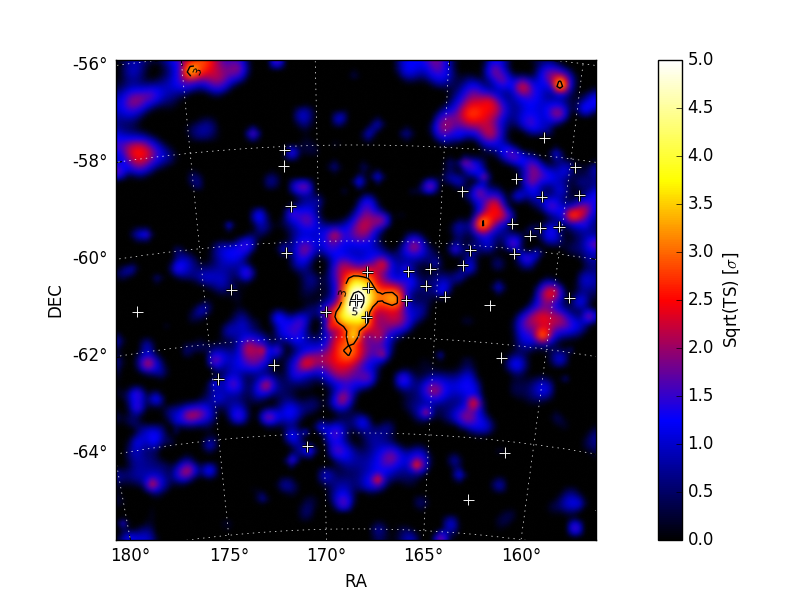}
\includegraphics[scale=0.35]{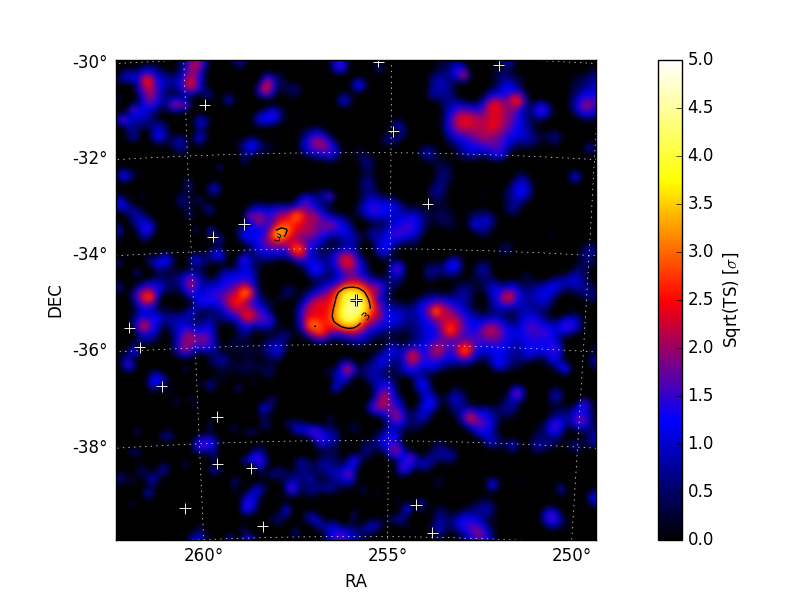}
\noindent
\caption{\small TS map in the time window with maximal TS for V679\,Car (left) and V1535\,Sco (right). The white crosses indicate the point sources used in the model, i.e. the 3FGL background sources and the nova in the center. The black lines indicate the three and five sigma contours.}
\label{fig:TSmap}
\end{center}
\end{figure*}

Besides the 6 already known gamma-ray emitting novae we find 2 candidates with $TS_{\textrm{max}} > 14.5$ (i.e.~3$\sigma$ before trials): V679\,Car\,2008 has a $TS_{\textrm{max}}$ of 17.6 and
V1535\,Sco\,2015 a $TS_{\textrm{max}}$ of 17.2 (see Fig.~\ref{fig:TSLC}). Those values barely reach the trials corrected 2$\sigma$ level. Because these two candidates show the highest TS in the individual source analysis, they also dominate the sum of TS. The TS map for both candidates in the 15-day time window with maximal TS is shown in Fig.~\ref{fig:TSmap}. 

Spectroscopic observations identified V679\,Car as a classical nova. Its spectrum is dominated by broad emission lines of the Balmer series and of Fe II emission features~\citep{IAUC8999}. 
V1535\,Sco is a symbiotic system~\citep[][see also Sec.~\ref{subsec:recsymnovae}]{2015MNRAS.454.1297S}, similar to V407\,Cyg. Hard, absorbed X-rays and synchrotron radio emission were detected at the early phase of the outburst indicating that the nova is produced by a white dwarf embedded within the wind of a red-giant companion~\citep {ATel7060,ATel7085}.
\citet{2017ApJ...842...73L} show that the measured X-ray emission indicates the presence of strong shocks during the first two weeks of the nova's evolution, which is expected for a nova ejecta expanding into a thick wind
from a giant companion. A rebrightening in X-rays after $\sim50$\,days indicates the existence of a second shock possibly produced in collisions between multiple outflows within the ejecta. No gamma-ray emission was found coincident with the second shock.

\begin{figure*}[htb!]
\begin{center}
\includegraphics[scale=0.35]{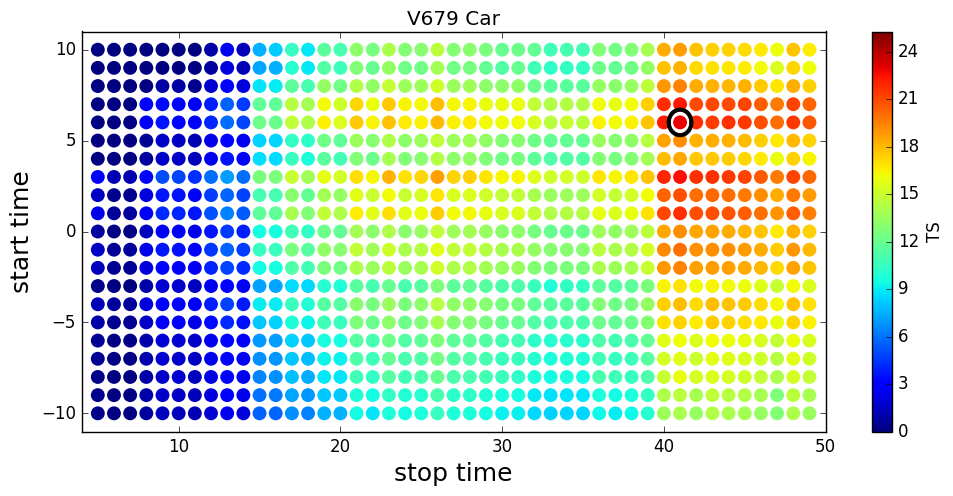}
\includegraphics[scale=0.35]{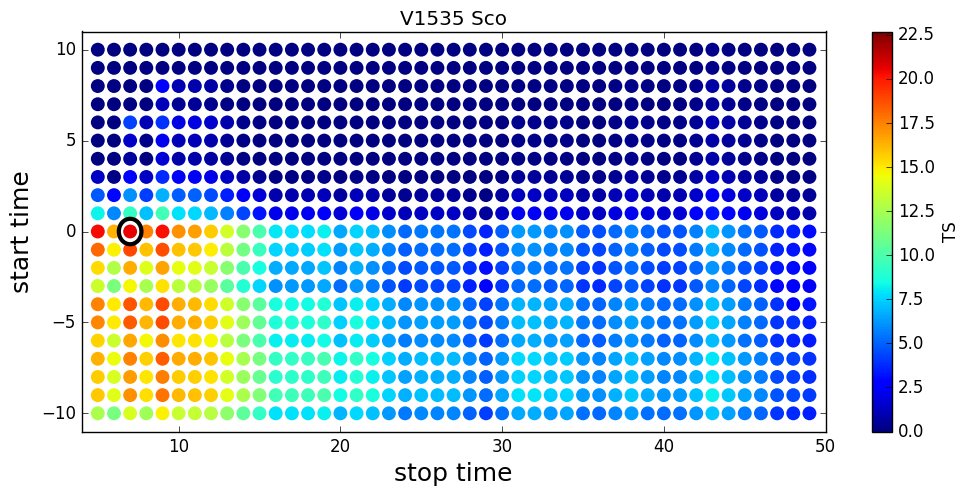}
\noindent
\caption{\small TS as a function of start and stop time in days relative to the peak time for V679\,Car (left) and V1535\,Sco (right). The maximal TS is marked with a black circle.}
\label{fig:duration}
\end{center}
\end{figure*}

To obtain the duration of the nova candidates we repeat the analysis for various start and stop times in one-day time steps. We define the duration from the time window, which yields the largest TS. Figure~\ref{fig:duration} shows the TS distribution as a function of start and stop time. For V679\,Car we obtain a duration of 35\,days and for V1535\,Sco of 7\,days.

\begin{figure*}[htb!]
\begin{center}
\includegraphics[scale=0.4]{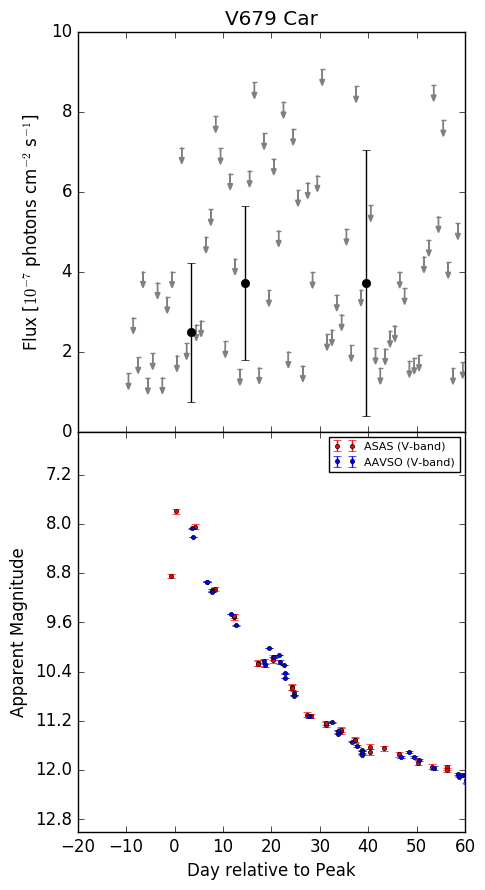}
\includegraphics[scale=0.4]{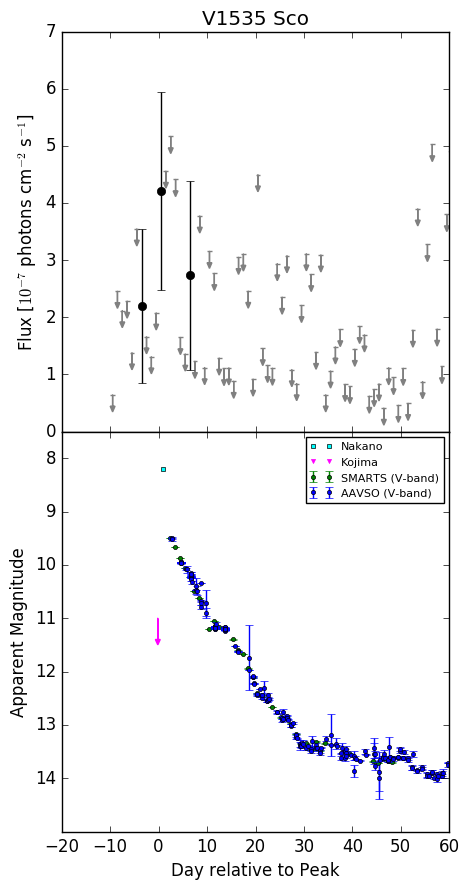}
\noindent
\caption{\small Optical (bottom) and gamma-ray (top) light curves of  V679\,Car (left) and V1535\,Sco (right). The gamma-ray data is binned in one-day bins. For days with $TS<4$ we show 95\% upper limits.}
\label{fig:LCCand}
\end{center}
\end{figure*}

\begin{figure*}[htb!]
\begin{center}
\includegraphics[scale=0.35]{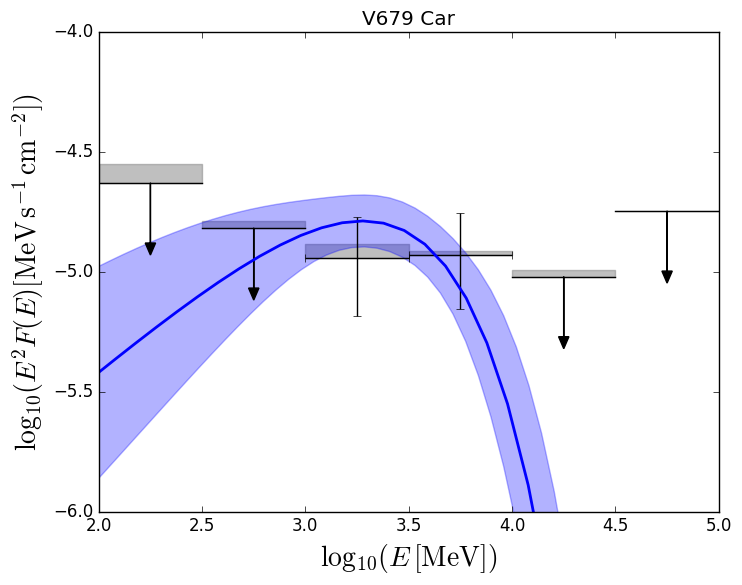}
\includegraphics[scale=0.35]{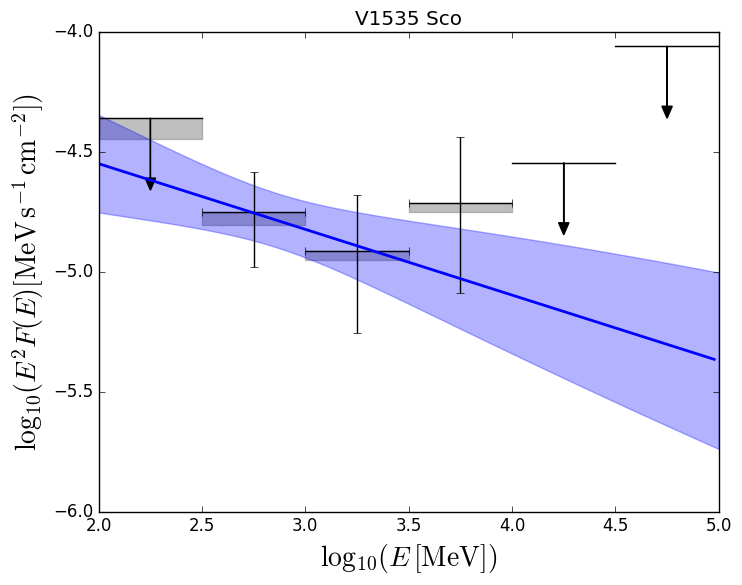}
\noindent
\caption{\small SED for V679\,Car (left) and V1535\,Sco (right) using the duration obtained from Fig.~\ref{fig:duration}. Both SEDs suffer from small statistics. The SED points derived with the standard Galactic diffuse model are shown as black crosses and a power-law with exponential cutoff fit to that data is shown in blue. For V1535\,Sco the energy cutoff is not well constrained by the fit and we thus present the results of a simple power-law fit. The systematic uncertainties introduced by modeling of the Galactic diffuse emission are estimated by repeating the analysis with alternative diffuse models. The envelope of the results using the alternative models are shown as grey bands for each energy bin.}
\label{fig:SEDCand}
\end{center}
\end{figure*}

The gamma-ray and optical light curve of the two candidate sources are shown in Fig.~\ref{fig:LCCand}. The SED suffers from low statistics and is shown in Fig.~\ref{fig:SEDCand}. Table~\ref{tab:newNovae} lists the spectral fit parameters for the two candidates. The fit was applied in the energy range from 100 MeV to 100 GeV using the time window found in Fig.~\ref{fig:duration}. Due to small statistics the errors on the fit parameters are large. Within the errors the spectral parameters are similar to the spectra of the 6 known gamma-ray novae.

We estimate the distance of V679\,Car and V1535\,Sco using two steps.
First we use the maximum magnitude versus rate of decline (MMRD) relationship that provides an estimate of the maximum absolute magnitude as a function of $t_2$~\citep{1995ApJ...452..704D,2000AJ....120.2007D}. We note that other authors \citep[e.g.][]{2011ApJ...735...94K} found the MMRD to be an oversimplification.
Second we use the extinction derived from the reddening E(B-V) measurements (SMARTS). With maximum magnitudes of 7.6 and 8.2\,mag, $t_2  = 14$ and $6$ days and extinctions of 3.72 and 2.81\,mag, we estimated distances of $(2.9 \pm 0.7)$ and $(7.3 \pm 1.7)$\,kpc for V679\,Car and V1535\,Sco, respectively. Note that our distance estimate of V1535\,Sco is compatible with the Galactic Center distance of $\sim8.5$\,kpc, which was also suggested by~\citet{2017ApJ...842...73L}.

\begin{figure}[htb!]
\begin{center}
\includegraphics[scale=0.4]{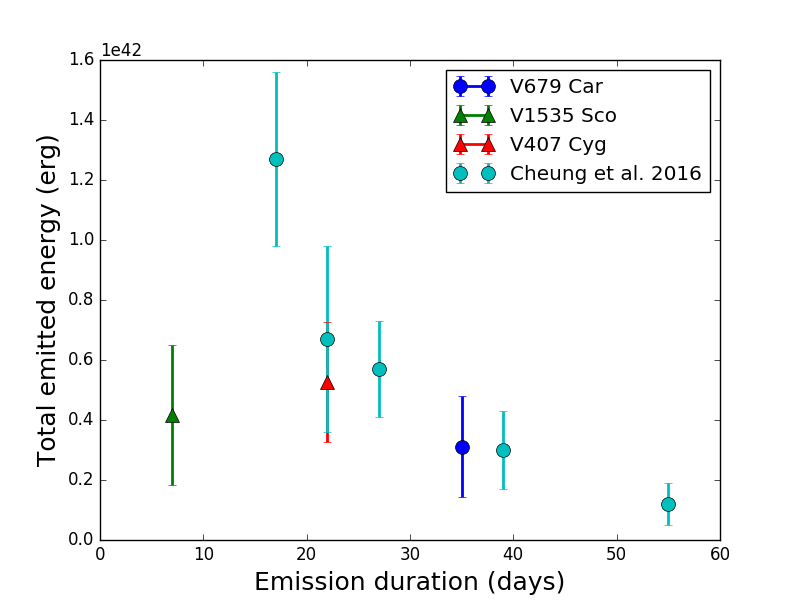}
\noindent
\caption{\small Total emitted energy above $100$\,MeV as a function of the duration of the gamma-ray emission. Classical novae are marked with a circle and symbiotic ones with a triangle. The classical novae V679\,Car is in agreement with the suggested relation between the total emission duration and the total energy of classical novae, while the symbiotic nova V1535\,Sco emitted less energy than expected from the relation. }
\label{fig:durationEnergy}
\end{center}
\end{figure}

We calculate the total number of gamma-ray photons and the total energy emitted by the two novae candidates. Note that the number of photons depends strongly on the spectral shape, which is not well constrained at low energies. The total energy is more robust to uncertainties in the spectral shape at low energies. \citet{2016ApJ...826..142C} suggest a relation between the total emission duration and the total energy of classical novae. While the classical nova V679\,Car is in agreement with this relation, the symbiotic nova V1535\,Sco emitted less energy than expected from the relation (see Fig.~\ref{fig:durationEnergy}).

\begin{table*}
\caption{New Gamma-ray Nova Candidates}
\label{tab:newNovae}
\centering
\begin{tabular}{lccccccccc}
\hline\hline
Nova                &  $t_{\textrm{start}}$ &  $t_{\textrm{stop}}$ & Duration  & TS   &  Index               & Cutoff               & Distance            &  Photons           & Energy\\
                        & [MET]                       & [MET]                      & [Days]     &         &                          & [GeV]               & [kpc]                  & [$10^{44}$]       & [$10^{41}$\,erg]\\
\hline\\
V679\,Car	       &  249868800              & 252892800             & 35           &  24.9 &   $1.3\pm0.8$  &   $2.6\pm2.0$   & $ 2.9 \pm 0.7$  &  $3.0\pm2.3$    & $2.3 \pm 1.5$ \\
V1535\,Sco	& 445305600              & 445910400             & 7             & 20.9  &   $2.3\pm0.3$  & -- &  $ 7.3 \pm 1.7$ &    $8.5\pm5.4$  & $5.7 \pm 3.3$ \\   
\hline
\end{tabular}
\tablefoot{Durations of the $>100$\,MeV gamma-ray emission, total number of photons, and total energy emitted by the two nova candidates as derived from LAT data analysis. The energy cutoff for V1535\,Sco is not constrained by the fit. We repeated the fit with a simple power-law fit without a cutoff and find that the TS value decreases only insignificantly from 21.2 to 20.9.} 
\end{table*}

\subsection{Correlation of Gamma-ray Flux and Optical Light Curve Properties}
\begin{figure*}[htb!]
\begin{center}
\includegraphics[scale=0.4]{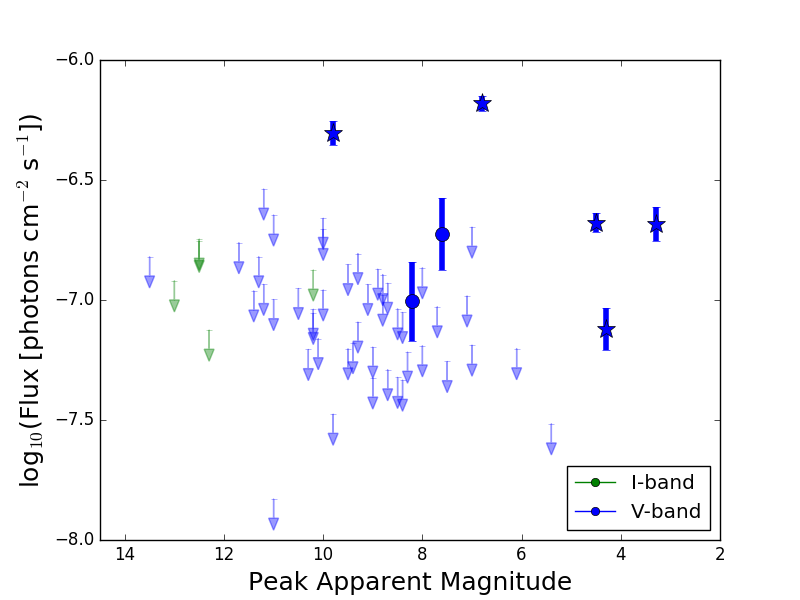}
\includegraphics[scale=0.4]{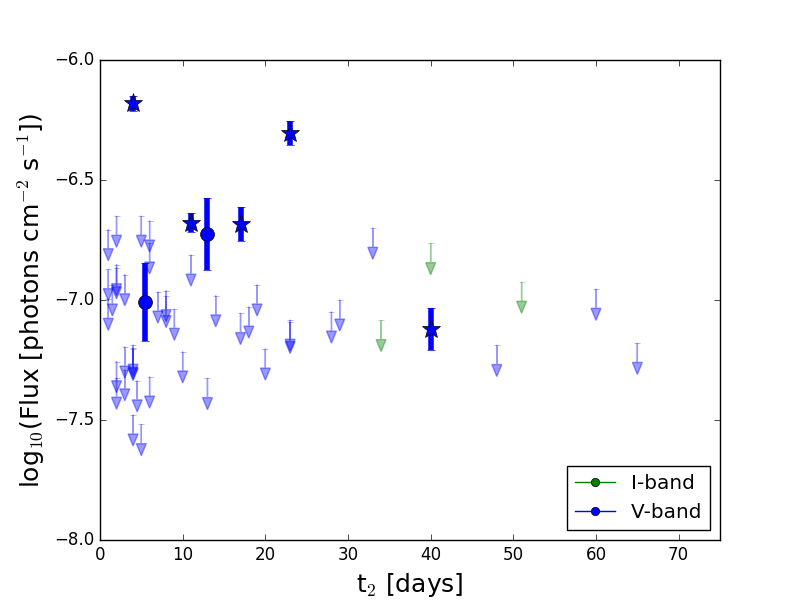}
\noindent
\caption{\small Correlation of the gamma-ray flux with the optical peak apparent magnitude (left) and $t_2$ (right): Sources with $TS>14.5$ (corresponding to $3\sigma$) are shown as blue circles, while already known gamma-ray novae are indicated as stars. Note that V959\,Mon is not included, because the peak was not covered by optical observations. Less significant sources are shown as upper limits indicated by arrows. The peak magnitude is measured in V-band for most cases (shown in blue), but for a few cases only I-band data was available (shown in green).}
\label{fig:GammaOptCorr}
\end{center}
\end{figure*}

We study the correlation of gamma-ray and optical flux with our nova sample. Eighteen sources with a large uncertainty on the optical apparent peak magnitude were excluded from this study. Fig.~\ref{fig:GammaOptCorr} (left) shows the gamma-ray flux as a function of the optical apparent peak magnitude. No correlation is found. Note that the magnitudes are not corrected for extinction and can thus only be interpreted as upper limits on the real magnitude and lower limits on the real flux~\citep{2015MNRAS.450.2739M}. Two bright novae T\,Pyx and KT\,Eri with optical apparent peak magnitudes of 6.1 and 5.4 respectively were not detected in gamma-rays. 

Fig.~\ref{fig:GammaOptCorr} (right) shows $t_2$ as a function of the optical apparent peak magnitude. No correlation is found. The 18 sources with a large uncertainty on the optical apparent peak magnitude were excluded and additional five sources were excluded because the poorly sampled light curve did not allow to estimate $t_2$.

\subsection{Flux-Distance Relationship}
Some of the novae in our catalog have a distance estimate reported in the literature. We list those in Tab.~\ref{tab:distance}. The most reliable distance estimates stem from parallax measurements of the angular expansion of the nova shell several years after its outburst. However, such a measurement is only available for two novae in our sample, V959\,Mon~\citep{Linford} and V339\,Del~\citep{2014Natur.515..234S}. Note that \citet{Linford} combined the radio-interferometric expansion rate with the optical spectroscopic velocity measurement, which might lead to uncertainties if the radio and optical wavelength data are produced in different emission regions. Another method relies on the Galactic reddening-distance relation for the line-of-sight of the nova and its independent reddening measure~\citep{Oezdoenmez}.

\begin{figure}[htb!]
\begin{center}
\includegraphics[scale=0.4]{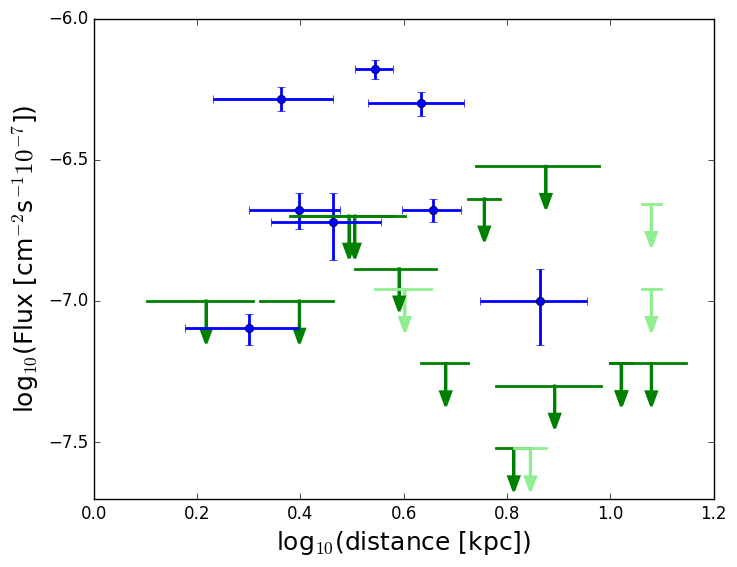}
\noindent
\caption{\small Nova flux (blue) or flux upper limit (green) as a function of distance for a subsample of our catalog with estimated distances. Light green values indicate lower limits of the distance.}
\label{fig:GammaOptCorr2}
\end{center}
\end{figure}

The gamma-ray flux as a function of distance is displayed in Fig.~\ref{fig:GammaOptCorr2}. Most gamma-ray novae are near-by with distance estimates $\le 4.5$\,kpc with the exception of the gamma-ray nova candidate V1535\,Sco at a distance of $7.3\pm1.7$\,kpc. Note that the distance for V1535\,Sco was estimated with the MMRD method, whose justification was questioned by~\citet{2011ApJ...735...94K}. We note that ~\citet{2015ApJ...809..160F} find a larger distance of $>6.5$\,kpc for V1324\,Sco compared to $4.3\pm0.9$\,kpc from \citet{Oezdoenmez} used here. Adopting this larger distance estimate would add V1324\,Sco to the exception of gamma-ray bright but distant novae.

Not all near-by novae have been detected in gamma rays. V2672\,Oph, V496\,Sct and V2674\,Oph are closer than 4\,kpc, but show no gamma-ray flux at the level of $2\times10^{-7}$\,cm$^{-2}$\,s$^{-1}$. Also the luminous red nova, V1309\,Sco (see Sec.~\ref{subsec:LRN}), at a distance of $2.5\pm0.4$\,kpc was not detected in gamma rays indicating that the gamma-ray flux is smaller than $10^{-7}$\,cm$^{-2}$\,s$^{-1}$.

\begin{figure*}[htb!]
\begin{center}
\includegraphics[scale=0.4]{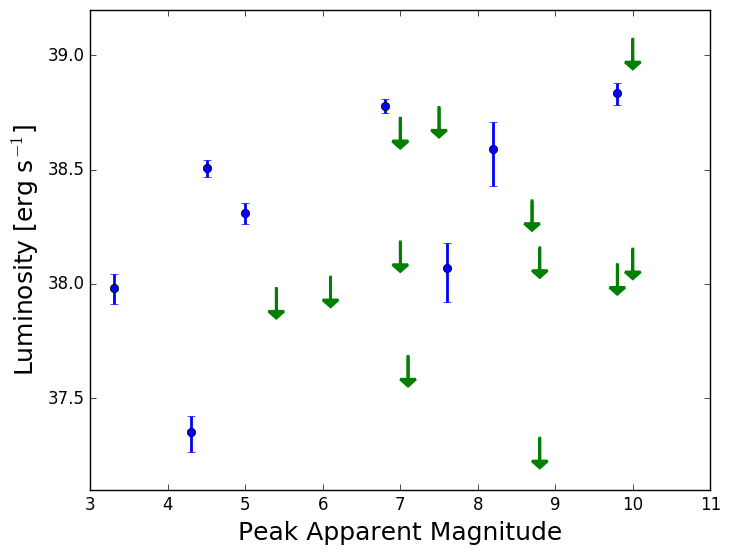}
\includegraphics[scale=0.4]{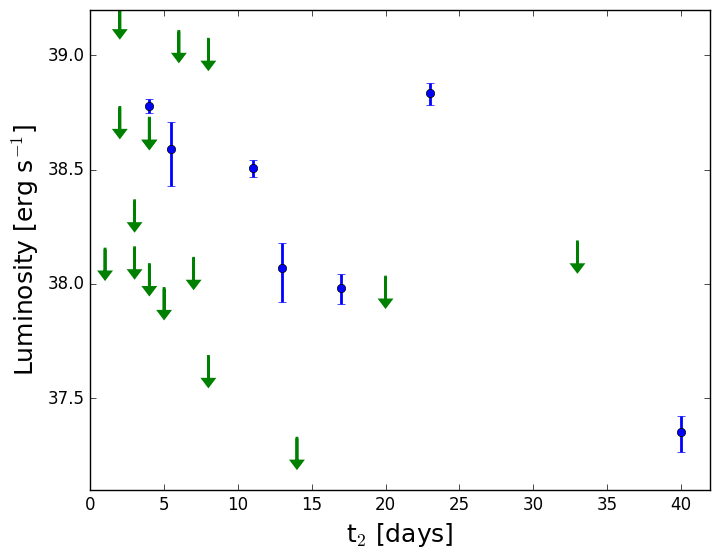}
\noindent
\caption{\small Gamma-ray luminosity as a function of the optical peak apparent magnitude (left) and $t_2$ (right).}
\label{fig:LgVsMag}
\end{center}
\end{figure*}

Using the subset of novae with a distance estimate we investigated the gamma-ray luminosity, $L_\gamma$ as a function of the optical peak apparent magnitude and $t_2$ (see Fig.~\ref{fig:LgVsMag}).

\begin{table*}
\caption{Distance estimates of a subset of novae from our catalog.}
\label{tab:distance}
\centering
\begin{tabular}{lccc}
\hline\hline
Name & Distance & Flux                                             & Reference\\
           & [kpc]       & [$10^{-7}$\,cm$^{-2}$\,s$^{-1}$] & \\
            \hline\\
V1309 Sco & $2.5\pm0.4$ & <1.0 & \citet{Oezdoenmez}\\
V1721 Aql & $7.5\pm2.0$ & <3.0 & \citet{Oezdoenmez}\\
V679 Car & $2.9\pm0.7$ & $1.9\pm0.5$ & MMRD, this work\\
V5583 Sgr & $10.5\pm0.5$ & <0.6 & \citet{Schwarz2011}\\
V2672 Oph & $3.12\pm0.69$ & <2.0 & \citet{Oezdoenmez}\\
KT Eri & $6.5\pm0.5$ & <0.3 & \citet{ATel2327}\\
V496 Sct & $3.2\pm0.8$ & <2.0 & \citet{Oezdoenmez}\\
V1722 Aql & >12 & <1.1 & \citet{Munari2010a}\\
U Sco & $12.0\pm2.0$ & <0.6 & \citet{Schaefer2010}\\
V2674 Oph & $1.65\pm0.38$ & <1.0 & \citet{Oezdoenmez}\\
V407 Cyg & $3.5\pm0.3$ & $6.6\pm0.5$ & \citet{Oezdoenmez}\\
V1723 Aql & $5.7\pm0.4$ & <2.3 & \citet{2016MNRAS.457..887W}\\
V5588 Sgr & >4 & <1.1 & \citet{Munari2015a}\\
T Pyx & $4.8\pm0.5$ & <0.6 & \citet{2013ApJ...770L..33S}\\
V5589 Sgr & $3.9\pm0.7$ & <1.3 & \citet{2016MNRAS.460.2687W}\\
V1324 Sco & $4.3\pm0.9$ & $5.0\pm0.5$ & \citet{Oezdoenmez}\\
V959 Mon & $2.3\pm0.6$ & $5.2\pm0.5$ & \citet{Linford}\\
V809 Cep & >7 & <0.3 & \citet{Munari2014}\\
V339 Del & $4.54\pm0.59$ & $2.1\pm0.2$ & \citet{2014Natur.515..234S}\\
V1830 Aql & >12 & <2.2 & \citet{Munari2014}\\
V1369 Cen & $2.5\pm0.5$ & $2.1\pm0.3$ & \citet{ATel6413}\\
V745 Sco & $7.8\pm1.8$ & <0.5 & \citet{Schaefer2010}\\
V1535 Sco & $7.3\pm1.7$ & $1.0\pm0.3$ & MMRD, this work\\
V5668 Sgr & $2.0\pm0.5$ & $0.8\pm0.1$ & \citet{Banerjee2015}\\
V5583 Sgr & $10.5\pm0.5$ & <0.6 & \citet{Schwarz2011}\\
\hline 
\end{tabular}
\tablefoot{A different distance for V1324\,Sco of $<6.5$\,kpc was estimated by~\citet{2015ApJ...809..160F}. \citet{2016MNRAS.460.2687W} estimate a distance range of 3.2 to 4.6\,kpc for V5589\,Sgr; for simplicity we have adopted a distance of $3.9\pm0.7$\,kpc for this source.}
\end{table*}

\subsection{Recurrent and Symbiotic Novae}
\label{subsec:recsymnovae}

Symbiotic novae have an evolved companion (e.g., red giant) with a dense wind as opposed to a main-sequence companion for classical novae. Recurrent novae~\citep[see][for a review]{2010ApJS..187..275S,2015AcPPP...2..246M} have undergone several outbursts over the past century (with typical time intervals between outbursts of 10 to 100 years). 
All novae might be recurrent given enough time. The time between outbursts is thought to be smaller for more massive white dwarfs, which makes recurrent novae candidate progenitors of Type Ia supernova~\citep{2011arXiv1109.5799P}.

The first \Fermi-LAT gamma-ray detection of a nova was in the symbiotic-like nova, V407\,Cyg in 2010. Because the dense wind of the evolved companion star is interacting with the ejecta of the exploding white dwarf, the LAT emission duration can be more directly related to the specific parameters of the binary systems (e.g., separation).  Though such systems are relatively rare, with $\sim$10 known symbiotic novae known, LAT gamma-ray observations of other symbiotic systems with outbursts during the \textit{Fermi} mission could thus be examined separately from the classical novae. Our sample contains three additional confirmed symbiotic novae: V745\,Sco, V1534\,Sco~\citep{2014ATel.6032....1J} and V1535\,Sco~\citep{ATel7060}.  For three novae in our sample several historical outbursts have been observed. Those recurrent systems are: V745\,Sco (also a symbiotic system), T\,Pyx and U\,Sco, which likely have main sequence companions~\citep{1994inbi.conf.....S}. Gamma-ray light curves of the recurrent and symbiotic novae are displayed in Fig.~\ref{fig:LCV745Sco} for V745\,Sco and for the other sources in Appendix~\ref{app:LC}.

In the symbiotic recurrent novae, V745\,Sco, with outburst on 2014 Feb 6.694~\citep[CBET 3803,][]{CBET3803}, preliminary 2 and 3\,$\sigma$ significances on Feb 6th and 7th were reported \citep{2014ATel.5879....1C}. This apparent short gamma-ray emission duration compared to V407\,Cyg could be due to the fast onset of deceleration in the ejecta \citep{ban14}. With the new Pass 8 analysis, we confirm the low significances with TS = 6.5 and 5.6 on these days (see Fig.~\ref{fig:LCV745Sco}). Note, for V745\,Sco, the Galactic center biased pointing strategy that (re)commenced on Feb 5.0 after a ToO pointing toward M82/SN2014J gave optimized exposure at the nova position after the optical nova detection. Also the position of V1534\,Sco 2014 benefited from this exposure profile, although $\sim$10 days after the novae (end of March 31, 2014) a ToO pointing of 3C279 for 350 ksec decreased the exposure at the nova position~\citep{ATel6036}.

\begin{figure*}[htb!]
\begin{center}
\includegraphics[scale=0.4]{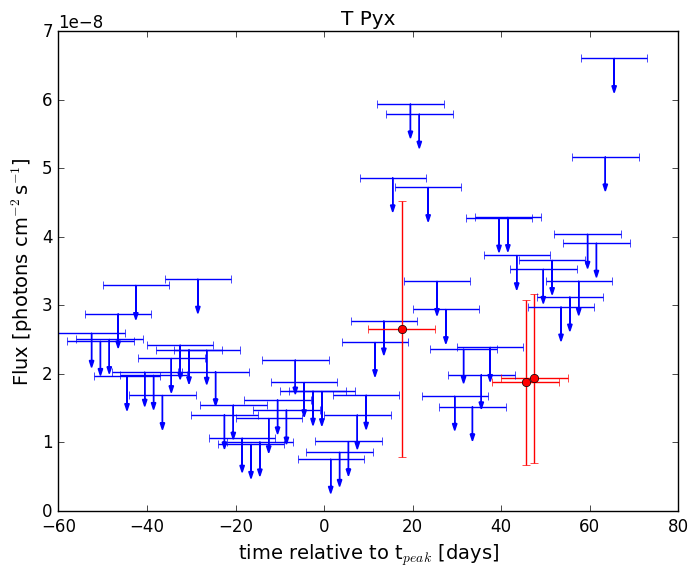}
\includegraphics[scale=0.4]{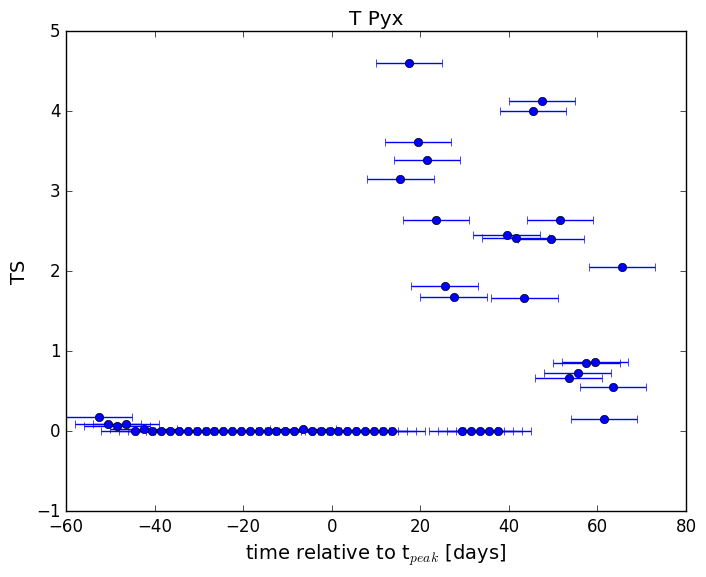}
\noindent
\caption{\small Sliding time window results for T\,Pyx extended to 60 days. The left panel shows the TS value for each tested time window, while the right panel shows the measured flux above 100\,MeV (TS$>4$, red) or 95\% flux upper limits (blue).}
\label{fig:TPyxSliding}
\end{center}
\end{figure*}

The recurrent nova T\,Pyx 2011 reached its optical peak on May 12, 2011 at 6.1\,mag~\citep[see][]{cho14} and is thus one of the brightest amongst non-detected in gamma rays, it shows a broad optical peak, which occurred unusually late, about 1 month after the optical discovery (April 14th). Due to the long duration of T\,Pyx we have decided to extend the sliding time window analysis to 60 instead of 20 days, but did not find any excess of gamma rays at later times either (see Fig.~\ref{fig:TPyxSliding}). The recurrent nova U\,Sco peaks at 7.5\,mag on January 28, 2010, but shows a narrow peak with $t_2$ of only 2 days. Another relevant parameter in discussing the LAT-detectability of these systems are their distances. Despite their bright optical peaks their estimated distance is quite large with $12.0\pm2.0$ kpc  and $4.8\pm0.5$ kpc for U\,Sco and T\,Pyx respectively~\citep{Schaefer2010,2013ApJ...770L..33S}. If their gamma-ray luminosities are scaled to any of the LAT-detected novae, we thus expect U\,Sco to be $7-36$ and T\,Pyx to be $1-6$ times fainter in the LAT band.

\begin{figure}[htb!]
\begin{center}
\includegraphics[scale=0.6]{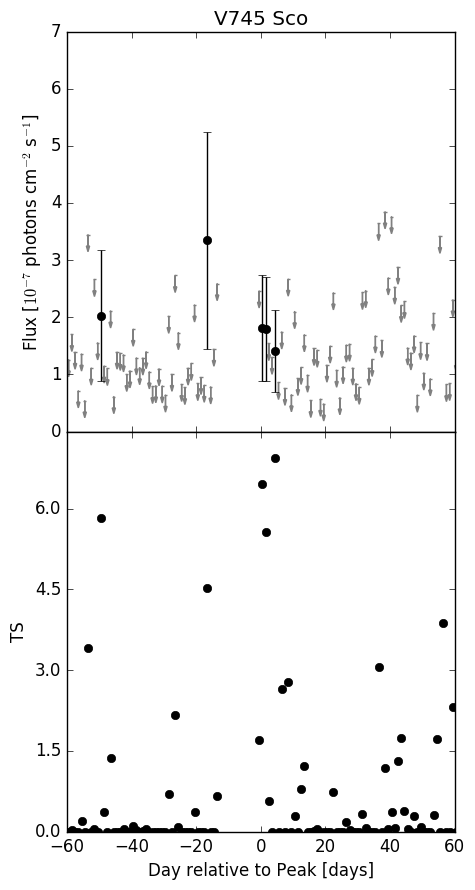}
\noindent
\caption{\small Upper (lower) panel: Flux (TS) vs. time relative to $t_{peak}$ for V745\,Sco in 1-day bins. No data is available from this position in the sky a few days before the peak due to a pointed observation of the LAT to observe M82.}
\label{fig:LCV745Sco}
\end{center}
\end{figure}

\subsection{Luminous Red Novae}
\label{subsec:LRN}
Amongst the novae that have been optically discovered since the start of the \textit{Fermi} mission, V1309 Sco 2008, which was not detected in gamma rays, deserves individual attention. It differs from the novae discussed and can be classified as a luminous red novae, a subclass of eruptive stellar transients that are less luminous than traditional supernovae, but more luminous than classical novae. Those transients might fill the luminosity gap between novae (absolute magnitude ranging  between -4 and -10 mag) and supernovae (absolute magnitude range between -15 and -22 mag)~\citep{2013IAUS..281....9K}. Another well studied example is V838 Mon 2002~\citep{2002A&A...389L..51M}.

Several hypotheses have been proposed to explain the nature of outbursts in different luminous red novae. The most common model is a stellar merger~\cite[see e.g.][]{2014ApJ...786...39N}\footnote{Note that this implies that luminous red novae are not thermonuclear explosions, i.e. technically not novae. However, removing V1309 Sco from our analysis does not significantly change our results.}. Stellar mergers are common with rates of $\sim0.1$/year for events more luminous than V1309 Sco~\citep{2014MNRAS.443.1319K}. 

\citet{2015AAS...22541505M} find that the contact binary star system KIC 9832227 at a distance of 565\,pc shows an orbital period spiraling exponentially down to zero -- similar to V1309 Sco prior to its outburst~\citep{2011A&A...528A.114T}. They predict that KIC 9832227 will merge and produce a red nova eruption in the year 2022.2 with an uncertainty of 0.7 years~\citep{2017ApJ...840....1M}. In case \textit{Fermi} is still operating at this time it would be a unique opportunity to study a stellar merger 4.4 times closer than V1309 Sco.

\section{Study of the Gamma-ray Emissivity Distribution}
\label{sec:pop} 

In the following we perform a population study with Monte Carlo simulations to set constraints on the properties of the gamma-ray emission of novae. \citet{2017MNRAS.465.1218M} initiated a similar population study with a simple flat gamma-ray emissivity model and were able to reproduce the rate of gamma-ray novae observed during 8 years of \Fermi-LAT mission (6 novae in 8 years were included in their analysis). The method we have developed slightly differs and we compare several emissivity models (listed below) taking into account the new data listed in Appendix~\ref{appendixA}. This subsection presents the method and the results that have to be considered as preliminary as the number of detected gamma-ray novae is low.

The aim is to reproduce the optical peak apparent magnitude, $m_{\textrm{max}}$, and gamma-ray flux, $F_{\gamma}$, distribution of novae detected in optical and in gamma rays (see Fig.~\ref{fig:GammaOptCorr}), assuming a gamma-ray emissivity (mean number of photons emitted per second in 15 days) distribution and taking into account the measured peak apparent magnitude distribution of all novae detected in optical from our list (see Fig.~\ref{fig:novaDist}). We build parametrized models and fit their parameters to the observed distributions using a maximum likelihood method. Several emissivity models are tested and their maximum likelihood values are compared to each other.

The observed distributions $O_{\textrm{all}}(m_{\textrm{max}})$ and $O_{\gamma}(m_{\textrm{max}}, F_{\gamma})$ are built with the list of novae presented in Appendix~\ref{appendixA}. However, novae discovered in the I band and V959\,Mon (that has not been observed at peak magnitude) are not included in the $O_{\textrm{all}}(m_{\textrm{max}})$ distribution. V959\,Mon is included in the $O_{\gamma}(m_{\textrm{max}}, F_{\gamma})$ distribution with an observed peak apparent magnitude of 5 as adopted in ~\citet[][see references herein]{2014Sci...345..554A}, as well as the two nova candidates in order to improve the already low statistics. In the following we assume that the gamma-ray emissivity model of symbiotic and classical novae is the same. The luminous red nova V1309 Sco has been excluded from the analysis (note that including it does not change the results significantly).

The modeled distributions $M_{\textrm{all}}(m_{\textrm{max}})$ and $M_{\gamma}(m_{\textrm{max}}, F_{\gamma})$ are generated with Monte Carlo methods. The spatial distribution of novae in our Galaxy is based on the model proposed by \citet{1991ApJ...378..131K} (see also \citet{2008A&A...485..223S} and \citet{2000MNRAS.319..350J} for the method). Their absolute optical magnitude at maximum are distributed as a Gaussian distribution with a mean value of $-7.5$ and a standard deviation of $0.8$ \citep[as in][]{2017ApJ...834..196S} and their apparent maximum magnitude are calculated with the extinction law of \citet{2017ApJ...834..196S}. 
With this spatial distribution and a nova rate of $\sim 50$\,yr$^{-1}$~\citep{2017ApJ...834..196S}, the nova rate obtained in the bulge is $\sim 16$\,yr$^{-1}$ which is in agreement with the rate of (13.8 $\pm$ 2.6)\,yr$^{-1}$ measured by \citet{OGLEREF2} with OGLE observations. 
The decision whether a simulated nova with an apparent peak magnitude $m_{\textrm{max}}$ is detected is taken according to a parametrized probability function (see below). 

The set of those simulated novae is used to calculate the $M_{\textrm{all}}(m_{\textrm{max}})$ distribution. By fitting the model to the observed distribution $O_{\textrm{all}}(m_{\textrm{max}})$ we obtain the likelihood $\lambda_{\textrm{all}}$. Actually, we compute the function $\Lambda_{\textrm{all}} = -2 {\rm ln}(\lambda_{\textrm{all}})$ which depends on the Galactic nova rate and on the parameters of the probability function.
For each simulated nova, the mean gamma-ray flux in a 15-day time window, $F_{\gamma}$, is calculated with the emissivity model and its distance. The decision whether a nova is detected in gamma-ray is obtained with a bootstrap method using the upper-limit fluxes listed in Appendix~\ref{appendixA}. It has to be noted that the probability to detect the gamma-ray flux from a nova does not depend on its Galactic coordinate as in \citet{2017MNRAS.465.1218M}, since we do not observe significant correlation of upper-limit fluxes with Galactic longitudes with our list of novae. 
The modeled distribution $M_{\gamma}(m_{\textrm{max}}, F_{\gamma})$ is built with the set of simulated novae that are detected in optical and in gamma rays. It allows to obtain the likelihood function $\Lambda_{\gamma}$ taking into account the observed distribution $O_{\gamma}(m_{\textrm{max}}, F_{\gamma})$. $\Lambda_{\gamma}$ depends on the Galactic nova rate, the parameters of the probability function and the parameter of the emissivity model. The combination of the two likelihood functions yields to the total likelihood function: $\Lambda_{\tot} = \Lambda_{\textrm{all}} +\Lambda_{\gamma}$. The first term takes into account the uncertainties in the Galactic rate of novae and the probability to detect them. The second term evaluates the quality of the fit of the emissivity model to the data.

As the Galactic nova rate and their probability of detection in optical are highly uncertain~\citep[e.g. see][]{2017ApJ...834..196S}, the parameters of a detection probability law are fitted to the optical peak apparent magnitude distribution, $O_{\textrm{all}}(m_{\textrm{max}})$ for each gamma-ray emissivity model. For simplicity, we assume that the probability to detect a nova follows a decaying exponential law of the peak magnitude such that the probability to detect bright novae is maximum and the probability decreases when the brightness decreases. The parametrized detection probability law is the analytical function $p_0 \times e^{-(m_{\textrm{max}} - p_1)/p_2}$ where $p_0$ is a completeness factor to take into account that even bright novae can be missed in optical~\citep[see][]{2017ApJ...834..196S}, $p_1$ a magnitude shift and $p_2$ a magnitude cut-off. When $m_{\textrm{max}} < p_1$, the function value is $p_0$. There is a degeneracy between the Galactic nova rate and the completeness factor since the same optical peak apparent magnitude distribution can be obtained with a high or low nova rate and a low or high completeness factor. Therefore, in the combined likelihood analyses, the parameter which is fit to the data is $\nu_{\rm novae} \times p_0$, 
the product of the completeness factor ($p_0$) with the Galactic nova rate ($\nu_{\rm novae}$).

As the origin of the gamma-ray emission is not yet known, we consider several emissivity distributions modeled by simple functions. Excepted the Constant model (see below), the emissivity distributions are chosen such as to reproduce the wide range of measured emissivities and to search for a possible link with the maximum absolute magnitude of novae.
The emissivity models tested in this analysis are ($p_\gamma$ for each model is in units of $10^{38}$\,ph/s): 

\begin{itemize}
\item Constant: the mean emissivity in 15 days is the same for all novae. The parameter $p_{\gamma,\textrm{Constant}}$ is the mean emissivity.
\item Uniform: the mean emissivity is randomly distributed according to a uniform distribution from zero to the maximum value, $p_{\gamma,\textrm{Uniform}}$.
\item 10Gauss: the logarithm of the mean emissivity is distributed as a Gaussian with a mean value $\log_{10}$($p_{\gamma,\textrm{10Gauss}}$) and a standard deviation of one (see the end of this section for a discussion on the standard deviation value). 
\item Gaussian: the mean emissivity is distributed as an absolute value of a Gaussian distribution centered on zero ph/s. The parameter $p_{\gamma,\textrm{Gauss}}$ is the standard deviation of the Gaussian.
\item PLslope2: the mean emissivity is distributed as a power-law distribution with a slope of 2 (see the discussion on the slope value at the end of this section). The parameter $p_{\gamma,\textrm{PL2}}$ is the minimum emissivity of the power law.
\item CorrelMv: the mean emissivity is (correlated with) proportional to the maximum luminosity in optical. The emissivity is equal to $p_{\gamma,\textrm{CorrelMv}} \times 10^{(|M_{\textrm{max}}|-7.5)/2.5}$. The parameter $p_{\gamma,\textrm{CorrelMv}}$ is the mean emissivity when the maximum absolute magnitude is $-7.5$.
\item AnticorMv: the mean emissivity is (anticorrelated with) inversely proportional to the maximum luminosity in optical (e.g. see Fig.~\ref{fig:LgVsMag}). The emissivity is equal to $p_{\gamma,\textrm{AnticorMv}} \times 10^{-(|M_{\textrm{max}}|-7.5)/2.5}$, with $M_{\textrm{max}}$ the absolute maximum magnitude. The parameter $p_{\gamma,\textrm{AnticorMv}}$ is the mean emissivity when the maximum absolute magnitude of the nova is $-7.5$. We discuss alternative AnticorMv models at the end of this section. 
\end{itemize}

\begin{figure}[htb!]
\begin{center}
\includegraphics[scale=0.35]{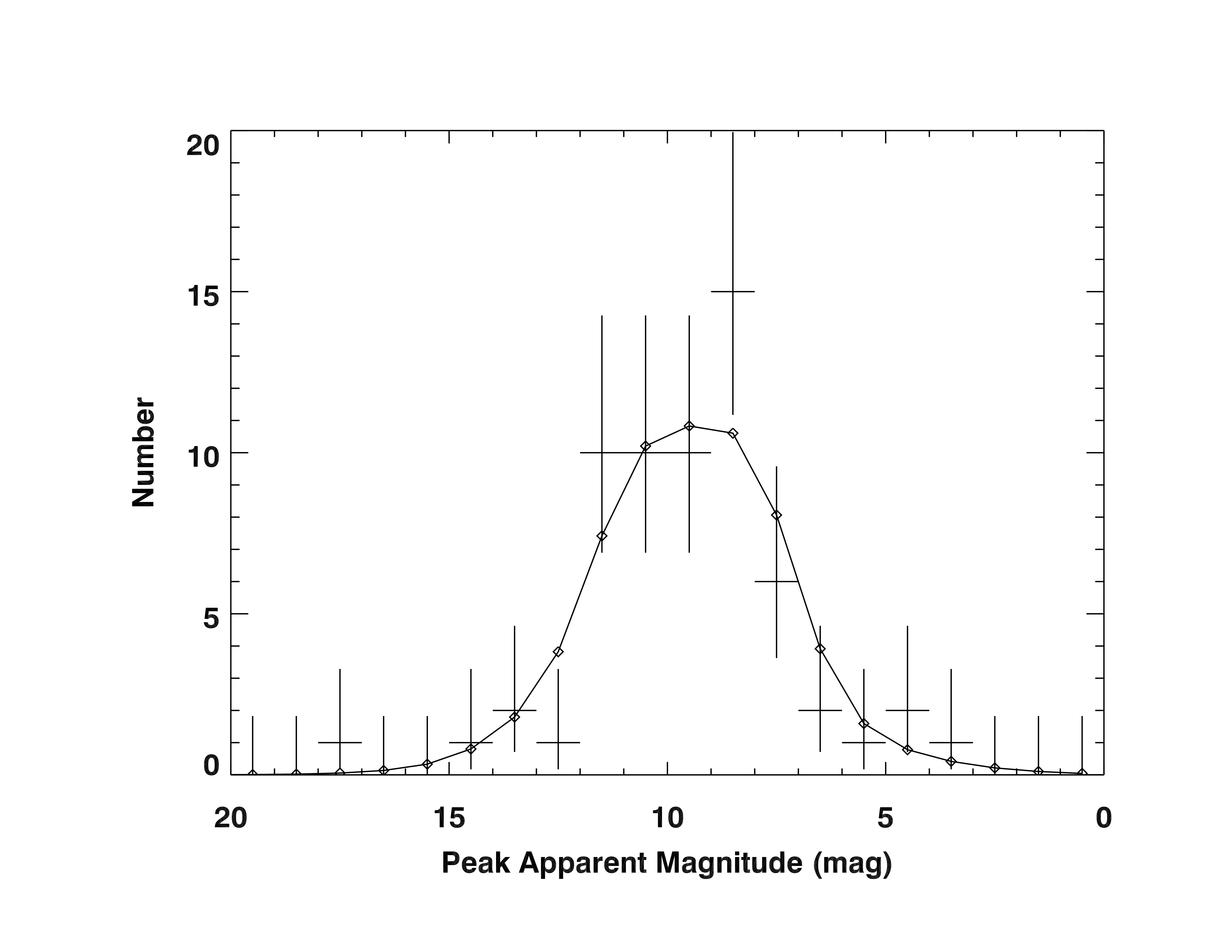}
\noindent
\caption{\small Measured distribution of optical peak apparent magnitudes and best fitting PLslope2 model. The Poissonian (1-$\sigma$) error bars are calculated with the method described by~\citet{1986ApJ...303..336G}.}
\label{fig:mvdistrib}
\end{center}
\end{figure}

\begin{figure*}[!t]
\begin{center}
\includegraphics[scale=0.33]{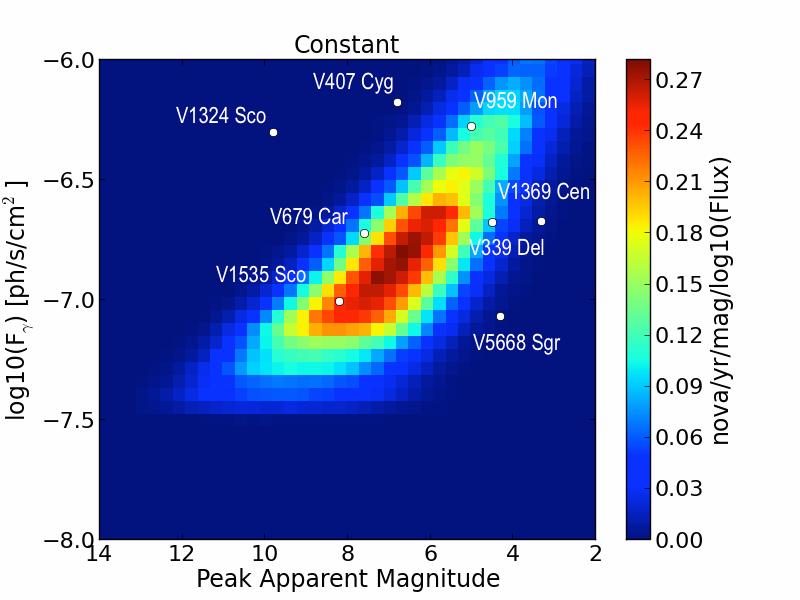}
\includegraphics[scale=0.33]{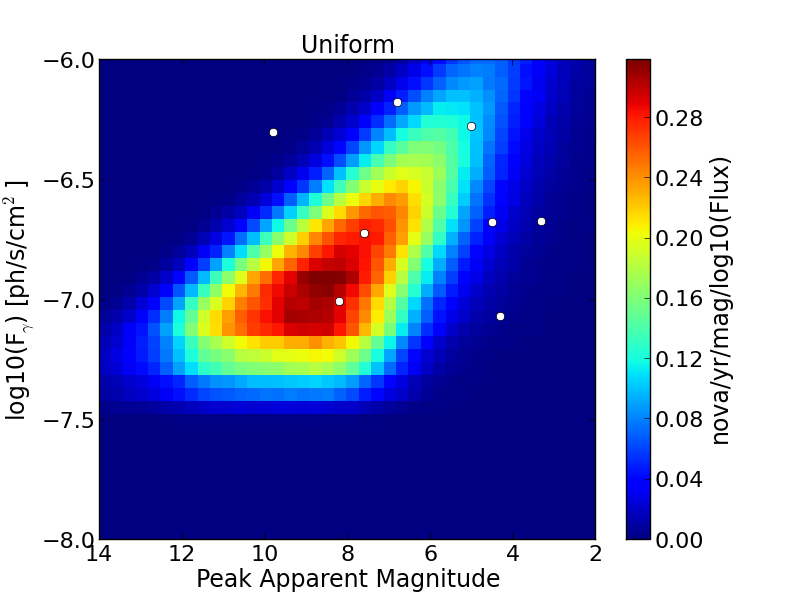}
\includegraphics[scale=0.33]{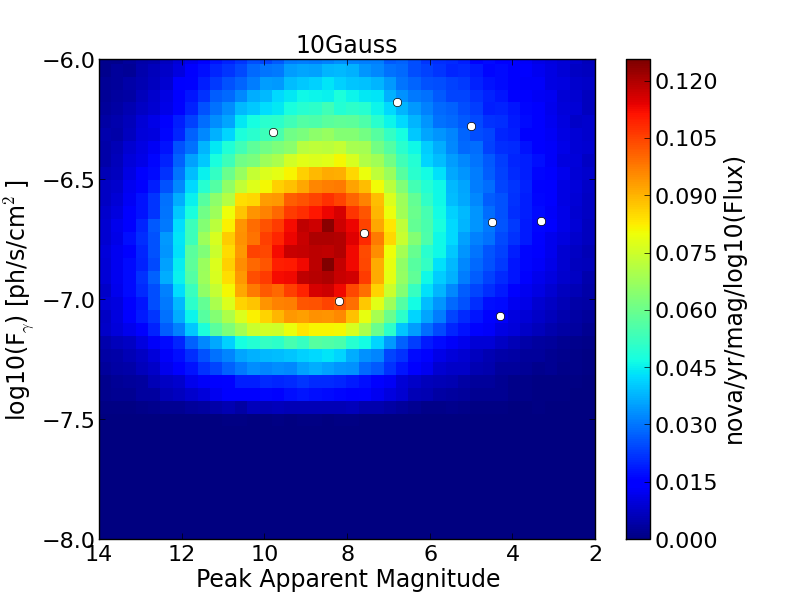}
\includegraphics[scale=0.33]{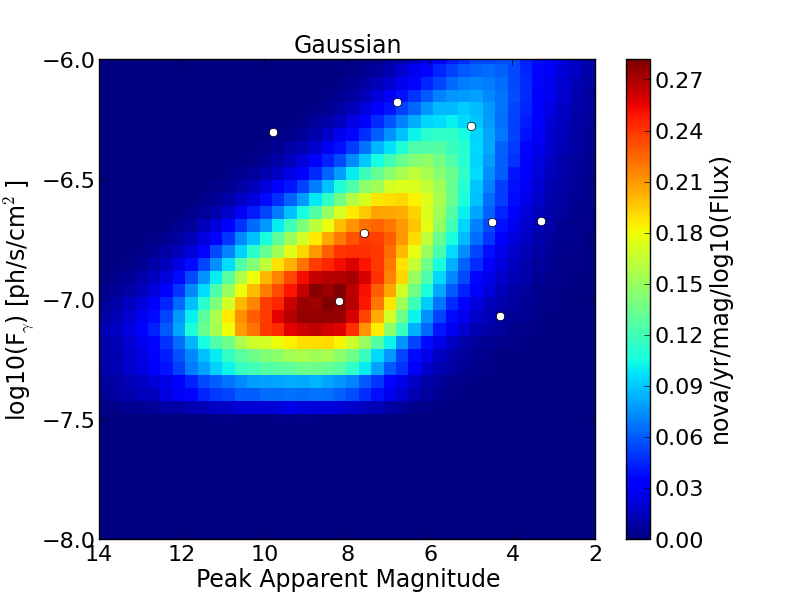}
\includegraphics[scale=0.33]{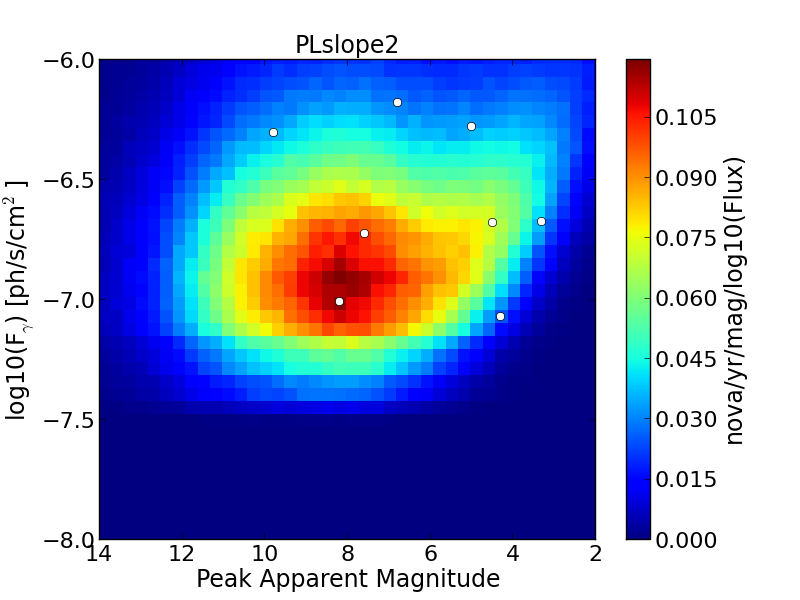}
\includegraphics[scale=0.33]{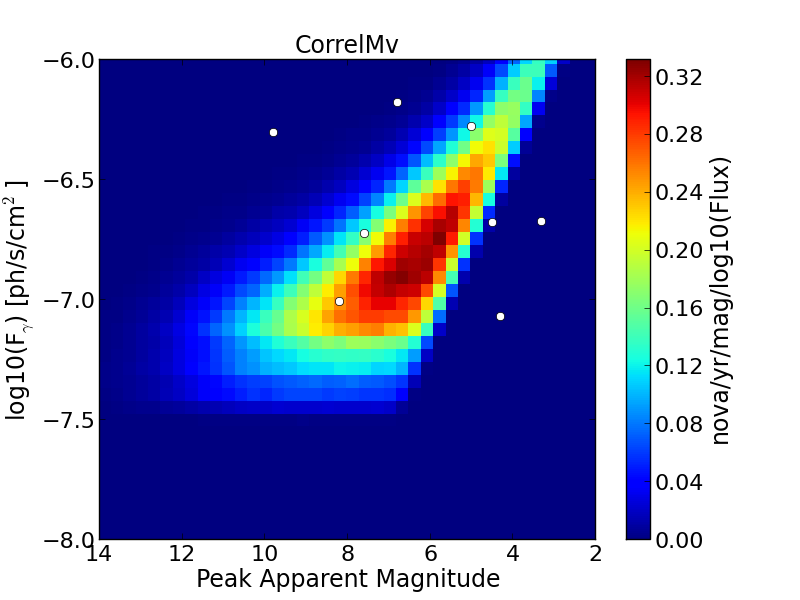}
\includegraphics[scale=0.33]{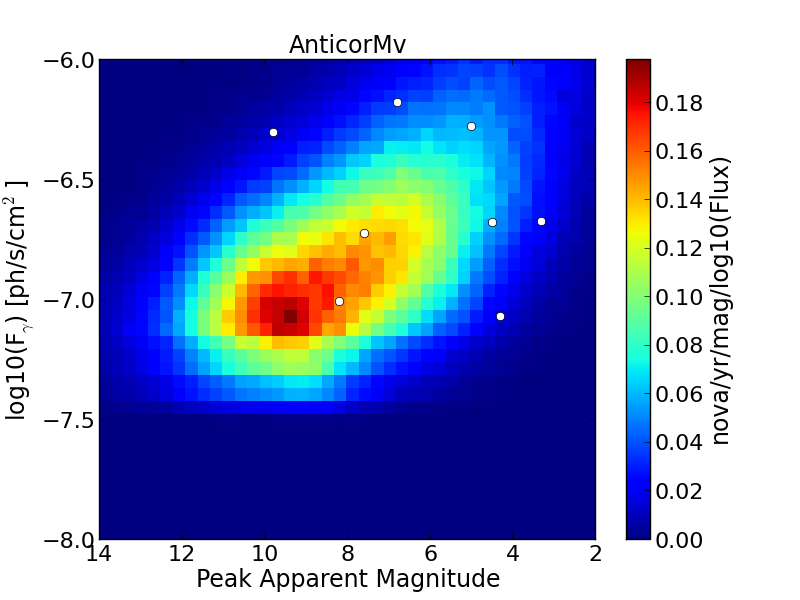}
\includegraphics[scale=0.33]{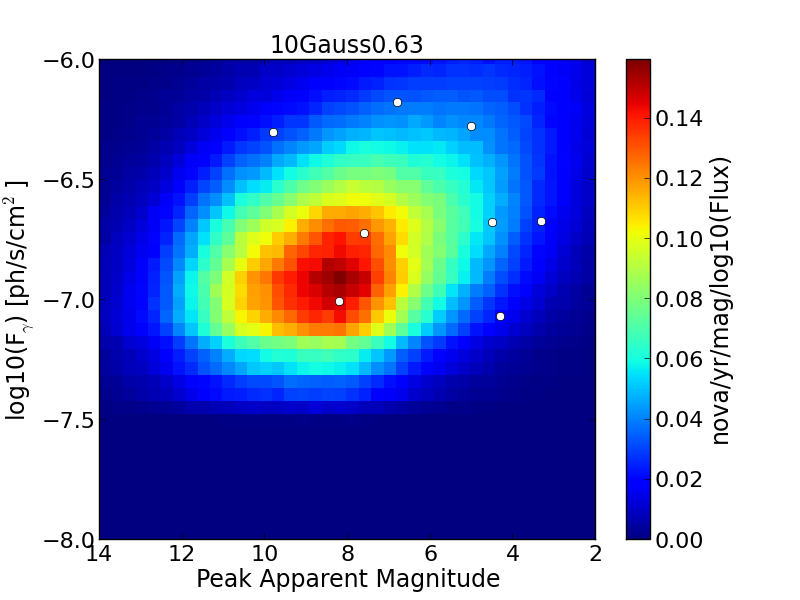}
\caption{Best fitting gamma-ray flux -- optical peak apparent magnitude distributions of several gamma-ray emissivity models (see text). The observed gamma-ray novae are overplotted for comparison.
\label{fig:2ddistrib}}
\end{center}
\end{figure*}

Table \ref{Tab:MLRmodels} presents the results of the likelihood analyses with the best fit emissivity parameters, mean rate of novae detected in optical and in gamma rays and mean rate of novae detected in gamma rays only. The later mean rate is however highly uncertain taking into account the large uncertainty on the Galactic nova rate. Figure \ref{fig:mvdistrib} shows the measured distribution of peak apparent magnitudes and the corresponding best fit distribution of the PLslope2 model obtained by minimizing $\chi^2$ ($\Lambda_{\textrm{all}}$ value of 53.5 for a number of degrees of freedom of 48). 
The best fitting emissivity distribution models are PLslope2, AnticorMv and 10Gauss. The difference in $\Lambda_{\tot}$ values of these 3 models are not large enough (significances $\lesssim$ 2 $\sigma$) to favor one of them over the others. The difference in $\Lambda_{\tot}$ of the 4 other models compared to PLslope2 model are $\gtrsim$ 16 (i.e. significance $> 4 \sigma$). They cannot explain the observed distribution and can be rejected. Note that the constant model (which is equivalent to a standard candle model) can be already rejected by simply comparing the gamma-ray emissivities of novae derived from the measured mean fluxes and distances (see Table~\ref{tab:distance} -- the difference in emissivity goes up to a factor of $\sim$ 30). The rate of novae detected in optical and in gamma rays listed in Table~\ref{Tab:MLRmodels} are in agreement with the observed rate of $\sim$ 1 yr$^{-1}$. The rate of novae detected in gamma rays only, obtained with the 3 best fit models, ranges from 1.4 yr$^{-1}$ to 5.0 yr$^{-1}$, taking into account uncertainties. These novae are not discovered in optical either because they are not observable (completeness effect -- e.g. as the case of V959\,Mon that was too close to the Sun to be discovered when it was at its peak apparent magnitude) or their peak  apparent magnitude is too faint to be discovered as it is the case of novae in the bulge region. With the 3 best fitting models, the detection rate of the gamma-ray emission from novae not discovered in optical within $\vert l \vert < $ 9 deg ranges from 0.5 to 0.8 yr$^{-1}$ (without taking into account statistical uncertainty of the fit). 
The best fitting peak apparent magnitude and gamma-ray flux distributions of analyzed models are presented in Figure~\ref{fig:2ddistrib} with the observed gamma-ray novae. The distributions of the worse models show a rather strong correlation between the gamma-ray flux and the peak apparent magnitude which is not the case for the 3 best fitting models. The extent of the later distributions are broader than the worse modeled ones and favor large gamma-ray flux from faint peak apparent magnitude novae. Novae with peak apparent magnitude as faint as $10-12$\,mag can be detected in gamma rays\footnote{e.g. $\sim 15$\% of novae detected in optical and in gamma rays have a peak apparent magnitude in the $10-12$ range with the PLslope2 model}, in agreement with \citet{2017MNRAS.465.1218M} which estimated that novae with R-band magnitude $\leq12$ and distance $\sim8$\,kpc are good candidates for a detection with the \Fermi-LAT.

\begin{table*}[!t]
\caption{Results of the maximum likelihood analyses. The columns rate and missed rate correspond to rate of novae detected in optical and in gamma rays and rate of novae detected in gamma rays only, respectively.}
\centering
\begin{tabular}{lcccc}
\hline
\hline
Model       & $\Lambda_{\tot}$ & Parameter                        & Rate (yr$^{-1}$)              & Missed rate (yr$^{-1}$) \\
\hline\\
Constant & 177.0 & 1.50$^{+0.62}_{-0.38}$ & 0.78$^{+0.46}_{-0.21}$ & 2.02$^{+0.98}_{-0.60}$ \\
Uniform & 165.2 & 4.75$^{+0.93}_{-0.80}$ & 1.33$^{+0.36}_{-0.25}$ & 4.13$^{+1.02}_{-1.01}$ \\
10Gauss & 150.1 & 0.19$^{+0.26}_{-0.09}$ & 0.78$^{+0.56}_{-0.28}$ & 2.59$^{+2.40}_{-0.93}$ \\
Gaussian & 158.9 & 2.50$^{+1.08}_{-0.39}$ & 1.16$^{+0.52}_{-0.34}$ & 3.34$^{+1.92}_{-1.30}$ \\
PLslope2 & 146.5 & 0.30$^{+0.19}_{-0.13}$ & 0.79$^{+0.39}_{-0.26}$ & 2.63$^{+1.15}_{-0.91}$ \\ 
CorrelMv & 203.2 & 1.30$^{+0.28}_{-0.12}$ & 0.99$^{+0.26}_{-0.21}$ & 2.67$^{+0.67}_{-0.49}$ \\
AnticorMv & 150.1 & 1.30$^{+0.51}_{-0.52}$ & 0.90$^{+0.44}_{-0.37}$ & 2.76$^{+1.38}_{-1.40}$ \\
\hline
\end{tabular}
\label{Tab:MLRmodels}
\end{table*}

We investigated alternative versions of the best fitting models. Maximum likelihood analyses were performed for several slopes of the emissivity power law model. The best fitting slope obtained is 2.01$^{+0.25}_{-0.23}$, which is compatible to the slope of 2 used in the initial analysis (PLslope2 model). Similarly, the analysis made with the gamma-ray emissivity model inversely proportional to the power-law of the maximum luminosity in optical (i.e. AnticorMv$^{\rm slope}$ model), with the slope as a free parameter, yields a best fitting slope value of 1.26$^{+0.67}_{-0.39}$, which is compatible to the slope of 1 used in the initial analysis (AnticorMv model).

The 10Gauss model uses a fixed standard deviation of 1. When the standard deviation is a free parameter, the maximum likelihood analysis results in a slightly better adjustment with a best fit standard deviation of 0.63$^{+0.19}_{-0.14}$ and a $\Lambda_{\tot}$ of 147.7 (see the distribution for the model 10Gauss0.63 in figure \ref{fig:2ddistrib}).

We also tested an emissivity model close to the one used by \citet{2017MNRAS.465.1218M} which assumes a gamma-ray emissivity uniformly distributed between the minimum and maximum values of the detected gamma-ray nova emissivities. This model is similar to the Uniform model but with a non zero value for the lowest emissivity. The likelihood analysis yields a too high rate of novae detected in optical and in gamma rays of 2.6 yr$^{-1}$ and a $\Lambda_{\tot}$ of 180.9 which corresponds to a difference of $\sim$6 $\sigma$ compared to the PLslope2 $\Lambda_{\tot}$ value. 

The results are not significantly changed (i.e. $\Delta\Lambda_{\tot} \lesssim 3$) when other extinction models (e.g. the extinction model of~\citet{1997AJ....114.2043H} as used in~\citet{2008A&A...485..223S}) are applied in the analyses or when a mean maximum absolute magnitude of $-7.2$ is used instead of $-7.5$ 
\citep[see the discussion in][]{2017ApJ...834..196S}; the best fitting completeness corrected novae rate ($\nu_{\rm novae} \times p_0$) changes slightly but the other detection probability law parameters ($p_1$ and $p_2$) remain unchanged.

The optical peak apparent magnitude of V959\,Mon was not measured as this nova was too close to the Sun during its outburst. We check that the peak apparent magnitude uncertainty does not impact too much the results of the analyses by fitting some emissivity models with an optical peak apparent magnitude of 3 instead of 5. The results do not change with the PLslope2 model. This is because the optical peak apparent magnitude shift results in a position of the nova in the $m_{max}$-$F_{\gamma}$ diagram where the distribution value is similar (see Fig. \ref{fig:mvdistrib}). With the AnticorMv model the $\Lambda_{\tot}$ increases from 150.1 to 151.7 but the best fit parameters do not change significantly.

\section{Summary and Conclusion}
\label{sec:conclusion}
We have systematically searched for gamma-ray emission from 75 novae. We confirmed the six already known gamma-ray novae and find two additional candidates, which are found at $3\sigma$ significance (not including trial factors) but barely reach a 2$\sigma$ significance after trials correction. V679\,Car is a classical nova, while V1535\,Sco is a symbiotic system. Their spectral characteristics and their duration are similar to the previous gamma-ray detected ones. If we consider the two candidates as detected, the observed rate of LAT-detected novae is about $\sim1$ per year (8 novae in 7.4\,years). 

The paper presents the results of the analysis of novae discovered with their optical flux. It is not excluded that some novae, not discovered in optical (e.g.~due to extinction), emit enough gamma-ray flux to be detected with \Fermi-LAT. If they are bright enough they should be found by the \textit{Fermi} all-sky variability analysis (FAVA)~\citep{2016arXiv161203165A}, which searches for gamma-ray flares on the time-scale of one week. Five flares of the second FAVA catalog have been associated with the gamma-ray bright novae studied in this paper. It is possible that some of the unidentified FAVA flares are produced by novae, which were missed by optical surveys.

We provide the measured gamma-ray flux or flux upper limits for the non-detections. We find an indication at $3\sigma$ significance for a sub-threshold population of dim novae. No correlation of optical peak magnitude and gamma-ray flux was found. Non-detections in gamma rays can be caused by a too large distance, by absorption of the high-energy emission or are due to absence of particle acceleration in the nova. The provided gamma-ray flux upper limits will be useful for future modeling of physical processes taking place in novae.

We compare our measurements to simulated gamma-ray emissivity models and find that a power-law distribution with a slope of 2, an emissivity distribution inversely proportional to the maximum luminosity in optical and a log-normal distribution of the emissivity match the data best. A constant emissivity (i.e. assuming novae are standard candles) can be rejected. In general the best fitting distributions are not extended enough compared to the observed distributions. The location of detected gamma-ray novae are spread throughout the $m_{\textrm{max}}$-$F_{\gamma}$ diagram. This suggests that the true emissivity distribution would/should be more complicated than the tested ones. 
Moreover, we assumed that the emissivities of all gamma-ray novae originate from the same distribution while it is not excluded that they differ with the chemical composition of the white dwarf (CO vs. ONe) and the type of nova (e.g. classical vs. symbiotic, see discussion in section~\ref{subsec:recsymnovae}).

If this is the case, such analysis has to be made independently for each type of nova. Therefore, we cannot yet conclude on the gamma-ray emissivity distribution of novae.

Physical models of the gamma-ray emission combined with novae properties would allow to build more reliable emissivity distributions that will be better constrained in the future with a larger sample of observed gamma-ray novae (i.e. larger statistics) and a finer/improved analysis (e.g. modeled distributions including the Galactic coordinates $l, b$ of simulated novae and more elaborated optical detection probability law). 

The gamma-ray properties of the small number of LAT-detected novae (1 symbiotic and 5 classical; see Table~\ref{tab:knownNovae}) plus the two new candidates presented here  (1 symbiotic and 1 classical; Table~\ref{tab:newNovae}) and distances are not measured well enough to claim a firm difference in their emissivities. Observationally, their gamma-ray emission properties (spectra, light curves) are similar, which could suggest a common gamma-ray emission origin. However, as noted in \citet{2014Sci...345..554A}, small differences exist that could imply different emission mechanisms: (i) the power-law index of the V407 Cyg spectrum is smaller than the classical nova ones but it is compatible with the one of V959 Mon taking into account the statistical uncertainties, and (ii) the gamma-ray onset of both symbiotic novae is coincident with the optical peak magnitude (see discussion in Section~\ref{subsec:recsymnovae}) while there is a delay found in three of the classical novae, with one exception where the gamma-ray onset of the classical nova V1324 Sco occurred before the optical peak. On the modelling side, the gamma-ray emission of the symbiotic nova V407 Cyg could be explained with an interaction between the ejecta and the dense wind of the secondary~\citep{2010Sci...329..817A,2013A&A...551A..37M} while internal shocks are favored to explain the one of classical novae~\citep{2014Sci...345..554A,2014MNRAS.442..713M}. However, \citet{2017AASubmitted} were recently able to reproduce reasonably well the gamma-ray spectrum and light curve of V407 Cyg with an internal shock model. Taking into account their recent results and the similarities in gamma-ray emission properties, it is possible that the gamma-rays in symbiotic novae could be due to a combination of external and internal shocked emission. Further observations of gamma-ray novae would provide better statistics to separate the two populations of novae, as well as better measured characteristics to detect clear differences of their emissions.

Up to now no VHE ($>0.1$ TeV) gamma-ray emission was detected from novae~\citep{Ahnen:2015pha,2012ApJ...754...77A}. Future optical surveys such as the All-Sky Automated Survey for Supernovae~\citep[ASAS-SN;][]{2014AAS...22323603S} and the Zwicky Transient Facility~\citep[ZTF;][]{2014htu..conf...27B} will deliver more optical nova detections, which would allow an extended search for gamma-ray emission from those sources.

\section*{Acknowledgements}

The \textit{Fermi} LAT Collaboration acknowledges generous ongoing support
from a number of agencies and institutes that have supported both the
development and the operation of the LAT as well as scientific data analysis.
These include the National Aeronautics and Space Administration and the
Department of Energy in the United States, the Commissariat \`a l'Energie Atomique
and the Centre National de la Recherche Scientifique / Institut National de Physique
Nucl\'eaire et de Physique des Particules in France, the Agenzia Spaziale Italiana
and the Istituto Nazionale di Fisica Nucleare in Italy, the Ministry of Education,
Culture, Sports, Science and Technology (MEXT), High Energy Accelerator Research
Organization (KEK) and Japan Aerospace Exploration Agency (JAXA) in Japan, and
the K.~A.~Wallenberg Foundation, the Swedish Research Council and the
Swedish National Space Board in Sweden.
 
Additional support for science analysis during the operations phase is gratefully
acknowledged from the Istituto Nazionale di Astrofisica in Italy and the Centre
National d'\'Etudes Spatiales in France. This work performed in part under DOE
Contract DE-AC02-76SF00515.

We acknowledge with thanks the variable
star observations from the AAVSO International
Database contributed by observers worldwide and used
in this research.

A.~F.~was supported by the Initiative and Networking Fund of the Helmholtz Association.

Work by C.C.C. at NRL is supported in part by NASA DPR S-15633-Y and Fermi Guest Investigator program 14-FERMI14-0005.

\onecolumn
\appendix

\section{List of Novae}
\label{appendixA}
The following table contains all 75 novae from the catalog including their optical peak time and optical peak apparent magnitude. It further presents the results of the sliding time window search for gamma-ray emission: the maximal TS of all tested time windows is shown together with the gamma-ray flux (or 95\% flux upper limit) in that 15-day time window. $\Delta T$ indicates the difference of the central time of the time window and the peak time. The given peak magnitudes are predominantly Visual or in V-band, unless otherwise indicated.
\begin{landscape}
\begin{tiny}
\begin{longtable*}{ L{0.07\linewidth} p{0.03\linewidth}  p{0.04\linewidth} p{0.07\linewidth} p{0.03\linewidth} p{0.04\linewidth} p{0.04\linewidth} p{0.03\linewidth} p{0.1\linewidth}  L{0.3\linewidth} }
\hline
\hline
Name & $l$     & $b$         & Peak Time &$\Delta T$  & Peak & $t_2$  & TS$_{max}$ & Flux                             & Reference\\
     & [$^\circ$] & [$^\circ$] &           & [days]     &  mag & [days] &    & [$10^{-7}$\,cm$^{-2}\,$s$^{-1}$] & \\
\hline
V1212 Cen & 313.95 & -3.47 & 2008-08-30 & 15.5 & 8.4 & 28.0 & 3.1 & $<$0.9 & CBET 1497 \citep{CBET1497}\\
V1309 Sco & 359.79 & -3.13 & 2008-09-06 & 11.5 & 7.1 & 8.0 & 4.1 & $<$1.0 & CBET 1496 \citep{CBET1496}, IAUC 8972 \citep{IAUC8972}\\
V1721 Aql & 40.97 & -0.08 & 2008-09-22 & -8.5 & 14.0 & 6.0 & 2.8 & $<$3.0 & IAUC 8989 \citep{IAUC8989}\\
QY Mus & 305.33 & -4.86 & 2008-09-30 & -8.5 & 8.0 & 48.0 & 0.3 & $<$0.6 & IAUC 8990 \citep{IAUC8990}\\
XMMU J115113.3-623730 & 296.07 & -0.56 & 2008-11-23 & -12.5 & 10.5 & - & 1.5 & $<$1.0 & ATel 2746 \citep{ATel2746}\\
V679 Car & 291.47 & -0.55 & 2008-11-26 & 13.5 & 7.6 & 13.0 & 17.6 & 1.9 $\pm$ 0.5 & IAUC 8999 \citep{IAUC8999}\\
V5580 Sgr & 4.70 & -6.54 & 2008-11-30 & 21.5 & 7.8 & - & 2.5 & $<$0.7 & CBET 1591 \citep{CBET1591}, IAUC 9004 \citep{IAUC9004}, IAUC 9005 \citep{IAUC9005}\\
V5582 Sgr & 7.53 & 4.72 & 2009-02-26 & 19.5 & 10.3 & 4.0 & 0.2 & $<$0.6 & IAUC 9049 \citep{IAUC9049}\\
V5581 Sgr & 2.25 & 1.76 & 2009-04-21 & 3.5 & 11.7 & 6.0 & 8.0 & $<$1.7 & IAUC 9041 \citep{IAUC9041}, IAUC 9048 \citep{IAUC9048}\\
V1213 Cen & 307.29 & -1.43 & 2009-05-08 & -0.5 & 8.5 & 9.0 & 2.5 & $<$0.9 & CBET 1800 \citep{CBET1800}, IAUC 9043 \citep{IAUC9043}\\
V5583 Sgr & 358.10 & -6.39 & 2009-08-08 & 23.5 & 7.0 & 4.0 & 4.1 & $<$0.6 & CBET 1899 \citep{CBET1899}, CBET 1900 \citep{CBET1900}, IAUC 9061 \citep{IAUC9061}\\
V2672 Oph & 1.02 & 2.53 & 2009-08-17 & -0.5 & 10.0 & 1.0 & 7.1 & $<$2.0 & CBET 1910 \citep{CBET1910}, CBET 1911 \citep{CBET1911}, CBET 1912 \citep{CBET1912}, IAUC 9064 \citep{IAUC9064}\\
V5584 Sgr & 16.17 & -3.10 & 2009-10-31 & 25.5 & 9.3 & 23.0 & 0.5 & $<$0.8 & CBET 1994 \citep{CBET1994}, CBET 1995 \citep{CBET1995}, CBET 1999 \citep{CBET1999}, IAUC 9089 \citep{IAUC9089}\\
KT Eri & 207.99 & -32.02 & 2009-11-15 & 25.5 & 5.4 & 5.0 & 0.7 & $<$0.3 & CBET 2050 \citep{CBET2050}, CBET 2053 \citep{CBET2053}, CBET 2055 \citep{CBET2055}, IAUC 9098 \citep{IAUC9098}\citep{2010ApJ...724..480H}\\
V496 Sct & 25.28 & -1.77 & 2009-11-18 & -0.5 & 7.0 & 33.0 & 10.0 & $<$2.0 & CBET 2008 \citep{CBET2008}, CBET 2034 \citep{CBET2034}, IAUC 9093 \citep{IAUC9093}\\
V1722 Aql & 49.08 & 2.01 & 2009-12-17 & 3.5 & 10.0 & 8.0 & 3.3 & $<$1.1 & CBET 2076 \citep{CBET2076}, IAUC 9100 \citep{IAUC9100}\\
V2673 Oph & 5.49 & 4.96 & 2010-01-17 & -12.5 & 8.3 & 10.0 & 0.4 & $<$0.6 & CBET 2128 \citep{CBET2128}, CBET 2139 \citep{CBET2139}, IAUC 9111 \citep{IAUC9111}\\
V5585 Sgr & 2.33 & -4.17 & 2010-01-20 & 1.5 & 8.5 & 8.0 & 6.1 & $<$1.0 & CBET 2140 \citep{CBET2140}, CBET 2142 \citep{CBET2142}, IAUC 9112 \citep{IAUC9112}\\
U Sco & 357.67 & 21.87 & 2010-01-28 & 3.5 & 7.5 & 2.0 & 2.0 & $<$0.6 & IAUC 9111 \citep{IAUC9111}\\
V2674 Oph & 357.84 & 3.58 & 2010-02-21 & 19.5 & 8.8 & 14.0 & 0.4 & $<$1.0 & CBET 2176 \citep{CBET2176}, CBET 2179 \citep{CBET2179}, CBET 2185 \citep{CBET2185}, IAUC 9119 \citep{IAUC9119}\\
V1310 Sco & 348.50 & 2.19 & 2010-02-22 & 17.5 & 10.2 & 1.0 & 0.4 & $<$1.0 & CBET 2183 \citep{CBET2183}, CBET 2186 \citep{CBET2186}, IAUC 9120 \citep{IAUC9120}\\
OGLE-2010-NOVA-03 & 359.58 & -5.18 & 2010-03-05 & -6.5 & 12.0 (I) & 53.0 & 0.2 & $<$0.6 & \citep{OGLEREF2}\\
V407 Cyg & 86.98 & -0.48 & 2010-03-10 & 6.7 & 6.8 & 4.0 & 526.7 & 6.6 $\pm$ 0.5 & CBET 2199 \citep{CBET2199}, CBET 2204 \citep{CBET2204}, CBET 2205 \citep{CBET2205}, CBET 2210 \citep{CBET2210}, IAUC 9130 \citep{IAUC9130}\\
WISE J181834.00-284919.6 & 3.65 & -6.22 & 2010-03-11 & -0.5 & 11.0 (I) & - & 1.0 & $<$0.9 & ATel 4268 \citep{ATel4268}, ATel 4296 \citep{ATel4296}\\
V1311 Sco & 346.52 & 3.39 & 2010-04-25 & 5.5 & 8.0 & 2.0 & 6.1 & $<$1.4 & CBET 2262 \citep{CBET2262}, CBET 2265 \citep{CBET2265}, IAUC 9142 \citep{IAUC9142}\\
V5586 Sgr & 1.47 & -1.02 & 2010-04-27 & 1.5 & 10.0 & 6.0 & 4.3 & $<$2.1 & CBET 2261 \citep{CBET2261}, CBET 2264 \citep{CBET2264}, IAUC 9140 \citep{IAUC9140}\\
OGLE-2010-NOVA-01 & 5.66 & 1.05 & 2010-06-30 & 9.5 & 13.0 (I) & 7.0 & 1.3 & $<$1.3 & \citep{OGLEREF2}\\
V1723 Aql & 29.13 & -0.88 & 2010-09-11 & 13.5 & 12.4 & - & 3.3 & $<$2.3 & IAUC 9167 \citep{IAUC9167}\\
OGLE-2010-NOVA-02 & 358.83 & 1.49 & 2010-10-20 & -6.5 & 11.2 (I) & - & 5.9 & $<$1.6 & \citep{OGLEREF2}\\
V5587 Sgr & 4.82 & 2.36 & 2011-01-25 & 21.5 & 11.2 & 19.0 & 8.3 & $<$1.2 & CBET 2644 \citep{CBET2644}, IAUC 9196 \citep{IAUC9196}\\
OGLE-2011-NOVA-02 & 358.22 & 1.10 & 2011-02-10 & 13.5 & 13.0 (I) & 40.0 & 4.0 & $<$1.6 & \citep{OGLEREF2}\\
V5588 Sgr & 7.84 & -1.88 & 2011-04-07 & 21.5 & 11.4 & 7.0 & 0.0 & $<$1.1 & CBET 2679 \citep{CBET2679}, IAUC 9203 \citep{IAUC9203}\\
T Pyx & 257.21 & 9.71 & 2011-05-12 & 17.5 & 6.1 & 20.0 & 4.6 & $<$0.6 & CBET 2700 \citep{CBET2700}, IAUC 9205 \citep{IAUC9205}\\
V1312 Sco & 346.07 & 3.04 & 2011-06-01 & 7.5 & 9.5 & 4.0 & 1.8 & $<$0.6 & CBET 2735 \citep{CBET2735}, IAUC 9216 \citep{IAUC9216}\\
PR Lup & 319.98 & 3.66 & 2011-08-14 & -4.5 & 8.5 & 6.0 & 0.5 & $<$0.5 & CBET 2796 \citep{CBET2796}, IAUC 9228 \citep{IAUC9228}\\
V1313 Sco & 341.31 & 4.12 & 2011-09-07 & -7.0 & 9.1 & 1.5 & 4.3 & $<$1.2 & CBET 2813 \citep{CBET2813},  IAUC 9216 \citep{IAUC9216}\\
V5590 Sgr & 4.23 & -4.04 & 2011-10-05 & 25.5 & 10.1 & - & 1.7 & $<$0.7 & CBET 3140 \citep{CBET3140},\citep{ OGLEREF1}\\
OGLE-2011-NOVA-01 & 357.72 & 1.97 & 2011-10-17 & -4.5 & 12.5 (I) & - & 4.9 & $<$1.8 & \citep{OGLEREF2}\\
V965 Per & 151.64 & -17.87 & 2011-11-07 & 15.5 & 17.5 & - & 4.6 & $<$0.4 & IAUC 9247 \citep{IAUC9247}\\
V834 Car & 290.18 & -4.28 & 2012-02-26 & 13.5 & 10.2 & 17.0 & 7.6 & $<$0.9 & CBET 3040 \citep{CBET3040}, IAUC 9251 \citep{IAUC9251}\\
V1368 Cen & 309.45 & 3.99 & 2012-03-24 & 23.0 & 8.4 & 4.5 & 1.1 & $<$0.5 & CBET 3073 \citep{CBET3073}\\
TCP J14250600-5845360 & 314.83 & 1.93 & 2012-04-05 & 8.7 & 9.3 & 11.0 & 9.3 & $<$1.5 & CBAT report\footnote{\url{http://www.cbat.eps.harvard.edu/unconf/followups/J14250600-5845360.html}}\\
V2676 Oph & 0.26 & 5.30 & 2012-04-20 & 25.5 & 10.5 & 60.0 & 2.0 & $<$1.1 & CBET 3072 \citep{CBET3072}\\
V5589 Sgr & 4.98 & 3.07 & 2012-04-21 & -7.2 & 8.8 & 3.0 & 8.1 & $<$1.3 & CBET 3089 \citep{CBET3089}\\
OGLE-2012-NOVA-01 & 2.73 & -1.25 & 2012-05-05 & 7.5 & 12.5 (I) & 40.0 & 3.4 & $<$1.7 & ATel 4323 \citep{ATel4323},\citep{ OGLEREF2}\\
V2677 Oph & 2.87 & 3.26 & 2012-05-20 & 5.0 & 9.5 & 2.0 & 1.0 & $<$1.4 & CBET 3124 \citep{CBET3124}\\
V1324 Sco & 357.43 & -2.87 & 2012-06-20 & -0.5 & 9.8 & 23.0 & 190.7 & 5.0 $\pm$ 0.5 & CBET 3136 \citep{CBET3136}\\
V959 Mon & 206.34 & 0.08 & 2012-06-24\footnote{Gamma-ray peak time from \Fermi-LAT light curve~\citep{2014Sci...345..554A}} & 3.5 & 5.0 & - & 195.0 & 5.2 $\pm$ 0.5 & CBET 3202 \citep{CBET3202}\\
V5591 Sgr & 7.22 & 2.54 & 2012-06-27 & 5.2 & 8.9 & 1.0 & 2.3 & $<$1.3 & CBET 3156 \citep{CBET3156}\\
OGLE-2011-BLG-1444 & 356.48 & -3.29 & 2012-07-01 & -6.5 & 10.2 (I) & - & 5.4 & $<$1.3 & \citep{OGLEREF1}\\
V5592 Sgr & 4.81 & -6.09 & 2012-07-07 & 0.9 & 7.7 & 18.0 & 0.9 & $<$0.9 & CBET 3166 \citep{CBET3166}, CBET 3177 \citep{CBET3177}\\
V5593 Sgr & 12.36 & -1.89 & 2012-07-23 & 11.5 & 11.0 & 29.0 & 2.6 & $<$1.0 & CBET 3182 \citep{CBET3182}, CBET 3184 \citep{CBET3184}\\
V1724 Aql & 32.78 & -0.40 & 2012-10-20 & 20.6 & 11.2 & - & 7.2 & $<$2.9 & CBET 3273 \citep{CBET3273}, CBET 3287 \citep{CBET3287}\\
V809 Cep & 110.65 & 0.40 & 2013-02-02 & -5.3 & 9.8 & 4.0 & -0.0 & $<$0.3 & CBET 3397 \citep{CBET3397}\\
OGLE-2013-NOVA-04 & 4.90 & -1.18 & 2013-04-14 & 15.5 & 13.0 (I) & 51.0 & 2.6 & $<$1.2 & \citep{OGLEREF2}\\
V1533 Sco & 352.62 & -1.74 & 2013-06-03 & -12.5 & 11.0 & 5.0 & 4.6 & $<$2.2 & CBET 3542 \citep{CBET3542}, CBET 3556 \citep{CBET3556}\\
V339 Del & 62.20 & -9.42 & 2013-08-16 & 9.5 & 4.5 & 11.0 & 363.9 & 2.1 $\pm$ 0.2 & CBET 3628 \citep{CBET3628}, IAUC9258 \citep{IAUC9258}\\
V1830 Aql & 37.10 & -0.99 & 2013-10-28 & 2.7 & 13.5 & 2.0 & 1.9 & $<$2.2 & CBET 3691 \citep{CBET3691}, CBET 3708 \citep{CBET3708}, IAUC9263 \citep{IAUC9263}\\
V556 Ser & 18.10 & 4.12 & 2013-11-24 & 5.5 & 11.7 & - & 2.9 & $<$0.8 & CBET 3724 \citep{CBET3724}, IAUC9264 \citep{IAUC9264}\\
V1369 Cen & 310.98 & 2.73 & 2013-12-14 & -1.2 & 3.3 & 17.0 & 94.2 & 2.1 $\pm$ 0.3 & CBET 3732 \citep{CBET3732}, IAUC9265 \citep{IAUC9265}\\
V5666 Sgr & 9.88 & -4.66 & 2014-01-26 & 0.7 & 8.7 & 3.0 & 0.5 & $<$0.8 & CBET 3802 \citep{CBET3802}\\
V745 Sco & 357.36 & -4.00 & 2014-02-06 & 10.7 & 8.7 & 3.0 & 4.3 & $<$0.5 & CBET 3803 \citep{CBET3803},\citep{ OGLEREF1}\\
OGLE-2014-NOVA-01 & 5.87 & 0.97 & 2014-02-16 & 23.5 & 15.0 (I) & 75.0 & 1.4 & $<$0.7 & \citep{OGLEREF2}\\
V962 Cep & 97.31 & 9.82 & 2014-03-14 & 19.5 & 11.0 & 13.0 & -0.0 & $<$0.5 & CBET 3825 \citep{CBET3825}\\
V1534 Sco & 354.33 & 3.99 & 2014-03-26 & -3.4 & 10.1 & 3.0 & 1.0 & $<$0.4 & CBET 3841 \citep{CBET3841},\citep{2017MNRAS.469.4341M}\\
V2659 Cyg & 70.53 & -3.29 & 2014-04-10 & 23.5 & 9.4 & 65.0 & 3.9 & $<$0.7 & CBET 3842 \citep{CBET3842}\\
V1535 Sco & 349.90 & 3.94 & 2015-02-11 & 0.7 & 8.2 & 5.5 & 17.2 & 1.0 $\pm$ 0.3 & CBET 4078 \citep{CBET4078},\citep{2017MNRAS.469.4341M}\\
V5667 Sgr & 5.80 & -4.04 & 2015-02-25 & 17.5 & 9.0 & 3.0 & 3.2 & $<$0.6 & CBET 4079 \citep{CBET4079}\\
V5668 Sgr & 5.38 & -9.87 & 2015-03-21 & 9.0 & 4.3 & 40.0 & 61.7 & 0.75 $\pm$ 0.14 & CBET 4080 \citep{CBET4080}\\
V5852 Sgr & 357.17 & -2.37 & 2015-04-03 & -10.5 & 12.3 (I) & 34.0 & 1.8 & $<$0.8 & \citep{OGLEREF3}\\
V2944 Oph & 6.64 & 8.58 & 2015-04-14 & -7.2 & 9.0 & 2.0 & 0.9 & $<$0.5 & CBET 4086 \citep{CBET4086}\\
V5669 Sgr & 2.57 & -3.06 & 2015-09-27 & 17.5 & 8.7 & 23.0 & 3.0 & $<$0.8 & CBET 4145 \citep{CBET4145}\\
V2949 Oph & 2.78 & 4.59 & 2015-10-12 & 13.5 & 11.7 & - & 1.7 & $<$0.7 & CBET 4150 \citep{CBET4150}\\
V1831 Aql & 49.84 & 0.32 & 2015-10-13 & 11.5 & 13.8 & - & 0.9 & $<$1.4 & ATel 8126 \citep{ATel8126}\\
V5850 Sgr & 12.63 & -2.65 & 2015-10-31 & -0.5 & 11.3 & - & 7.5 & $<$1.5 & CBET 4163 \citep{CBET4163}\\
\hline
\hline
\label{tab:novaeList}
\end{longtable*}
\end{tiny}
\end{landscape}

\section{Light Curves of Recurrent and Symbiotic Novae}
\label{app:LC}

\begin{figure}[htb!]
\begin{center}
\includegraphics[scale=0.6]{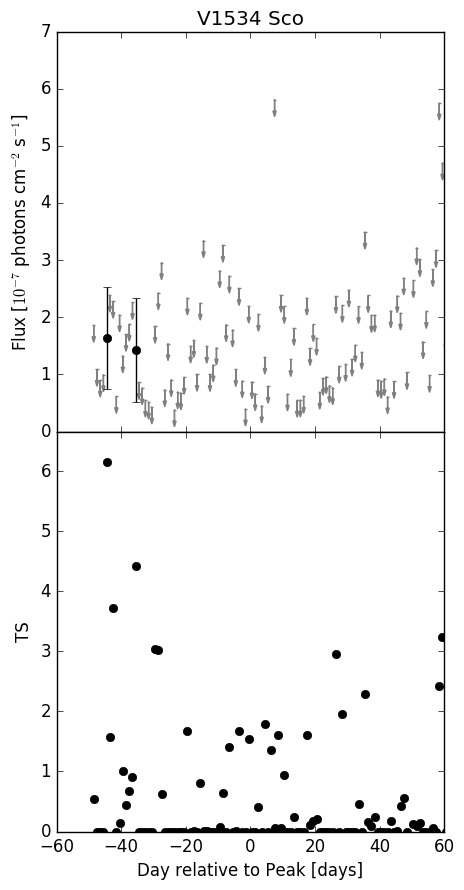}
\includegraphics[scale=0.6]{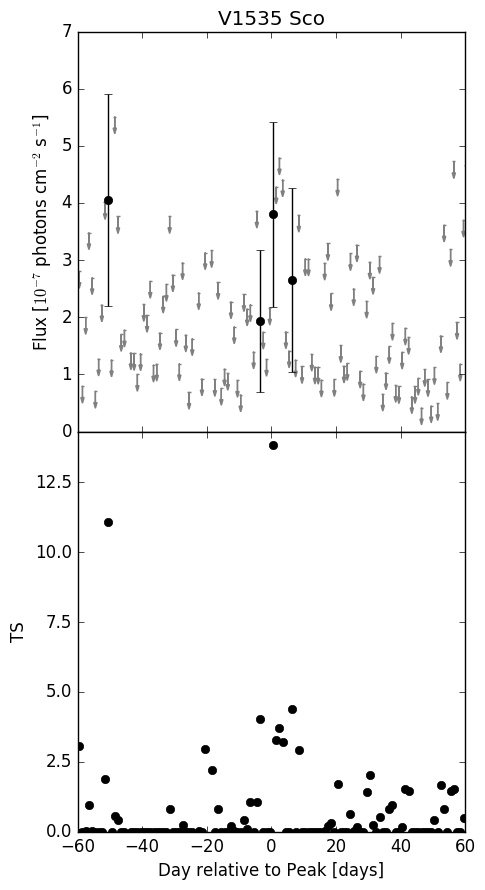}
\noindent
\caption{\small Upper (lower) panel: Flux (TS) vs. time relative to $t_{peak}$ for the symbiotic novae V1534\,Sco (left) and V1535\,Sco (right) in 1-day bins. }
\end{center}
\end{figure}

\begin{figure}[htb!]
\begin{center}
\includegraphics[scale=0.6]{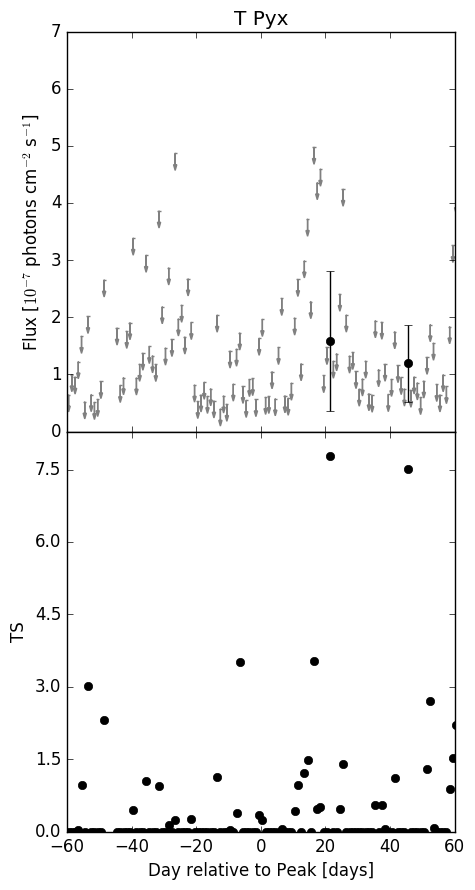}
\includegraphics[scale=0.6]{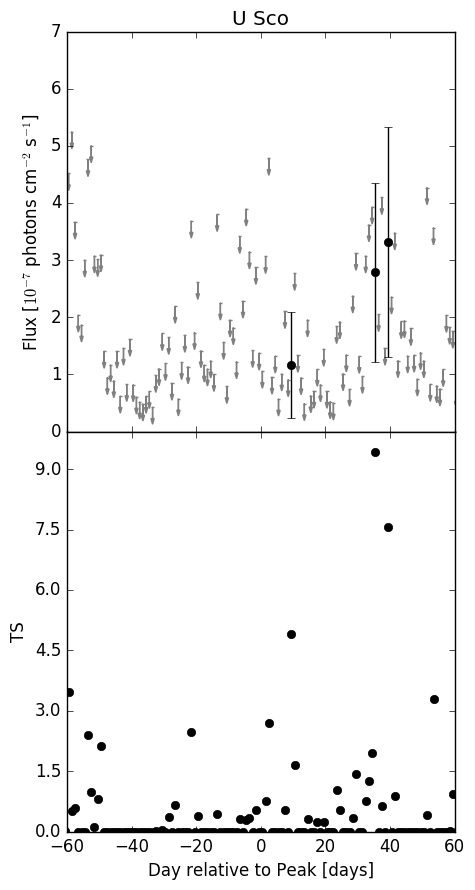}
\noindent
\caption{\small Upper (lower) panel: Flux (TS) vs. time relative to $t_{peak}$ for the recurrent novae T\,Pyx (left) and U\,Sco (right) in 1-day bins. }
\end{center}
\end{figure}

\newpage

{\footnotesize
\bibliography{Nova_paper} }

\begin{thebibliography}{189}
\expandafter\ifx\csname natexlab\endcsname\relax\def\natexlab#1{#1}\fi

\bibitem[{{Abdo} {et~al.}(2010){Abdo}, {Ackermann}, {Ajello}, {Atwood},
  {Baldini}, {Ballet}, {Barbiellini}, {Bastieri}, {Bechtol}, {Bellazzini}, \&
  et~al.}]{2010Sci...329..817A}
{Abdo}, A.~A., {Ackermann}, M., {Ajello}, M., {et~al.} 2010, Science, 329, 817

\bibitem[{{Abdollahi} {et~al.}(2017){Abdollahi}, {Ackermann}, {Ajello},
  {Albert}, {Baldini}, {Ballet}, {Barbiellini}, {Bastieri}, {Becerra Gonzalez},
  {Bellazzini}, {Bissaldi}, {Blandford}, {Bloom}, {Bonino}, {Bottacini},
  {Bregeon}, {Bruel}, {Buehler}, {Buson}, {Cameron}, {Caragiulo}, {Caraveo},
  {Cavazzuti}, {Cecchi}, {Chekhtman}, {Cheung}, {Chiaro}, {Ciprini}, {Conrad},
  {Costantin}, {Costanza}, {Cutini}, {D'Ammando}, {de Palma}, {Desai},
  {Desiante}, {Digel}, {Di Lalla}, {Di Mauro}, {Di Venere}, {Donaggio},
  {Drell}, {Favuzzi}, {Fegan}, {Ferrara}, {Focke}, {Franckowiak}, {Fukazawa},
  {Funk}, {Fusco}, {Gargano}, {Gasparrini}, {Giglietto}, {Giomi}, {Giordano},
  {Giroletti}, {Glanzman}, {Green}, {Grenier}, {Grove}, {Guillemot}, {Guiriec},
  {Hays}, {Horan}, {Jogler}, {J{\'o}hannesson}, {Johnson}, {Kocevski}, {Kuss},
  {La Mura}, {Larsson}, {Latronico}, {Li}, {Longo}, {Loparco}, {Lovellette},
  {Lubrano}, {Magill}, {Maldera}, {Manfreda}, {Mayer}, {Mazziotta},
  {Michelson}, {Mitthumsiri}, {Mizuno}, {Monzani}, {Morselli}, {Moskalenko},
  {Negro}, {Nuss}, {Ohsugi}, {Omodei}, {Orienti}, {Orlando}, {Paliya},
  {Paneque}, {Perkins}, {Persic}, {Pesce-Rollins}, {Petrosian}, {Piron},
  {Porter}, {Principe}, {Rain{\`o}}, {Rando}, {Razzano}, {Razzaque}, {Reimer},
  {Reimer}, {Sgr{\`o}}, {Simone}, {Siskind}, {Spada}, {Spandre}, {Spinelli},
  {Stawarz}, {Suson}, {Takahashi}, {Tanaka}, {Thayer}, {Thompson}, {Torres},
  {Torresi}, {Tosti}, {Troja}, {Vianello}, \& {Wood}}]{2016arXiv161203165A}
{Abdollahi}, S., {Ackermann}, M., {Ajello}, M., {et~al.} 2017, \apj, 846, 34

\bibitem[{{Acero} {et~al.}(2015){Acero}, {Ackermann}, {Ajello}, {Albert},
  {Atwood}, {Axelsson}, {Baldini}, {Ballet}, {Barbiellini}, {Bastieri},
  {Belfiore}, {Bellazzini}, {Bissaldi}, {Blandford}, {Bloom}, {Bogart},
  {Bonino}, {Bottacini}, {Bregeon}, {Britto}, {Bruel}, {Buehler}, {Burnett},
  {Buson}, {Caliandro}, {Cameron}, {Caputo}, {Caragiulo}, {Caraveo},
  {Casandjian}, {Cavazzuti}, {Charles}, {Chaves}, {Chekhtman}, {Cheung},
  {Chiang}, {Chiaro}, {Ciprini}, {Claus}, {Cohen-Tanugi}, {Cominsky}, {Conrad},
  {Cutini}, {D`Ammando}, {de Angelis}, {DeKlotz}, {de Palma}, {Desiante},
  {Digel}, {Di Venere}, {Drell}, {Dubois}, {Dumora}, {Favuzzi}, {Fegan},
  {Ferrara}, {Finke}, {Franckowiak}, {Fukazawa}, {Funk}, {Fusco}, {Gargano},
  {Gasparrini}, {Giebels}, {Giglietto}, {Giommi}, {Giordano}, {Giroletti},
  {Glanzman}, {Godfrey}, {Grenier}, {Grondin}, {Grove}, {Guillemot}, {Guiriec},
  {Hadasch}, {Harding}, {Hays}, {Hewitt}, {Hill}, {Horan}, {Iafrate}, {Jogler},
  {J{\'o}hannesson}, {Johnson}, {Johnson}, {Johnson}, {Johnson}, {Kamae},
  {Kataoka}, {Katsuta}, {Kuss}, {La Mura}, {Landriu}, {Larsson}, {Latronico},
  {Lemoine-Goumard}, {Li}, {Li}, {Longo}, {Loparco}, {Lott}, {Lovellette},
  {Lubrano}, {Madejski}, {Massaro}, {Mayer}, {Mazziotta}, {McEnery},
  {Michelson}, {Mirabal}, {Mizuno}, {Moiseev}, {Mongelli}, {Monzani},
  {Morselli}, {Moskalenko}, {Murgia}, {Nuss}, {Ohno}, {Ohsugi}, {Omodei},
  {Orienti}, {Orlando}, {Ormes}, {Paneque}, {Panetta}, {Perkins},
  {Pesce-Rollins}, {Piron}, {Pivato}, {Porter}, {Racusin}, {Rando}, {Razzano},
  {Razzaque}, {Reimer}, {Reimer}, {Reposeur}, {Rochester}, {Romani},
  {Salvetti}, {S{\'a}nchez-Conde}, {Saz Parkinson}, {Schulz}, {Siskind},
  {Smith}, {Spada}, {Spandre}, {Spinelli}, {Stephens}, {Strong}, {Suson},
  {Takahashi}, {Takahashi}, {Tanaka}, {Thayer}, {Thayer}, {Thompson},
  {Tibaldo}, {Tibolla}, {Torres}, {Torresi}, {Tosti}, {Troja}, {Van Klaveren},
  {Vianello}, {Winer}, {Wood}, {Wood}, {Zimmer}, \& {Fermi-LAT
  Collaboration}}]{2015ApJS..218...23A}
{Acero}, F., {Ackermann}, M., {Ajello}, M., {et~al.} 2015, \apjs, 218, 23

\bibitem[{{Acero} {et~al.}(2016{\natexlab{a}}){Acero}, {Ackermann}, {Ajello},
  {Baldini}, {Ballet}, {Barbiellini}, {Bastieri}, {Bellazzini}, {Bissaldi},
  {Blandford}, {Bloom}, {Bonino}, {Bottacini}, {Brandt}, {Bregeon}, {Bruel},
  {Buehler}, {Buson}, {Caliandro}, {Cameron}, {Caputo}, {Caragiulo}, {Caraveo},
  {Casandjian}, {Cavazzuti}, {Cecchi}, {Chekhtman}, {Chiang}, {Chiaro},
  {Ciprini}, {Claus}, {Cohen}, {Cohen-Tanugi}, {Cominsky}, {Condon}, {Conrad},
  {Cutini}, {D'Ammando}, {de Angelis}, {de Palma}, {Desiante}, {Digel}, {Di
  Venere}, {Drell}, {Drlica-Wagner}, {Favuzzi}, {Ferrara}, {Franckowiak},
  {Fukazawa}, {Funk}, {Fusco}, {Gargano}, {Gasparrini}, {Giglietto}, {Giommi},
  {Giordano}, {Giroletti}, {Glanzman}, {Godfrey}, {Gomez-Vargas}, {Grenier},
  {Grondin}, {Guillemot}, {Guiriec}, {Gustafsson}, {Hadasch}, {Harding},
  {Hayashida}, {Hays}, {Hewitt}, {Hill}, {Horan}, {Hou}, {Iafrate}, {Jogler},
  {J{\'o}hannesson}, {Johnson}, {Kamae}, {Katagiri}, {Kataoka}, {Katsuta},
  {Kerr}, {Kn{\"o}dlseder}, {Kocevski}, {Kuss}, {Laffon}, {Lande}, {Larsson},
  {Latronico}, {Lemoine-Goumard}, {Li}, {Li}, {Longo}, {Loparco}, {Lovellette},
  {Lubrano}, {Magill}, {Maldera}, {Marelli}, {Mayer}, {Mazziotta}, {Michelson},
  {Mitthumsiri}, {Mizuno}, {Moiseev}, {Monzani}, {Moretti}, {Morselli},
  {Moskalenko}, {Murgia}, {Nemmen}, {Nuss}, {Ohsugi}, {Omodei}, {Orienti},
  {Orlando}, {Ormes}, {Paneque}, {Perkins}, {Pesce-Rollins}, {Petrosian},
  {Piron}, {Pivato}, {Porter}, {Rain{\`o}}, {Rando}, {Razzano}, {Razzaque},
  {Reimer}, {Reimer}, {Renaud}, {Reposeur}, {Rousseau}, {Saz Parkinson},
  {Schmid}, {Schulz}, {Sgr{\`o}}, {Siskind}, {Spada}, {Spandre}, {Spinelli},
  {Strong}, {Suson}, {Tajima}, {Takahashi}, {Tanaka}, {Thayer}, {Thompson},
  {Tibaldo}, {Tibolla}, {Torres}, {Tosti}, {Troja}, {Uchiyama}, {Vianello},
  {Wells}, {Wood}, {Wood}, {Yassine}, {den Hartog}, \&
  {Zimmer}}]{2015arXiv150703633B}
{Acero}, F., {Ackermann}, M., {Ajello}, M., {et~al.} 2016{\natexlab{a}}, \apjs,
  224, 8

\bibitem[{{Acero} {et~al.}(2016{\natexlab{b}}){Acero}, {Ackermann}, {Ajello},
  {et~al.}}]{Acero:2016qlg}
{Acero}, F., {Ackermann}, M., {Ajello}, M., {et~al.} 2016{\natexlab{b}},
  Astrophys. J. Suppl., 223, 26

\bibitem[{{Ackermann} {et~al.}(2014){Ackermann}, {Ajello}, {Albert}, {Baldini},
  {Ballet}, {Barbiellini}, {Bastieri}, {Bellazzini}, {Bissaldi}, {Blandford},
  {Bloom}, {Bottacini}, {Brandt}, {Bregeon}, {Bruel}, {Buehler}, {Buson},
  {Caliandro}, {Cameron}, {Caragiulo}, {Caraveo}, {Cavazzuti}, {Charles},
  {Chekhtman}, {Cheung}, {Chiang}, {Chiaro}, {Ciprini}, {Claus},
  {Cohen-Tanugi}, {Conrad}, {Corbel}, {D'Ammando}, {de Angelis}, {den Hartog},
  {de Palma}, {Dermer}, {Desiante}, {Digel}, {Di Venere}, {do Couto e Silva},
  {Donato}, {Drell}, {Drlica-Wagner}, {Favuzzi}, {Ferrara}, {Focke},
  {Franckowiak}, {Fuhrmann}, {Fukazawa}, {Fusco}, {Gargano}, {Gasparrini},
  {Germani}, {Giglietto}, {Giordano}, {Giroletti}, {Glanzman}, {Godfrey},
  {Grenier}, {Grove}, {Guiriec}, {Hadasch}, {Harding}, {Hayashida}, {Hays},
  {Hewitt}, {Hill}, {Hou}, {Jean}, {Jogler}, {J{\'o}hannesson}, {Johnson},
  {Johnson}, {Kerr}, {Kn{\"o}dlseder}, {Kuss}, {Larsson}, {Latronico},
  {Lemoine-Goumard}, {Longo}, {Loparco}, {Lott}, {Lovellette}, {Lubrano},
  {Manfreda}, {Martin}, {Massaro}, {Mayer}, {Mazziotta}, {McEnery},
  {Michelson}, {Mitthumsiri}, {Mizuno}, {Monzani}, {Morselli}, {Moskalenko},
  {Murgia}, {Nemmen}, {Nuss}, {Ohsugi}, {Omodei}, {Orienti}, {Orlando},
  {Ormes}, {Paneque}, {Panetta}, {Perkins}, {Pesce-Rollins}, {Piron}, {Pivato},
  {Porter}, {Rain{\`o}}, {Rando}, {Razzano}, {Razzaque}, {Reimer}, {Reimer},
  {Reposeur}, {Saz Parkinson}, {Schaal}, {Schulz}, {Sgr{\`o}}, {Siskind},
  {Spandre}, {Spinelli}, {Stawarz}, {Suson}, {Takahashi}, {Tanaka}, {Thayer},
  {Thayer}, {Thompson}, {Tibaldo}, {Tinivella}, {Torres}, {Tosti}, {Troja},
  {Uchiyama}, {Vianello}, {Winer}, {Wolff}, {Wood}, {Wood}, {Wood},
  {Charbonnel}, {Corbet}, {De Gennaro Aquino}, {Edlin}, {Mason}, {Schwarz},
  {Shore}, {Starrfield}, {Teyssier}, \& {Fermi-LAT
  Collaboration}}]{2014Sci...345..554A}
{Ackermann}, M., {Ajello}, M., {Albert}, A., {et~al.} 2014, Science, 345, 554

\bibitem[{{Ackermann} {et~al.}(2015){Ackermann}, {Albert}, {Anderson},
  {Atwood}, {Baldini}, {Barbiellini}, {Bastieri}, {Bechtol}, {Bellazzini},
  {Bissaldi}, {Blandford}, {Bloom}, {Bonino}, {Bottacini}, {Brandt}, {Bregeon},
  {Bruel}, {Buehler}, {Caliandro}, {Cameron}, {Caputo}, {Caragiulo}, {Caraveo},
  {Cecchi}, {Charles}, {Chekhtman}, {Chiang}, {Chiaro}, {Ciprini}, {Claus},
  {Cohen-Tanugi}, {Conrad}, {Cuoco}, {Cutini}, {D'Ammando}, {de Angelis}, {de
  Palma}, {Desiante}, {Digel}, {Di Venere}, {Drell}, {Drlica-Wagner}, {Essig},
  {Favuzzi}, {Fegan}, {Ferrara}, {Focke}, {Franckowiak}, {Fukazawa}, {Funk},
  {Fusco}, {Gargano}, {Gasparrini}, {Giglietto}, {Giordano}, {Giroletti},
  {Glanzman}, {Godfrey}, {Gomez-Vargas}, {Grenier}, {Guiriec}, {Gustafsson},
  {Hays}, {Hewitt}, {Horan}, {Jogler}, {J{\'o}hannesson}, {Kuss}, {Larsson},
  {Latronico}, {Li}, {Li}, {Llena Garde}, {Longo}, {Loparco}, {Lubrano},
  {Malyshev}, {Mayer}, {Mazziotta}, {McEnery}, {Meyer}, {Michelson}, {Mizuno},
  {Moiseev}, {Monzani}, {Morselli}, {Murgia}, {Nuss}, {Ohsugi}, {Orienti},
  {Orlando}, {Ormes}, {Paneque}, {Perkins}, {Pesce-Rollins}, {Piron}, {Pivato},
  {Porter}, {Rain{\`o}}, {Rando}, {Razzano}, {Reimer}, {Reimer}, {Ritz},
  {S{\'a}nchez-Conde}, {Schulz}, {Sehgal}, {Sgr{\`o}}, {Siskind}, {Spada},
  {Spandre}, {Spinelli}, {Strigari}, {Tajima}, {Takahashi}, {Thayer},
  {Tibaldo}, {Torres}, {Troja}, {Vianello}, {Werner}, {Winer}, {Wood}, {Wood},
  {Zaharijas}, {Zimmer}, \& {Fermi-LAT Collaboration}}]{2015arXiv150302641F}
{Ackermann}, M., {Albert}, A., {Anderson}, B., {et~al.} 2015, PRL, 115, 231301

\bibitem[{Ahnen {et~al.}(2015)}]{Ahnen:2015pha}
Ahnen, M.~L. {et~al.} 2015, Astron. Astrophys., 582, A67

\bibitem[{{Aliu} {et~al.}(2012){Aliu}, {Archambault}, {Arlen}, {Aune},
  {Beilicke}, {Benbow}, {Bouvier}, {Bradbury}, {Buckley}, {Bugaev}, {Byrum},
  {Cannon}, {Cesarini}, {Ciupik}, {Collins-Hughes}, {Connolly}, {Cui},
  {Decerprit}, {Dickherber}, {Duke}, {Dumm}, {Dwarkadas}, {Errando}, {Falcone},
  {Feng}, {Finley}, {Finnegan}, {Fortson}, {Furniss}, {Galante}, {Gall},
  {Godambe}, {Griffin}, {Grube}, {Gyuk}, {Hanna}, {Holder}, {Huan}, {Hughes},
  {Humensky}, {Kaaret}, {Karlsson}, {Kertzman}, {Khassen}, {Kieda},
  {Krawczynski}, {Krennrich}, {Lang}, {Lee}, {Maier}, {Majumdar}, {McArthur},
  {McCann}, {Millis}, {Moriarty}, {Mukherjee}, {Nu{\~n}ez}, {Ong}, {Orr},
  {Otte}, {Pandel}, {Park}, {Perkins}, {Pohl}, {Prokoph}, {Quinn}, {Ragan},
  {Reyes}, {Reynolds}, {Roache}, {Rose}, {Ruppel}, {Saxon}, {Schroedter},
  {Sembroski}, {Skole}, {Smith}, {Staszak}, {Telezhinsky}, {Te{\v s}i{\'c}},
  {Theiling}, {Thibadeau}, {Tsurusaki}, {Tyler}, {Varlotta}, {Vincent},
  {Vivier}, {Wakely}, {Ward}, {Weekes}, {Weinstein}, {Weisgarber}, {Welsing},
  {Williams}, \& {Zitzer}}]{2012ApJ...754...77A}
{Aliu}, E., {Archambault}, S., {Arlen}, T., {et~al.} 2012, \apj, 754, 77

\bibitem[{{Atwood} {et~al.}(2009){Atwood}, {Abdo}, {Ackermann},
  {et~al.}}]{2009ApJ...697.1071A}
{Atwood}, W.~B., {Abdo}, A.~A., {Ackermann}, M., {et~al.} 2009, \apj, 697, 1071

\bibitem[{{Atwood} {et~al.}(2013){Atwood}, {Albert}, {Baldini}, {Tinivella},
  {Bregeon}, {Pesce-Rollins}, {Sgr{\`o}}, {Bruel}, {Charles}, {Drlica-Wagner},
  {Franckowiak}, {Jogler}, {Rochester}, {Usher}, {Wood}, {Cohen-Tanugi}, \&
  {S.~Zimmer for the Fermi-LAT Collaboration}}]{2013arXiv1303.3514A}
{Atwood}, W.~B., {Albert}, A., {Baldini}, L., {et~al.} 2013, ArXiv e-prints
  [\eprint[arXiv]{1303.3514}]

\bibitem[{{Ayani} {et~al.}(2009)}]{CBET1911}
{Ayani}, K. {et~al.} 2009, CBET, 1911

\bibitem[{{Ayani} {et~al.}(2010)}]{CBET2186}
{Ayani}, K. {et~al.} 2010, CBET, 2186

\bibitem[{{Ayani} {et~al.}(2012)}]{CBET3177}
{Ayani}, K. {et~al.} 2012, CBET, 3177

\bibitem[{{Aydi} {et~al.}(2016){Aydi}, {Mr{\'o}z}, {Whitelock}, {Mohamed},
  {Wyrzykowski}, {Udalski}, {Vaisanen}, {Nagayama}, {Dominik}, {Scholz},
  {Onozato}, {Williams}, {Hodgkin}, {Nishiyama}, {Yamagishi}, {Smith}, {Ryu},
  {Iwamatsu}, \& {Kawamata}}]{OGLEREF3}
{Aydi}, E., {Mr{\'o}z}, P., {Whitelock}, P.~A., {et~al.} 2016, \mnras, 461,
  1529

\bibitem[{{Banerjee} {et~al.}(2014){Banerjee}, {Joshi}, {Venkataraman},
  {Ashok}, {Marion}, {Hsiao}, \& {Raj}}]{ban14}
{Banerjee}, D.~P.~K., {Joshi}, V., {Venkataraman}, V., {et~al.} 2014, \apjl,
  785, L11

\bibitem[{{Banerjee} {et~al.}(2016){Banerjee}, {Srivastava}, {Ashok}, \&
  {Venkataraman}}]{Banerjee2015}
{Banerjee}, D.~P.~K., {Srivastava}, M.~K., {Ashok}, N.~M., \& {Venkataraman},
  V. 2016, \mnras, 455, L109

\bibitem[{{Bellm}(2014)}]{2014htu..conf...27B}
{Bellm}, E. 2014, in The Third Hot-wiring the Transient Universe Workshop, ed.
  P.~R. {Wozniak}, M.~J. {Graham}, A.~A. {Mahabal}, \& R.~{Seaman}, 27--33

\bibitem[{{Brown} {et~al.}(2011{\natexlab{a}})}]{CBET2796}
{Brown}, N.~J. {et~al.} 2011{\natexlab{a}}, CBET, 2796

\bibitem[{{Brown} {et~al.}(2011{\natexlab{b}})}]{IAUC9228}
{Brown}, N.~J. {et~al.} 2011{\natexlab{b}}, IAUC, 9228

\bibitem[{{Cheung} {et~al.}(2014){Cheung}, {Jean}, \&
  {Shore}}]{2014ATel.5879....1C}
{Cheung}, C.~C., {Jean}, P., \& {Shore}, S.~N. 2014, The Astronomer's Telegram,
  5879

\bibitem[{{Cheung} {et~al.}(2016{\natexlab{a}}){Cheung}, {Jean}, {Shore}, \&
  {Fermi Large Area Telescope Collaboration}}]{ATel9594}
{Cheung}, C.~C., {Jean}, P., {Shore}, S.~N., \& {Fermi Large Area Telescope
  Collaboration}. 2016{\natexlab{a}}, The Astronomer's Telegram, 9594

\bibitem[{{Cheung} {et~al.}(2016{\natexlab{b}}){Cheung}, {Jean}, {Shore},
  {Stawarz}, {Corbet}, {Kn{\"o}dlseder}, {Starrfield}, {Wood}, {Desiante},
  {Longo}, {Pivato}, \& {Wood}}]{2016ApJ...826..142C}
{Cheung}, C.~C., {Jean}, P., {Shore}, S.~N., {et~al.} 2016{\natexlab{b}}, \apj,
  826, 142

\bibitem[{{Chomiuk} {et~al.}(2014){Chomiuk}, {Nelson}, {Mukai}, {Sokoloski},
  {Rupen}, {Page}, {Osborne}, {Kuulkers}, {Mioduszewski}, {Roy}, {Weston}, \&
  {Krauss}}]{cho14}
{Chomiuk}, L., {Nelson}, T., {Mukai}, K., {et~al.} 2014, \apj, 788, 130

\bibitem[{{Ciprini} \& {Becerra Gonzalez}(2014)}]{ATel6036}
{Ciprini}, S. \& {Becerra Gonzalez}, J. 2014, ATel, 6036

\bibitem[{{Della Valle} \& {Livio}(1995)}]{1995ApJ...452..704D}
{Della Valle}, M. \& {Livio}, M. 1995, \apj, 452, 704

\bibitem[{{Downes} \& {Duerbeck}(2000)}]{2000AJ....120.2007D}
{Downes}, R.~A. \& {Duerbeck}, H.~W. 2000, \aj, 120, 2007

\bibitem[{{Elenin} {et~al.}(2009)}]{CBET1900}
{Elenin}, L. {et~al.} 2009, CBET, 1900

\bibitem[{{Finzell} {et~al.}(2015){Finzell}, {Chomiuk}, {Munari}, \&
  {Walter}}]{2015ApJ...809..160F}
{Finzell}, T., {Chomiuk}, L., {Munari}, U., \& {Walter}, F.~M. 2015, \apj, 809,
  160

\bibitem[{{Fujikawa} {et~al.}(2012){Fujikawa}, {Yamaoka}, \&
  {Nakano}}]{CBET3202}
{Fujikawa}, S., {Yamaoka}, H., \& {Nakano}, S. 2012, CBET, 3202

\bibitem[{{Gehrels}(1986)}]{1986ApJ...303..336G}
{Gehrels}, N. 1986, \apj, 303, 336

\bibitem[{{Greiner} {et~al.}(2010)}]{ATel2746}
{Greiner}, J. {et~al.} 2010, ATel, 2746

\bibitem[{{Guido} {et~al.}(2009)}]{CBET2053}
{Guido}, E. {et~al.} 2009, CBET, 2053

\bibitem[{{Hakkila} {et~al.}(1997){Hakkila}, {Myers}, {Stidham}, \&
  {Hartmann}}]{1997AJ....114.2043H}
{Hakkila}, J., {Myers}, J.~M., {Stidham}, B.~J., \& {Hartmann}, D.~H. 1997,
  \aj, 114, 2043

\bibitem[{{Hounsell} {et~al.}(2010){Hounsell}, {Bode}, {Hick}, {Buffington},
  {Jackson}, {Clover}, {Shafter}, {Darnley}, {Mawson}, {Steele}, {Evans},
  {Eyres}, \& {O'Brien}}]{2010ApJ...724..480H}
{Hounsell}, R., {Bode}, M.~F., {Hick}, P.~P., {et~al.} 2010, \apj, 724, 480

\bibitem[{{Itagaki} {et~al.}(2015)}]{CBET4145}
{Itagaki}, K. {et~al.} 2015, CBET, 4145

\bibitem[{{Jean} {et~al.}(2000){Jean}, {Hernanz}, {G{\'o}mez-Gomar}, \&
  {Jos{\'e}}}]{2000MNRAS.319..350J}
{Jean}, P., {Hernanz}, M., {G{\'o}mez-Gomar}, J., \& {Jos{\'e}}, J. 2000,
  \mnras, 319, 350

\bibitem[{{Joshi} {et~al.}(2014){Joshi}, {Banerjee}, {Venkataraman}, \&
  {Ashok}}]{2014ATel.6032....1J}
{Joshi}, V., {Banerjee}, D.~P.~K., {Venkataraman}, V., \& {Ashok}, N.~M. 2014,
  ATel, 6032

\bibitem[{{Joshi} {et~al.}(2010)}]{CBET2210}
{Joshi}, V. {et~al.} 2010, CBET, 2210

\bibitem[{{Kabashima} {et~al.}(2009)}]{IAUC9089}
{Kabashima}, F. {et~al.} 2009, IAUC, 9089

\bibitem[{{Kajikawa} {et~al.}(2013)}]{CBET3556}
{Kajikawa}, T. {et~al.} 2013, CBET, 3556

\bibitem[{{Kasliwal}(2013)}]{2013IAUS..281....9K}
{Kasliwal}, M.~M. 2013, in IAU Symposium, Vol. 281, Binary Paths to Type Ia
  Supernovae Explosions, ed. R.~{Di Stefano}, M.~{Orio}, \& M.~{Moe}, 9--16

\bibitem[{{Kasliwal} {et~al.}(2011){Kasliwal}, {Cenko}, {Kulkarni}, {Ofek},
  {Quimby}, \& {Rau}}]{2011ApJ...735...94K}
{Kasliwal}, M.~M., {Cenko}, S.~B., {Kulkarni}, S.~R., {et~al.} 2011, \apj, 735,
  94

\bibitem[{{Kent} {et~al.}(1991){Kent}, {Dame}, \&
  {Fazio}}]{1991ApJ...378..131K}
{Kent}, S.~M., {Dame}, T.~M., \& {Fazio}, G. 1991, \apj, 378, 131

\bibitem[{{Kinugasa} {et~al.}(2009{\natexlab{a}})}]{IAUC9041}
{Kinugasa}, K. {et~al.} 2009{\natexlab{a}}, IAUC, 9041

\bibitem[{{Kinugasa} {et~al.}(2009{\natexlab{b}})}]{CBET1995}
{Kinugasa}, K. {et~al.} 2009{\natexlab{b}}, CBET, 1995

\bibitem[{{Kinugasa} {et~al.}(2009{\natexlab{c}})}]{CBET2076}
{Kinugasa}, K. {et~al.} 2009{\natexlab{c}}, CBET, 2076

\bibitem[{{Kochanek} {et~al.}(2014){Kochanek}, {Adams}, \&
  {Belczynski}}]{2014MNRAS.443.1319K}
{Kochanek}, C.~S., {Adams}, S.~M., \& {Belczynski}, K. 2014, \mnras, 443, 1319

\bibitem[{{Korotkiy} {et~al.}(2012)}]{CBET3089}
{Korotkiy}, S. {et~al.} 2012, CBET, 3089

\bibitem[{{Kozlowski} {et~al.}(2012)}]{ATel4323}
{Kozlowski}, S. {et~al.} 2012, ATel, 4323

\bibitem[{{Li} {et~al.}(2016){Li}, {Chomiuk}, {Strader}, {Cheung}, {Jean}, \&
  {Shore}}]{ATel9736}
{Li}, K.-L., {Chomiuk}, L., {Strader}, J., {et~al.} 2016, The Astronomer's
  Telegram, 9771

\bibitem[{{Liller} {et~al.}(2008{\natexlab{a}})}]{IAUC8990}
{Liller}, W. {et~al.} 2008{\natexlab{a}}, IAUC, 8990

\bibitem[{{Liller} {et~al.}(2008{\natexlab{b}})}]{CBET1591}
{Liller}, W. {et~al.} 2008{\natexlab{b}}, CBET, 1591

\bibitem[{{Liller} {et~al.}(2008{\natexlab{c}})}]{IAUC9004}
{Liller}, W. {et~al.} 2008{\natexlab{c}}, IAUC, 9004

\bibitem[{{Liller} {et~al.}(2008{\natexlab{d}})}]{IAUC9005}
{Liller}, W. {et~al.} 2008{\natexlab{d}}, IAUC, 9005

\bibitem[{{Liller} {et~al.}(2010)}]{CBET2264}
{Liller}, W. {et~al.} 2010, CBET, 2264

\bibitem[{{Linford} {et~al.}(2017){Linford}, {Chomiuk}, {Nelson}, {Finzell},
  {Walter}, {Sokoloski}, {Mukai}, {Mioduszewski}, {van der Horst}, {Weston}, \&
  {Rupen}}]{2017ApJ...842...73L}
{Linford}, J.~D., {Chomiuk}, L., {Nelson}, T., {et~al.} 2017, \apj, 842, 73

\bibitem[{{Linford} {et~al.}(2015){Linford}, {Ribeiro}, {Chomiuk}, {Nelson},
  {Sokoloski}, {Rupen}, {Mukai}, {O'Brien}, {Mioduszewski}, \&
  {Weston}}]{Linford}
{Linford}, J.~D., {Ribeiro}, V.~A.~R.~M., {Chomiuk}, L., {et~al.} 2015, \apj,
  805, 136

\bibitem[{{Lipunov} {et~al.}(2011)}]{IAUC9247}
{Lipunov}, V. {et~al.} 2011, IAUC, 9247

\bibitem[{{Maehara} {et~al.}(2010{\natexlab{a}})}]{CBET2139}
{Maehara}, H. {et~al.} 2010{\natexlab{a}}, CBET, 2139

\bibitem[{{Maehara} {et~al.}(2010{\natexlab{b}})}]{CBET2142}
{Maehara}, H. {et~al.} 2010{\natexlab{b}}, CBET, 2142

\bibitem[{{Maehara} {et~al.}(2010{\natexlab{c}})}]{CBET2199}
{Maehara}, H. {et~al.} 2010{\natexlab{c}}, CBET, 2199

\bibitem[{{Maehara} {et~al.}(2010{\natexlab{d}})}]{CBET2205}
{Maehara}, H. {et~al.} 2010{\natexlab{d}}, CBET, 2205

\bibitem[{{Martin} \& {Dubus}(2013)}]{2013A&A...551A..37M}
{Martin}, P. \& {Dubus}, G. 2013, \aap, 551, A37

\bibitem[{{Martin} {et~al.}(2017){Martin}, {Dubus}, {Jean}, {Tatischeff}, \&
  {Dosne}}]{2017AASubmitted}
{Martin}, P., {Dubus}, G., {Jean}, P., {Tatischeff}, V., \& {Dosne}, C. 2017,
  submitted in \aap

\bibitem[{{Mattox} {et~al.}(1996){Mattox}, {Bertsch}, {Chiang}, {Dingus},
  {Digel}, {Esposito}, {Fierro}, {Hartman}, {Hunter}, {Kanbach}, {Kniffen},
  {Lin}, {Macomb}, {Mayer-Hasselwander}, {Michelson}, {von Montigny},
  {Mukherjee}, {Nolan}, {Ramanamurthy}, {Schneid}, {Sreekumar}, {Thompson}, \&
  {Willis}}]{1996ApJ...461..396M}
{Mattox}, J.~R., {Bertsch}, D.~L., {Chiang}, J., {et~al.} 1996, \apj, 461, 396

\bibitem[{{Metzger} {et~al.}(2015){Metzger}, {Finzell}, {Vurm}, {Hasco{\"e}t},
  {Beloborodov}, \& {Chomiuk}}]{2015MNRAS.450.2739M}
{Metzger}, B.~D., {Finzell}, T., {Vurm}, I., {et~al.} 2015, \mnras, 450, 2739

\bibitem[{{Metzger} {et~al.}(2014){Metzger}, {Hasco{\"e}t}, {Vurm},
  {Beloborodov}, {Chomiuk}, {Sokoloski}, \& {Nelson}}]{2014MNRAS.442..713M}
{Metzger}, B.~D., {Hasco{\"e}t}, R., {Vurm}, I., {et~al.} 2014, \mnras, 442,
  713

\bibitem[{{Molnar} {et~al.}(2017){Molnar}, {Van Noord}, {Kinemuchi},
  {Smolinski}, {Alexander}, {Cook}, {Jang}, {Kobulnicky}, {Spedden}, \&
  {Steenwyk}}]{2017ApJ...840....1M}
{Molnar}, L.~A., {Van Noord}, D.~M., {Kinemuchi}, K., {et~al.} 2017, \apj, 840,
  1

\bibitem[{{Molnar} {et~al.}(2015){Molnar}, {Van Noord}, {Steenwyk}, {Spedden},
  \& {Kinemuchi}}]{2015AAS...22541505M}
{Molnar}, L.~A., {Van Noord}, D.~M., {Steenwyk}, S.~D., {Spedden}, C.~J., \&
  {Kinemuchi}, K. 2015, in AAS, Vol. 225, 415.05

\bibitem[{{Morris} {et~al.}(2017){Morris}, {Cotter}, {Brown}, \&
  {Chadwick}}]{2017MNRAS.465.1218M}
{Morris}, P.~J., {Cotter}, G., {Brown}, A.~M., \& {Chadwick}, P.~M. 2017,
  \mnras, 465, 1218

\bibitem[{{Mr{\'o}z} {et~al.}(2014){Mr{\'o}z}, {Poleski}, {Udalski},
  {Soszy{\'n}ski}, {Szyma{\'n}ski}, {Kubiak}, {Pietrzy{\'n}ski}, {Wyrzykowski},
  {Ulaczyk}, {Koz{\l}owski}, {Pietrukowicz}, \& {Skowron}}]{OGLEREF1}
{Mr{\'o}z}, P., {Poleski}, R., {Udalski}, A., {et~al.} 2014, \mnras, 443, 784

\bibitem[{{Mr{\'o}z} {et~al.}(2015){Mr{\'o}z}, {Udalski}, {Poleski},
  {Soszy{\'n}ski}, {Szyma{\'n}ski}, {Pietrzy{\'n}ski}, {Wyrzykowski},
  {Ulaczyk}, {Koz{\l}owski}, {Pietrukowicz}, \& {Skowron}}]{OGLEREF2}
{Mr{\'o}z}, P., {Udalski}, A., {Poleski}, R., {et~al.} 2015, \apjs, 219, 26

\bibitem[{{Mukai}(2015)}]{2015AcPPP...2..246M}
{Mukai}, K. 2015, Acta Polytechnica CTU Proceedings, 2, 246

\bibitem[{{Munari} {et~al.}(2017){Munari}, {Hambsch}, \&
  {Frigo}}]{2017MNRAS.469.4341M}
{Munari}, U., {Hambsch}, F.-J., \& {Frigo}, A. 2017, \mnras, 469, 4341

\bibitem[{{Munari} {et~al.}(2015){Munari}, {Henden}, {Banerjee}, {Ashok},
  {Righetti}, {Dallaporta}, \& {Cetrulo}}]{Munari2015a}
{Munari}, U., {Henden}, A., {Banerjee}, D.~P.~K., {et~al.} 2015, \mnras, 447,
  1661

\bibitem[{{Munari} {et~al.}(2002){Munari}, {Henden}, {Kiyota}, {Laney},
  {Marang}, {Zwitter}, {Corradi}, {Desidera}, {Marrese}, {Giro}, {Boschi}, \&
  {Schwartz}}]{2002A&A...389L..51M}
{Munari}, U., {Henden}, A., {Kiyota}, S., {et~al.} 2002, \aap, 389, L51

\bibitem[{{Munari} {et~al.}(2010{\natexlab{a}}){Munari}, {Henden}, {Valisa},
  {Dallaporta}, \& {Righetti}}]{Munari2010a}
{Munari}, U., {Henden}, A., {Valisa}, P., {Dallaporta}, S., \& {Righetti},
  G.~L. 2010{\natexlab{a}}, \pasp, 122, 898

\bibitem[{{Munari} {et~al.}(2014){Munari}, {Ochner}, {Dallaporta}, {Valisa},
  {Graziani}, {Righetti}, {Cherini}, {Castellani}, {Cetrulo}, \&
  {Englaro}}]{Munari2014}
{Munari}, U., {Ochner}, P., {Dallaporta}, S., {et~al.} 2014, \mnras, 440, 3402

\bibitem[{{Munari} {et~al.}(2009{\natexlab{a}})}]{CBET1912}
{Munari}, U. {et~al.} 2009{\natexlab{a}}, CBET, 1912

\bibitem[{{Munari} {et~al.}(2009{\natexlab{b}})}]{CBET1999}
{Munari}, U. {et~al.} 2009{\natexlab{b}}, CBET, 1999

\bibitem[{{Munari} {et~al.}(2009{\natexlab{c}})}]{CBET2034}
{Munari}, U. {et~al.} 2009{\natexlab{c}}, CBET, 2034

\bibitem[{{Munari} {et~al.}(2010{\natexlab{b}})}]{CBET2185}
{Munari}, U. {et~al.} 2010{\natexlab{b}}, CBET, 2185

\bibitem[{{Munari} {et~al.}(2012)}]{CBET3184}
{Munari}, U. {et~al.} 2012, CBET, 3184

\bibitem[{{Nakano} {et~al.}(2010{\natexlab{a}})}]{CBET2128}
{Nakano}, H. {et~al.} 2010{\natexlab{a}}, CBET, 2128

\bibitem[{{Nakano} {et~al.}(2015{\natexlab{a}})}]{CBET4086}
{Nakano}, P. {et~al.} 2015{\natexlab{a}}, CBET, 4086

\bibitem[{{Nakano} {et~al.}(2015{\natexlab{b}}){Nakano}, {Kojima}, \&
  {Maehara}}]{CBET4078}
{Nakano}, S., {Kojima}, T., \& {Maehara}, H. 2015{\natexlab{b}}, CBET, 4078

\bibitem[{{Nakano} {et~al.}(2008{\natexlab{a}})}]{CBET1496}
{Nakano}, S. {et~al.} 2008{\natexlab{a}}, CBET, 1496

\bibitem[{{Nakano} {et~al.}(2008{\natexlab{b}})}]{IAUC8972}
{Nakano}, S. {et~al.} 2008{\natexlab{b}}, IAUC, 8972

\bibitem[{{Nakano} {et~al.}(2009{\natexlab{a}})}]{CBET1910}
{Nakano}, S. {et~al.} 2009{\natexlab{a}}, CBET, 1910

\bibitem[{{Nakano} {et~al.}(2009{\natexlab{b}})}]{IAUC9064}
{Nakano}, S. {et~al.} 2009{\natexlab{b}}, IAUC, 9064

\bibitem[{{Nakano} {et~al.}(2009{\natexlab{c}})}]{CBET2008}
{Nakano}, S. {et~al.} 2009{\natexlab{c}}, CBET, 2008

\bibitem[{{Nakano} {et~al.}(2009{\natexlab{d}})}]{IAUC9093}
{Nakano}, S. {et~al.} 2009{\natexlab{d}}, IAUC, 9093

\bibitem[{{Nakano} {et~al.}(2010{\natexlab{b}})}]{CBET2176}
{Nakano}, S. {et~al.} 2010{\natexlab{b}}, CBET, 2176

\bibitem[{{Nakano} {et~al.}(2010{\natexlab{c}})}]{IAUC9119}
{Nakano}, S. {et~al.} 2010{\natexlab{c}}, IAUC, 9119

\bibitem[{{Nakano} {et~al.}(2010{\natexlab{d}})}]{CBET2265}
{Nakano}, S. {et~al.} 2010{\natexlab{d}}, CBET, 2265

\bibitem[{{Nakano} {et~al.}(2011{\natexlab{a}})}]{CBET2644}
{Nakano}, S. {et~al.} 2011{\natexlab{a}}, CBET, 2644

\bibitem[{{Nakano} {et~al.}(2011{\natexlab{b}})}]{IAUC9196}
{Nakano}, S. {et~al.} 2011{\natexlab{b}}, IAUC, 9196

\bibitem[{{Nakano} {et~al.}(2012{\natexlab{a}})}]{CBET3140}
{Nakano}, S. {et~al.} 2012{\natexlab{a}}, CBET, 3140

\bibitem[{{Nakano} {et~al.}(2012{\natexlab{b}})}]{CBET3072}
{Nakano}, S. {et~al.} 2012{\natexlab{b}}, CBET, 3072

\bibitem[{{Nakano} {et~al.}(2012{\natexlab{c}})}]{CBET3182}
{Nakano}, S. {et~al.} 2012{\natexlab{c}}, CBET, 3182

\bibitem[{{Nakano} {et~al.}(2013{\natexlab{a}})}]{CBET3628}
{Nakano}, S. {et~al.} 2013{\natexlab{a}}, CBET, 3628

\bibitem[{{Nakano} {et~al.}(2013{\natexlab{b}})}]{IAUC9258}
{Nakano}, S. {et~al.} 2013{\natexlab{b}}, IAUC, 9258

\bibitem[{{Nakano} {et~al.}(2013{\natexlab{c}})}]{CBET3691}
{Nakano}, S. {et~al.} 2013{\natexlab{c}}, CBET, 3691

\bibitem[{{Nakano} {et~al.}(2013{\natexlab{d}})}]{IAUC9263}
{Nakano}, S. {et~al.} 2013{\natexlab{d}}, IAUC, 9263

\bibitem[{{Nakano} {et~al.}(2013{\natexlab{e}})}]{CBET3724}
{Nakano}, S. {et~al.} 2013{\natexlab{e}}, CBET, 3724

\bibitem[{{Nakano} {et~al.}(2013{\natexlab{f}})}]{IAUC9264}
{Nakano}, S. {et~al.} 2013{\natexlab{f}}, IAUC, 9264

\bibitem[{{Nakano} {et~al.}(2014)}]{CBET3802}
{Nakano}, S. {et~al.} 2014, CBET, 3802

\bibitem[{{Nandez} {et~al.}(2014){Nandez}, {Ivanova}, \&
  {Lombardi}}]{2014ApJ...786...39N}
{Nandez}, J.~L.~A., {Ivanova}, N., \& {Lombardi}, Jr., J.~C. 2014, \apj, 786,
  39

\bibitem[{{Nelson} {et~al.}(2015)}]{ATel7085}
{Nelson}, T. {et~al.} 2015, ATel, 7085

\bibitem[{{Nishimura} {et~al.}(2015)}]{CBET4163}
{Nishimura}, H. {et~al.} 2015, CBET, 4163

\bibitem[{{Nishiyama} {et~al.}(2009{\natexlab{a}})}]{CBET1899}
{Nishiyama}, K. {et~al.} 2009{\natexlab{a}}, CBET, 1899

\bibitem[{{Nishiyama} {et~al.}(2009{\natexlab{b}})}]{IAUC9061}
{Nishiyama}, K. {et~al.} 2009{\natexlab{b}}, IAUC, 9061

\bibitem[{{Nishiyama} {et~al.}(2009{\natexlab{c}})}]{CBET1994}
{Nishiyama}, K. {et~al.} 2009{\natexlab{c}}, CBET, 1994

\bibitem[{{Nishiyama} {et~al.}(2009{\natexlab{d}})}]{IAUC9100}
{Nishiyama}, K. {et~al.} 2009{\natexlab{d}}, IAUC, 9100

\bibitem[{{Nishiyama} {et~al.}(2010{\natexlab{a}})}]{CBET2183}
{Nishiyama}, K. {et~al.} 2010{\natexlab{a}}, CBET, 2183

\bibitem[{{Nishiyama} {et~al.}(2010{\natexlab{b}})}]{IAUC9120}
{Nishiyama}, K. {et~al.} 2010{\natexlab{b}}, IAUC, 9120

\bibitem[{{Nishiyama} {et~al.}(2010{\natexlab{c}})}]{IAUC9130}
{Nishiyama}, K. {et~al.} 2010{\natexlab{c}}, IAUC, 9130

\bibitem[{{Nishiyama} {et~al.}(2010{\natexlab{d}})}]{CBET2262}
{Nishiyama}, K. {et~al.} 2010{\natexlab{d}}, CBET, 2262

\bibitem[{{Nishiyama} {et~al.}(2010{\natexlab{e}})}]{IAUC9142}
{Nishiyama}, K. {et~al.} 2010{\natexlab{e}}, IAUC, 9142

\bibitem[{{Nishiyama} {et~al.}(2010{\natexlab{f}})}]{CBET2261}
{Nishiyama}, K. {et~al.} 2010{\natexlab{f}}, CBET, 2261

\bibitem[{{Nishiyama} {et~al.}(2010{\natexlab{g}})}]{IAUC9140}
{Nishiyama}, K. {et~al.} 2010{\natexlab{g}}, IAUC, 9140

\bibitem[{{Nishiyama} {et~al.}(2010{\natexlab{h}})}]{IAUC9167}
{Nishiyama}, K. {et~al.} 2010{\natexlab{h}}, IAUC, 9167

\bibitem[{{Nishiyama} {et~al.}(2011{\natexlab{a}})}]{CBET2679}
{Nishiyama}, K. {et~al.} 2011{\natexlab{a}}, CBET, 2679

\bibitem[{{Nishiyama} {et~al.}(2011{\natexlab{b}})}]{IAUC9203}
{Nishiyama}, K. {et~al.} 2011{\natexlab{b}}, IAUC, 9203

\bibitem[{{Nishiyama} {et~al.}(2012{\natexlab{a}})}]{CBET3166}
{Nishiyama}, K. {et~al.} 2012{\natexlab{a}}, CBET, 3166

\bibitem[{{Nishiyama} {et~al.}(2012{\natexlab{b}})}]{CBET3273}
{Nishiyama}, K. {et~al.} 2012{\natexlab{b}}, CBET, 3273

\bibitem[{{Nishiyama} {et~al.}(2013{\natexlab{a}})}]{CBET3397}
{Nishiyama}, K. {et~al.} 2013{\natexlab{a}}, CBET, 3397

\bibitem[{{Nishiyama} {et~al.}(2013{\natexlab{b}})}]{CBET3542}
{Nishiyama}, K. {et~al.} 2013{\natexlab{b}}, CBET, 3542

\bibitem[{{Nishiyama} {et~al.}(2014{\natexlab{a}})}]{CBET3825}
{Nishiyama}, K. {et~al.} 2014{\natexlab{a}}, CBET, 3825

\bibitem[{{Nishiyama} {et~al.}(2014{\natexlab{b}})}]{CBET3841}
{Nishiyama}, K. {et~al.} 2014{\natexlab{b}}, CBET, 3841

\bibitem[{{Nishiyama} {et~al.}(2014{\natexlab{c}})}]{CBET3842}
{Nishiyama}, K. {et~al.} 2014{\natexlab{c}}, CBET, 3842

\bibitem[{{Nishiyama} {et~al.}(2015{\natexlab{a}})}]{CBET4079}
{Nishiyama}, K. {et~al.} 2015{\natexlab{a}}, CBET, 4079

\bibitem[{{Nishiyama} {et~al.}(2015{\natexlab{b}})}]{CBET4150}
{Nishiyama}, K. {et~al.} 2015{\natexlab{b}}, CBET, 4150

\bibitem[{{{\"O}zd{\"o}nmez} {et~al.}(2016){{\"O}zd{\"o}nmez}, {G{\"u}ver},
  {Cabrera-Lavers}, \& {Ak}}]{Oezdoenmez}
{{\"O}zd{\"o}nmez}, A., {G{\"u}ver}, T., {Cabrera-Lavers}, A., \& {Ak}, T.
  2016, \mnras, 461, 1177

\bibitem[{{Patat}(2013)}]{2011arXiv1109.5799P}
{Patat}, F. 2013, in IAU Symposium, Vol. 281, Binary Paths to Type Ia
  Supernovae Explosions, ed. R.~{Di Stefano}, M.~{Orio}, \& M.~{Moe}, 291--298

\bibitem[{{Payne-Gaposchkin}(1964)}]{1964gano.book.....P}
{Payne-Gaposchkin}, C. 1964, {The Galactic Novae}

\bibitem[{{Pojmanski} {et~al.}(2008)}]{CBET1497}
{Pojmanski}, G. {et~al.} 2008, CBET, 1496

\bibitem[{{Pojmanski} {et~al.}(2009{\natexlab{a}})}]{CBET1800}
{Pojmanski}, G. {et~al.} 2009{\natexlab{a}}, CBET, 1800

\bibitem[{{Pojmanski} {et~al.}(2009{\natexlab{b}})}]{IAUC9043}
{Pojmanski}, G. {et~al.} 2009{\natexlab{b}}, IAUC, 9043

\bibitem[{{Ragan}(2009)}]{ATel2327}
{Ragan}, E. 2009, ATel, 2327

\bibitem[{{Rudy} {et~al.}(2009)}]{CBET2055}
{Rudy}, R.~J. {et~al.} 2009, CBET, 2055

\bibitem[{{Rudy} {et~al.}(2012)}]{CBET3287}
{Rudy}, R.~J. {et~al.} 2012, CBET, 3287

\bibitem[{{Samus}(2013)}]{CBET3708}
{Samus}, N.~N. 2013, CBET, 3708

\bibitem[{{Samus} {et~al.}(2009)}]{IAUC9048}
{Samus}, N.~N. {et~al.} 2009, IAUC, 9048

\bibitem[{{Sato} {et~al.}(2010)}]{CBET2204}
{Sato}, H. {et~al.} 2010, CBET, 2204

\bibitem[{{Schaefer}(2010{\natexlab{a}})}]{Schaefer2010}
{Schaefer}, B.~E. 2010{\natexlab{a}}, \apjs, 187, 275

\bibitem[{{Schaefer}(2010{\natexlab{b}})}]{2010ApJS..187..275S}
{Schaefer}, B.~E. 2010{\natexlab{b}}, \apjs, 187, 275

\bibitem[{{Schaefer} {et~al.}(2010)}]{IAUC9111}
{Schaefer}, B.~E. {et~al.} 2010, IAUC, 9111

\bibitem[{{Schaefer} {et~al.}(2014){Schaefer}, {Brummelaar}, {Gies},
  {Farrington}, {Kloppenborg}, {Chesneau}, {Monnier}, {Ridgway}, {Scott},
  {Tallon-Bosc}, {McAlister}, {Boyajian}, {Maestro}, {Mourard}, {Meilland},
  {Nardetto}, {Stee}, {Sturmann}, {Vargas}, {Baron}, {Ireland}, {Baines},
  {Che}, {Jones}, {Richardson}, {Roettenbacher}, {Sturmann}, {Turner},
  {Tuthill}, {van Belle}, {von Braun}, {Zavala}, {Banerjee}, {Ashok}, {Joshi},
  {Becker}, \& {Muirhead}}]{2014Natur.515..234S}
{Schaefer}, G.~H., {Brummelaar}, T.~T., {Gies}, D.~R., {et~al.} 2014, \nat,
  515, 234

\bibitem[{{Scholz} {et~al.}(2012)}]{ATel4268}
{Scholz}, R.-D. {et~al.} 2012, ATel, 4268

\bibitem[{{Schwarz} {et~al.}(2011){Schwarz}, {Ness}, {Osborne}, {Page},
  {Evans}, {Beardmore}, {Walter}, {Helton}, {Woodward}, {Bode}, {Starrfield},
  \& {Drake}}]{Schwarz2011}
{Schwarz}, G.~J., {Ness}, J.-U., {Osborne}, J.~P., {et~al.} 2011, \apjs, 197,
  31

\bibitem[{{Seach} {et~al.}(2010{\natexlab{a}})}]{CBET2140}
{Seach}, J. {et~al.} 2010{\natexlab{a}}, CBET, 2140

\bibitem[{{Seach} {et~al.}(2011{\natexlab{a}})}]{CBET2735}
{Seach}, J. {et~al.} 2011{\natexlab{a}}, CBET, 2735

\bibitem[{{Seach} {et~al.}(2011{\natexlab{b}})}]{IAUC9216}
{Seach}, J. {et~al.} 2011{\natexlab{b}}, IAUC, 9216

\bibitem[{{Seach} {et~al.}(2011{\natexlab{c}})}]{CBET2813}
{Seach}, J. {et~al.} 2011{\natexlab{c}}, CBET, 2813

\bibitem[{{Seach} {et~al.}(2012{\natexlab{a}})}]{CBET3040}
{Seach}, J. {et~al.} 2012{\natexlab{a}}, CBET, 3040

\bibitem[{{Seach} {et~al.}(2012{\natexlab{b}})}]{IAUC9251}
{Seach}, J. {et~al.} 2012{\natexlab{b}}, IAUC, 9251

\bibitem[{{Seach} {et~al.}(2012{\natexlab{c}})}]{CBET3073}
{Seach}, J. {et~al.} 2012{\natexlab{c}}, CBET, 3073

\bibitem[{{Seach} {et~al.}(2012{\natexlab{d}})}]{CBET3124}
{Seach}, J. {et~al.} 2012{\natexlab{d}}, CBET, 3124

\bibitem[{{Seach} {et~al.}(2013{\natexlab{a}})}]{CBET3732}
{Seach}, J. {et~al.} 2013{\natexlab{a}}, CBET, 3732

\bibitem[{{Seach} {et~al.}(2013{\natexlab{b}})}]{IAUC9265}
{Seach}, J. {et~al.} 2013{\natexlab{b}}, IAUC, 9265

\bibitem[{{Seach} {et~al.}(2015)}]{CBET4080}
{Seach}, J. {et~al.} 2015, CBET, 4080

\bibitem[{{Seach} {et~al.}(2010{\natexlab{b}})}]{IAUC9112}
{Seach}, S. {et~al.} 2010{\natexlab{b}}, IAUC, 9112

\bibitem[{{Senziani} {et~al.}(2008){Senziani}, {Skinner}, {Jean}, \&
  {Hernanz}}]{2008A&A...485..223S}
{Senziani}, F., {Skinner}, G.~K., {Jean}, P., \& {Hernanz}, M. 2008, \aap, 485,
  223

\bibitem[{{Shafter}(2017)}]{2017ApJ...834..196S}
{Shafter}, A.~W. 2017, \apj, 834, 196

\bibitem[{{Shappee} {et~al.}(2014){Shappee}, {Prieto}, {Stanek}, {Kochanek},
  {Holoien}, {Jencson}, {Basu}, {Beacom}, {Szczygiel}, {Pojmanski},
  {Brimacombe}, {Dubberley}, {Elphick}, {Foale}, {Hawkins}, {Mullins},
  {Rosing}, {Ross}, \& {Walker}}]{2014AAS...22323603S}
{Shappee}, B., {Prieto}, J., {Stanek}, K.~Z., {et~al.} 2014, in AAS Meeting
  Abstracts, Vol. 223, 236.03

\bibitem[{{Shappee} {et~al.}(2015)}]{ATel8126}
{Shappee}, B.~J. {et~al.} 2015, ATel, 8126

\bibitem[{{Shore}(2014)}]{ATel6413}
{Shore}, N.~S. 2014, ATel, 6413

\bibitem[{{Shore} {et~al.}(1994){Shore}, {Livio}, {van den Heuvel},
  {Nussbaumer}, \& {Orr}}]{1994inbi.conf.....S}
{Shore}, S.~N., {Livio}, M., {van den Heuvel}, E.~P.~J., {Nussbaumer}, H., \&
  {Orr}, A., eds. 1994, {Interacting binaries}

\bibitem[{{Sokoloski} {et~al.}(2013){Sokoloski}, {Crotts}, {Lawrence}, \&
  {Uthas}}]{2013ApJ...770L..33S}
{Sokoloski}, J.~L., {Crotts}, A.~P.~S., {Lawrence}, S., \& {Uthas}, H. 2013,
  \apjl, 770, L33

\bibitem[{{Srivastava} {et~al.}(2015){Srivastava}, {Ashok}, {Banerjee}, \&
  {Sand}}]{2015MNRAS.454.1297S}
{Srivastava}, M.~K., {Ashok}, N.~M., {Banerjee}, D.~P.~K., \& {Sand}, D. 2015,
  \mnras, 454, 1297

\bibitem[{{Sun} {et~al.}(2009)}]{IAUC9049}
{Sun}, G. {et~al.} 2009, IAUC, 9049

\bibitem[{{Tylenda} {et~al.}(2011){Tylenda}, {Hajduk}, {Kami{\'n}ski},
  {Udalski}, {Soszy{\'n}ski}, {Szyma{\'n}ski}, {Kubiak}, {Pietrzy{\'n}ski},
  {Poleski}, {Wyrzykowski}, \& {Ulaczyk}}]{2011A&A...528A.114T}
{Tylenda}, R., {Hajduk}, M., {Kami{\'n}ski}, T., {et~al.} 2011, \aap, 528, A114

\bibitem[{{Waagen} {et~al.}(2008)}]{IAUC8999}
{Waagen}, E.~O. {et~al.} 2008, IAUC, 8999

\bibitem[{{Waagen} {et~al.}(2011{\natexlab{a}})}]{CBET2700}
{Waagen}, E.~O. {et~al.} 2011{\natexlab{a}}, CBET, 2700

\bibitem[{{Waagen} {et~al.}(2011{\natexlab{b}})}]{IAUC9205}
{Waagen}, E.~O. {et~al.} 2011{\natexlab{b}}, IAUC, 9205

\bibitem[{{Waagen} {et~al.}(2014)}]{CBET3803}
{Waagen}, E.~O. {et~al.} 2014, CBET, 3803

\bibitem[{{Wagner} {et~al.}(2012)}]{CBET3136}
{Wagner}, R.~M. {et~al.} 2012, CBET, 3136

\bibitem[{{Walter}(2015)}]{ATel7060}
{Walter}, F. 2015, ATel, 7060

\bibitem[{{Walter} {et~al.}(2012){Walter}, {Battisti}, {Towers}, {Bond}, \&
  {Stringfellow}}]{2012PASP..124.1057W}
{Walter}, F.~M., {Battisti}, A., {Towers}, S.~E., {Bond}, H.~E., \&
  {Stringfellow}, G.~S. 2012, \pasp, 124, 1057

\bibitem[{{Weston} {et~al.}(2016{\natexlab{a}}){Weston}, {Sokoloski},
  {Chomiuk}, {Linford}, {Nelson}, {Mukai}, {Finzell}, {Mioduszewski}, {Rupen},
  \& {Walter}}]{2016MNRAS.460.2687W}
{Weston}, J.~H.~S., {Sokoloski}, J.~L., {Chomiuk}, L., {et~al.}
  2016{\natexlab{a}}, \mnras, 460, 2687

\bibitem[{{Weston} {et~al.}(2016{\natexlab{b}}){Weston}, {Sokoloski},
  {Metzger}, {Zheng}, {Chomiuk}, {Krauss}, {Linford}, {Nelson}, {Mioduszewski},
  {Rupen}, {Finzell}, \& {Mukai}}]{2016MNRAS.457..887W}
{Weston}, J.~H.~S., {Sokoloski}, J.~L., {Metzger}, B.~D., {et~al.}
  2016{\natexlab{b}}, \mnras, 457, 887

\bibitem[{{Wyrzykowski}(2012)}]{ATel4296}
{Wyrzykowski}, L. 2012, ATel, 4296

\bibitem[{{Yamaoka} {et~al.}(2008)}]{IAUC8989}
{Yamaoka}, H. {et~al.} 2008, IAUC, 8989

\bibitem[{{Yamaoka} {et~al.}(2009{\natexlab{a}})}]{CBET2050}
{Yamaoka}, H. {et~al.} 2009{\natexlab{a}}, CBET, 2050

\bibitem[{{Yamaoka} {et~al.}(2009{\natexlab{b}})}]{IAUC9098}
{Yamaoka}, H. {et~al.} 2009{\natexlab{b}}, IAUC, 9098

\bibitem[{{Yamaoka} {et~al.}(2010)}]{CBET2179}
{Yamaoka}, H. {et~al.} 2010, CBET, 2179

\bibitem[{{Yamaoka} {et~al.}(2012)}]{CBET3156}
{Yamaoka}, H. {et~al.} 2012, CBET, 3156

\end{thebibliography}

\end{document}